\documentclass{optica-article}

\journal{opticajournal} % for journals or Optica Open

\articletype{Research Article}

\usepackage{lineno} 
\usepackage{amsmath, dsfont}
\usepackage{bbm}
\usepackage[scr=boondox,  % heavily sloped
            cal=esstix]   % slightly sloped
           {mathalpha}

\DeclareMathAlphabet{\mathpzc}{OT1}{pzc}{m}{it}

\newcommand{\R}{\mathbb{R}}

\newcommand{\id}{\mathrm{d}}
\newcommand{\grad}{\nabla} % Gradient symbol
\newcommand{\laplace}{\bigtriangleup} % Laplacian
\usepackage{algorithm}
\usepackage{algorithmicx}
\usepackage{algpseudocode}
\usepackage[capitalise]{cleveref}
\usepackage{hyperref}

\usepackage{tikz}

\usepackage[export]{adjustbox}

\tikzset{>=latex} % for LaTeX arrow head
\usepackage{ifthen}
\usepackage{xcolor}
\usepackage{siunitx}

\definecolor{revision_nthupurple}{RGB}{127,15,133}
\definecolor{revision_black}{RGB}{0,0,0}

\usepackage[labelformat=simple]{subcaption}

\usepackage{booktabs}

\usepackage{multirow}

\begin{document}

\title{Breaking the Limitations with Sparse Inputs \\
by Variational Frameworks (BLIss) in Terahertz Super-Resolution 3D Reconstruction}

\vspace{-0.2cm}
\author{
Yiyao Zhang,\authormark{1,2,3} 
Ke Chen,\authormark{3,4} 
and 
Shang-Hua Yang\authormark{1,5,*}}

\address{
\authormark{1}Institute of Electronics Engineering, National Tsing Hua University, Hsinchu, 30013, Taiwan\\
\authormark{2}Department of Mathematical Sciences, University of Liverpool, Liverpool, L69 7ZL, UK\\
\authormark{3}Centre for Mathematical Imaging Techniques, University of Liverpool, Liverpool, L69 7ZL, UK\\
\authormark{4}Department of Mathematics and Statistics, University of Strathclyde, Glasgow, G1 1XH, UK\\
\authormark{5}Department of Electrical Engineering, National Tsing Hua University, Hsinchu, 30013, Taiwan
}

\email{\authormark{*}\href
{mailto:shanghua@ee.nthu.edu.tw}
{shanghua@ee.nthu.edu.tw}} %% email address is required; see note below about the corresponding author designation

% use {asbstract*} to suppress the copyright line. Copyright information will be added in production

% The abstract should be limited to approximately 100 words. 
% It should be an explicit summary of the paper that states the problem, the methods used, and the major results and conclusions. 
\vspace{-0.4cm}
\begin{abstract}Data acquisition, image processing, and image quality are the long-lasting issues for terahertz (THz) 3D reconstructed imaging. 
Existing methods are primarily designed for 2D scenarios, given the challenges associated with obtaining super-resolution (SR) data and the absence of an efficient SR 3D reconstruction framework in conventional computed tomography (CT). 
Here, we demonstrate BLIss, a new approach for THz SR 3D reconstruction with sparse 2D data input. 
BLIss seamlessly integrates conventional CT techniques and variational framework with the core of the adapted Euler-Elastica-based model. 
The quantitative 3D image evaluation metrics, including the standard deviation of Gaussian, mean curvatures, and the multi-scale structural similarity index measure (MS-SSIM), validate the superior smoothness and fidelity achieved with our variational framework approach compared with conventional THz CT modal. 
Beyond its contributions to advancing THz SR 3D reconstruction, BLIss demonstrates potential applicability in other imaging modalities, such as X-ray and MRI. 
This suggests extensive impacts on the broader field of imaging applications. 
\vspace{-0.5cm}
\end{abstract}
\vspace{-0.8cm}
%%%%%%%%%%%%%%%%%%%%%%%%%%  body  %%%%%%%%%%%%%%%%%%%%%%%%%%

\section{Introduction}
\label{sec:introduction}
\vspace{-0.2cm}
    Gaining increasing attention in recent years, terahertz (THz) imaging has emerged its worth as a versatile tool for various applications such as non-destructive inspection~\cite{Ok_2014_Food_Peanuts, Karaliunas_2018_Food_Oils, Li_2023_Food_PumpkinSeeds}, security screening~\cite{Cooper_2008_security, Cooper_2011_security, Cheng_2020_Security, Takida_2021_Security}, and non-contact testing~\cite{Bessou_2012_Testing_Bone, Jewariya_2013_Testing_Capsule, Krugener_2019_Testing_Wood, Tao_2020_Testing_Industry}. 
    The unique properties of THz radiation, including its low photon energy, make it suitable for safe use in bio-related scenarios~\cite{Mittleman_2018_THzImaging_Review, Leitenstorfer_2023_Review_Roadmap}. 
    Additionally, its capability to penetrate optically opaque materials allows extracting internal geometric and material information of tested objects~\cite{Hung_2022_CT_DL, Zhang_2022_CT_VM, Su_2023_IJCV_CT_DL, Su_2023_SPM_DL_CT, Zhang_2023_CLEO_CT_VM_EE}. 
    THz imaging includes a broad spectrum of modalities, with pulsed THz imaging emerging as one of the most prevalent and widely employed techniques in this field. 
    \textcolor{revision_black}{
    This is often associated with} a THz time-domain spectroscopy (THz-TDS) system~\cite{Hu_1995_TDS_Transmitted, Mittleman_1997_THzImaging_Reflected, Karpowicz_2005_Compare_TDS_FDI, Jansen_2010_THzImaging, Koch_2023_TDS_Review, Leitenstorfer_2023_Review_Roadmap}, enabling the acquisition of ultrafast responses within the domain of light-matter interactions across temporal, spatial, and spectral dimensions. 
    Recently, it has paved the way for a range of applications, including analysis of conformational behaviors~\cite{Markelz_2002_Conformational}, ultrafast molecular \textcolor{revision_black}{dynamics}~\cite{Nibali_2014_MolecularDynamics}, dielectric responses~\cite{Okada_2016_ElectrodynamicResponses}, and electrical properties~\cite{Shiraga_2016_DielectricConstant}. 

    Leveraging these advantages, pulsed THz imaging finds practical use in visualizing the internal structure of 3D objects. 
    Among the frequently employed pulsed THz imaging techniques, computed tomography (CT) represents an effective computational \textcolor{revision_black}{method}.
    This technique provides cross-sectional images (slices) reconstructed from a sequence of projected images taken at various projection angles by inverse Radon transform $\hat{\mathcal{R}}$ (IRT)~\cite{Herman_2009_CT, Jewariya_2013_Testing_Capsule}. 
    While X-ray CT has found extensive use in various real-world applications, its ionizing characteristics and high penetration capacity inherently restrict its utility, particularly in the visualization of composite materials, soft dielectrics, or liquids. 
    Furthermore, the limited differentiation of material responses within the X-ray spectrum poses challenges in identifying the chemical components present within objects~\cite{Brenner_2007_CT_XRay, Jewariya_2013_Testing_Capsule}. 
    In light of these limitations, the introduction of pulsed THz CT imaging, first proposed in 2002~\cite{Ferguson_2002_3DCT_first}, emerged as a promising supplement to traditional X-ray CT. 

    \textcolor{revision_black}{While pulsed THz CT imaging exhibits considerable potential as a supplement with non-ionizing characteristics and sensitive material differentiation, two primary concerns are (i) the protracted process of data acquisition with limited scanning resolution by THz-TDS; and (ii) Fraunhofer diffraction during the THz wave propagation through the object~\cite{Ferguson_2002_3DCT_first, Recur_2011_3DCT_Investigation}.}
    For clarification, the term \textit{resolution} typically refers to the number of pixels in digital imaging rather than diffraction-limited resolution. 
    This is because all datasets are collected by THz-TDS and processed under the pixel-based scanning resolution. 
    Concerning (i), THz-TDS has historically relied on mechanical time-delay scanning. 
    However, non-mechanical time-domain sampling methods, such as asynchronous optical sampling (ASOPS) introduced in 2005~\cite{Yasui_2005_ASOPS} and electronically controlled optical sampling (ECOPS) introduced in 2010~\cite{Kim_2010_ECOPS}, \textcolor{revision_black}{bring} a promising alternative. 
    These techniques can significantly reduce the data acquisition time for each single-cycle THz pulse, shifting from hour/minute scales to milliseconds. 
    Despite advancements, the implementation of ASOPS and ECOPS requires costly or bulky femtosecond lasers. 
    \textcolor{revision_black}{
    Moreover, projection data acquisition for THz 3D CT imaging still involves mechanical operations, such as raster scanning and object rotation, which constrain the resolution of projection image due to scanning resolution, impacting subsequent 3D reconstruction.}
    Recently, a high-throughput alternative is the use of terahertz focal-plane arrays (THz-FPAs)~\cite{Li_2023_FPA_Review}. 
    These detector arrays, designed based on field-effect transistors~\cite{Hadi_2012_FPA_FieldEffect}, microbolometers~\cite{Nemoto_2016_FPA_Microbolometer}, and plasmonic nanoantennas~\cite{Yardimci_2017_FPA_PlasmonicNanoantenna, Li_2023_FPA_PlasmonicNanoantennas}, offer the potential to bypass mechanical operations. 
    Nevertheless, the shift towards employing THz-FPAs introduces a unique set of challenges. 
    These arrays require scalable high-speed electronics, a considerable increase in computational power, and increased system complexity, all of which depend on the number of THz detectors in use, and notably, this transition could be cost-intensive. 
    Besides, while THz-FPAs can speed up data acquisition, the quality of reconstructed 3D images would \textcolor{revision_black}{be} affected by the uniformity of each device and, most importantly, the concern (ii) related to diffraction effects with noise. 

    Addressing these issues without costly upgrades to the imaging system, algorithm-based methods show promise in directly enhancing low-resolution (LR) images into super-resolution (SR) images, tackling concerns (i) and (ii). 
    For instance, by using prior knowledge of the degradation based on the specific point spread function (PSF) measured~\cite{Li_2010_PSF_Measured, Ding_2010_PSF_Measured, Popescu_2010_PSF_Measured} or simulated~\cite{Ahi_2017_PSF_Simulated, Ahi_2018_PSF_Simulated, Wong_2019_PSF_Simulated} from the THz imaging system, a variety of SR algorithms have been explored, including Richardson-Lucy algorithm~\cite{Li_2008_SR} and PSF deconvolution~\cite{Ahi_2019_PSF}. 
    These methods have demonstrated success in improving spatial resolution and reducing the diffractive limit, although challenges in rendering image edge and texture details remain. 
    Nowadays, deep-learning-based SR methods have been increasingly recognized as a prominent approach. 
    By establishing an end-to-end mapping correlation between LR and HR images, they have shown advantages in comprehensive learning ability and the overall effectiveness of image reconstruction after training. 
    Examples include the convolutional neural network (CNN) with its variants, the very deep super-resolution (VDSR) method, the super-resolution network for multiple degradations (SRMD), and generative adversarial networks (GAN)~\cite{Kim_2016_SR_VDCNN_CVPR, Li_2017_DL_CNN_THz, Zhang_2018_SRCNN_DL_CVPR, Long_2019_SR_CNN_THz_DL, Wan_2019_SR_GAN_THz_DL, Mao_2020_CNN_DL, Wang_2021_CCNN_THz, Hung_2022_CT_DL, Yang_2022_DL_CNN, Su_2023_SPM_DL_CT, Su_2023_IJCV_CT_DL}. 
    However, the application of deep-learning-based methods in THz single-image super-resolution (SISR) faces several challenges, including the complex THz image degradation process, limited efficiency in deblurring and denoising procedures, and the issue of gradient vanishing during deep network training. 
    \textcolor{revision_black}{
    In a nutshell, both algorithm-based and deep-learning-based methods demonstrate their potential in improving the spatial resolution of THz imaging, but their primary focus on 2D image processing often overlooks the global characteristics of the 3D post-reconstruction. 
    Moreover, the need for sufficient datasets and costly workstations with GPUs for model training presents another challenge.}

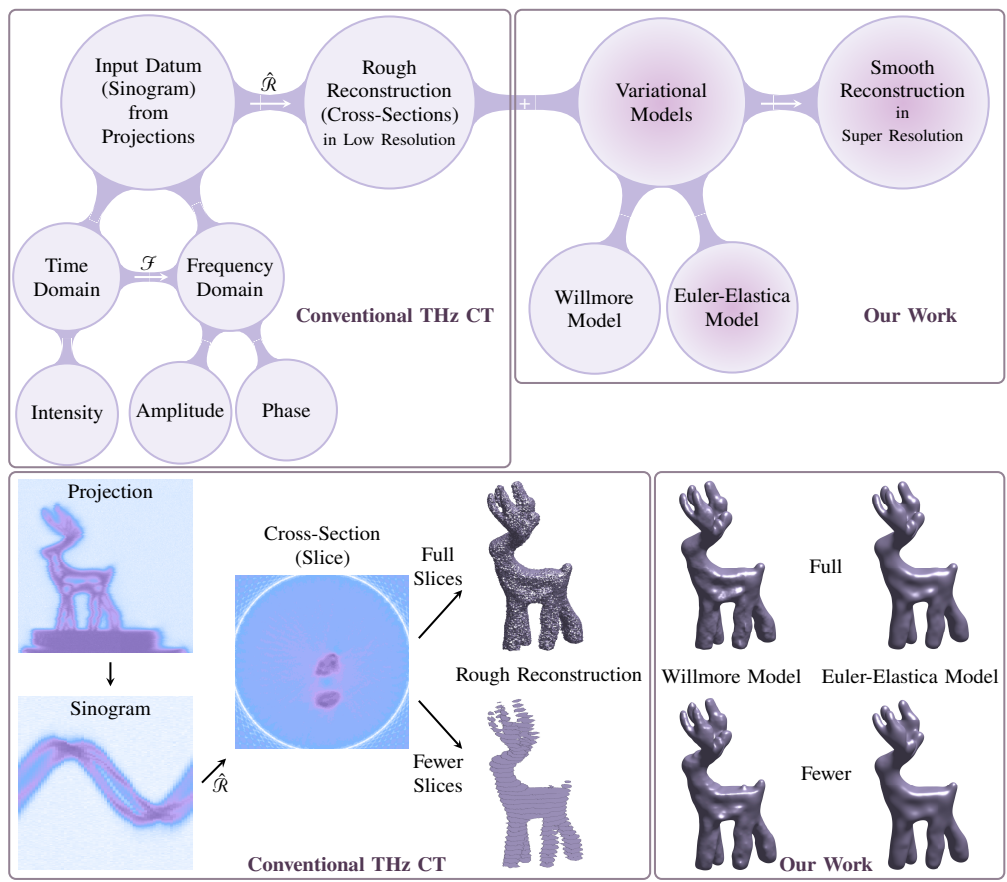
\begin{figure}[htbp]
    % \vspace{-0.35cm}
     \centering
        \begin{adjustbox}{width=1.15\textwidth}
        \hspace{-1.4cm}
        \usetikzlibrary{mindmap}

\definecolor{oppurple}{RGB}{180,167,214} % optica-purple
\definecolor{nthupurple}{RGB}{127,15,133} % nthupurple

% !TIKZEDT BOUNDINGBOX = -3 -3 8 5
\begin{tikzpicture}
% node 1.1
[ mindmap, concept color = oppurple!80, 
  minimum size = 1.5cm, font = \large, 
  text width = 2cm]
\node (left) [concept, 
        	     outer color = oppurple!20, 
        	     inner color = oppurple!20]
           {Time Domain}
% node 1.1.1
        child [ level distance = 2.85cm, concept color = oppurple!80, 
                   grow = -90,minimum size = 1.5cm, 
                   font = \large, text width = 2cm]
        { node (right) [concept, 
        	     outer color = oppurple!20, 
        	     inner color = oppurple!20]
                   {Intensity}}
% node 1
child [ level distance = 4cm, concept color = oppurple!80, 
           grow = 65, minimum size = 3.cm, 
           font = \large, text width = 3cm] 
        {  
        node [concept, 
        	     outer color = oppurple!20, 
        	     inner color = oppurple!20] 
                 {
                 Input Datum \\ 
                 (Sinogram) \\
                 from \\
                 Projections
                 } 
% node 1.2
        child [ level distance = 4cm, concept color = oppurple!80, 
                   grow = -65,minimum size = 1.5cm, 
                   font = \large, text width = 2cm]
        { node (right) [concept, 
        	     outer color = oppurple!20, 
        	     inner color = oppurple!20] 
                   {Frequency Domain}
                   % node 1.2.1
        		child [ level distance = 3cm, concept color = oppurple!80, 
                  		    grow = -110,minimum size = 1.5cm, 
                               font = \large, text width = 2cm]
        		{ node [concept, 
        	     outer color = oppurple!20, 
        	     inner color = oppurple!20] 
                 		      {Amplitude}}  
                 % node 1.2.2
        		child [ level distance = 3cm, concept color = oppurple!80, 
                  		    grow = -67.5,minimum size = 1.5cm, 
                               font = \large, text width = 2cm]
        		{ node [concept, 
        	     outer color = oppurple!20, 
        	     inner color = oppurple!20] 
                 		      {Phase}}
                   }
        % node 2
        child [level distance = 5cm, 
        	    concept color = oppurple!80, 
                  grow = 0,minimum size = 3.5cm, 
                  font = \large, text width = 3cm]
        { node[concept, outer color = oppurple!20, inner color = oppurple!20] 
                   {Rough \\ 
                   Reconstruction \\ 
                   (Cross-Sections) \\ \normalsize{in Low Resolution}}
        % node 3
        child [level distance = 5.65cm, 
                  concept color = oppurple!80, 
                  grow = 0, minimum size = 3.5cm, 
                  font = \large, text width = 3.cm]
        { node[concept, outer color = oppurple!20, inner color = nthupurple!30] 
                   {Variational Models} 
        % node 3.1
        child [ level distance = 4.5cm, concept color = oppurple!80, 
                   grow = -108.5, minimum size = 1.5cm, 
                   font = \large, text width = 2.5cm]
        { node[concept, outer color = oppurple!20, inner color = oppurple!20] 
                   {Willmore \\ Model} 
        } 
        % node 3.2
        child [ level distance = 4.5cm, concept color = oppurple!80, 
                   grow = -71.5, minimum size = 1.5cm, 
                   font = \large, text width = 2.5cm]
        { node[concept, outer color = oppurple!20, inner color = nthupurple!30] 
                   {Euler-Elastica Model} 
        } 
        % node 4
        child [ level distance = 5cm, concept color = oppurple!80, 
                   grow = 0, minimum size = 3.5cm, 
                   font = \large, text width = 3cm]
        { node[concept, outer color = oppurple!20, inner color = nthupurple!30] 
                   {Smooth Reconstruction \\\normalsize{in \\Super Resolution}} 
        } 
        } 
        }
        };
        
       \path 
       (left) 
       to 
       [circle connection bar switch color=from (oppurple!80) to (oppurple!80)] 
       (right);
       
       \draw [-stealth, white] (1.4,0) -- (2.1,0);
	\node at (1.7,0.3) {$\mathcal{F}$};
	
	\draw [-stealth, white] (3.8,3.6) -- (4.65,3.6);
	\node at (4.2,4.05) {$\hat{\mathcal{R}}$};
	
	% \node[text width = 20cm] at (11,-3.) {IRT: Inverse Radon Transform, FT: Fourier Transform};

    \node[white] at (9.5,3.6) {$+$};

    \draw [-stealth, white] (14.4,3.6) -- (15.25,3.6); % +0.85
    
    \draw[nthupurple!30!black!50, rounded corners] (-1.2,-3.95) rectangle (9.20,5.55);
	
    \node[nthupurple!30!black!80, text width = 10cm] at (6.8,-0.8) {\bf{Conventional THz CT}};

    \draw[nthupurple!30!black!50, rounded corners] (9.30,5.55) rectangle (19.45,-2.2);
    \node[nthupurple!30!black!80, text width = 10cm] at (17.5,-0.8) {\bf{Our Work}};

    \node[xshift=-10.8cm, yshift=-6cm] at (current page.south)
    {\includegraphics[width = 1.8\textwidth]{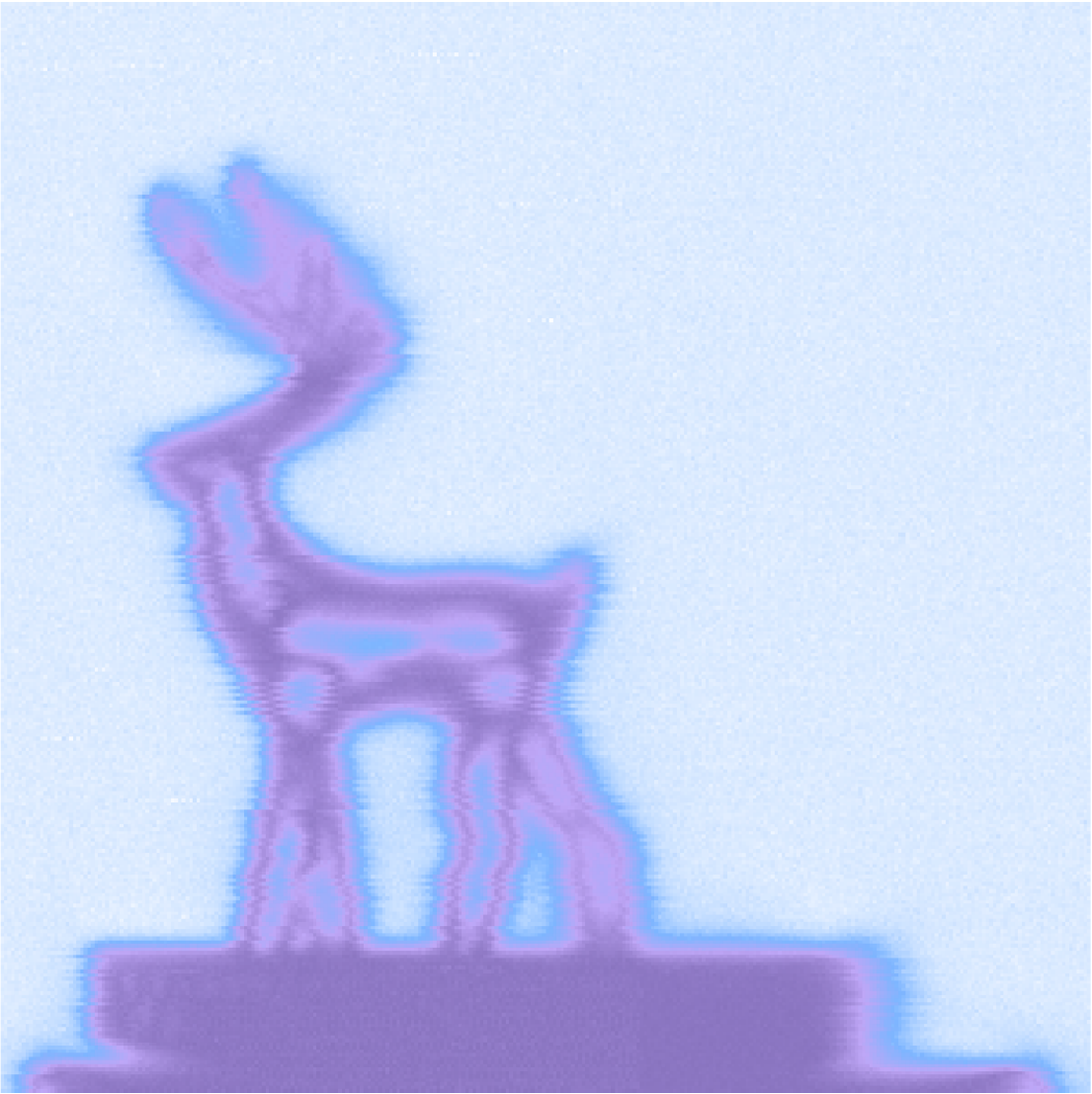}};
    \node at (0.9,-4.5) {Projection};

    \draw [-stealth] 
    (0.9,-8) -- (0.9,-8.5);
    
    \node[xshift=-10.8cm, yshift=-10.5cm] at (current page.south)
    {\includegraphics[width = 1.8\textwidth]{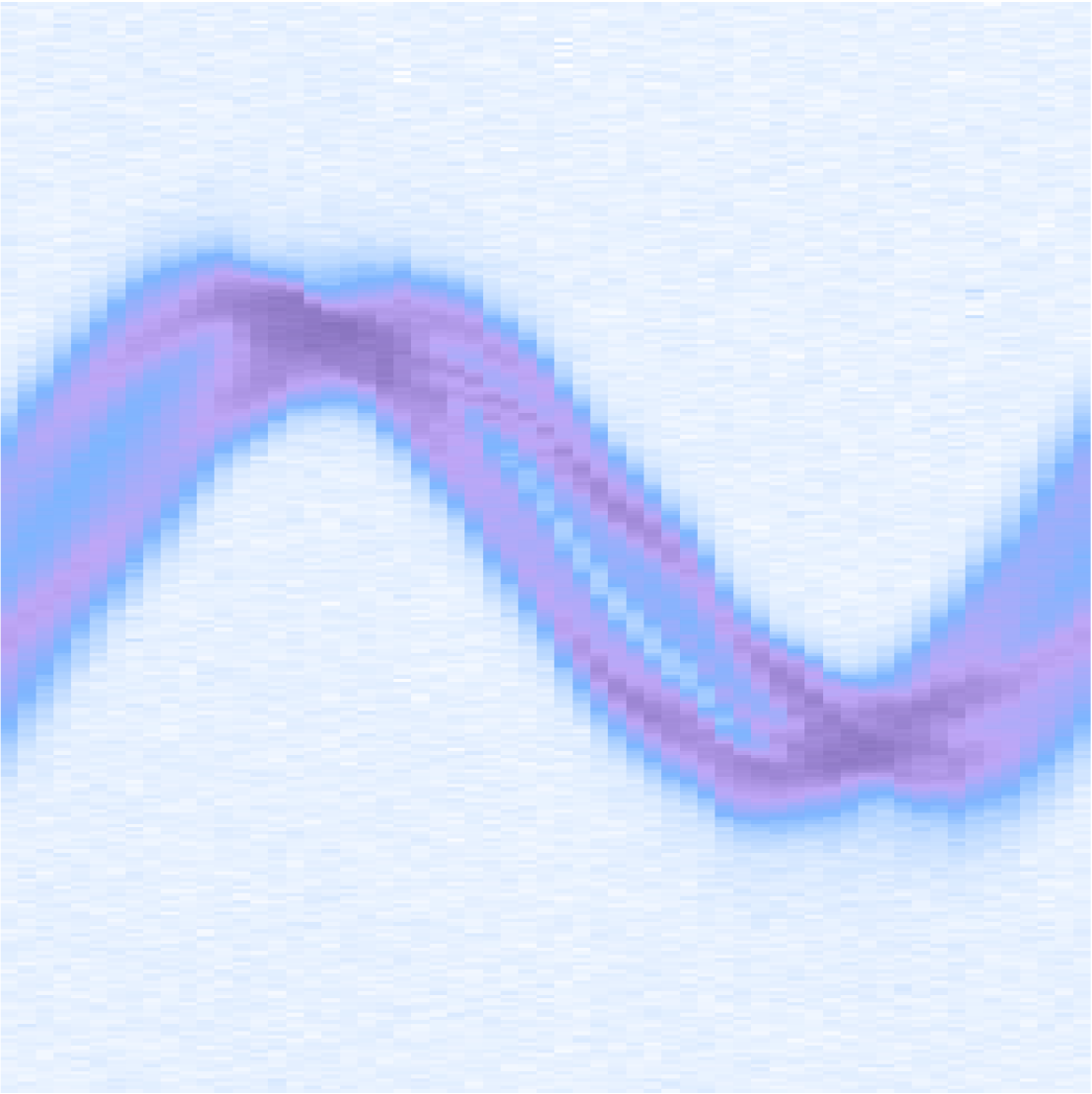}};
    \node at (0.9,-9) {Sinogram};

    \draw [-stealth] 
    (2.8,-10.5) -- (3.3,-10);
    \node at (3.2,-10.5) {$\hat{\mathcal{R}}$};

    \node[xshift=-6.3cm, yshift=-8cm] at (current page.south)
    {\includegraphics[width = 1.8\textwidth]{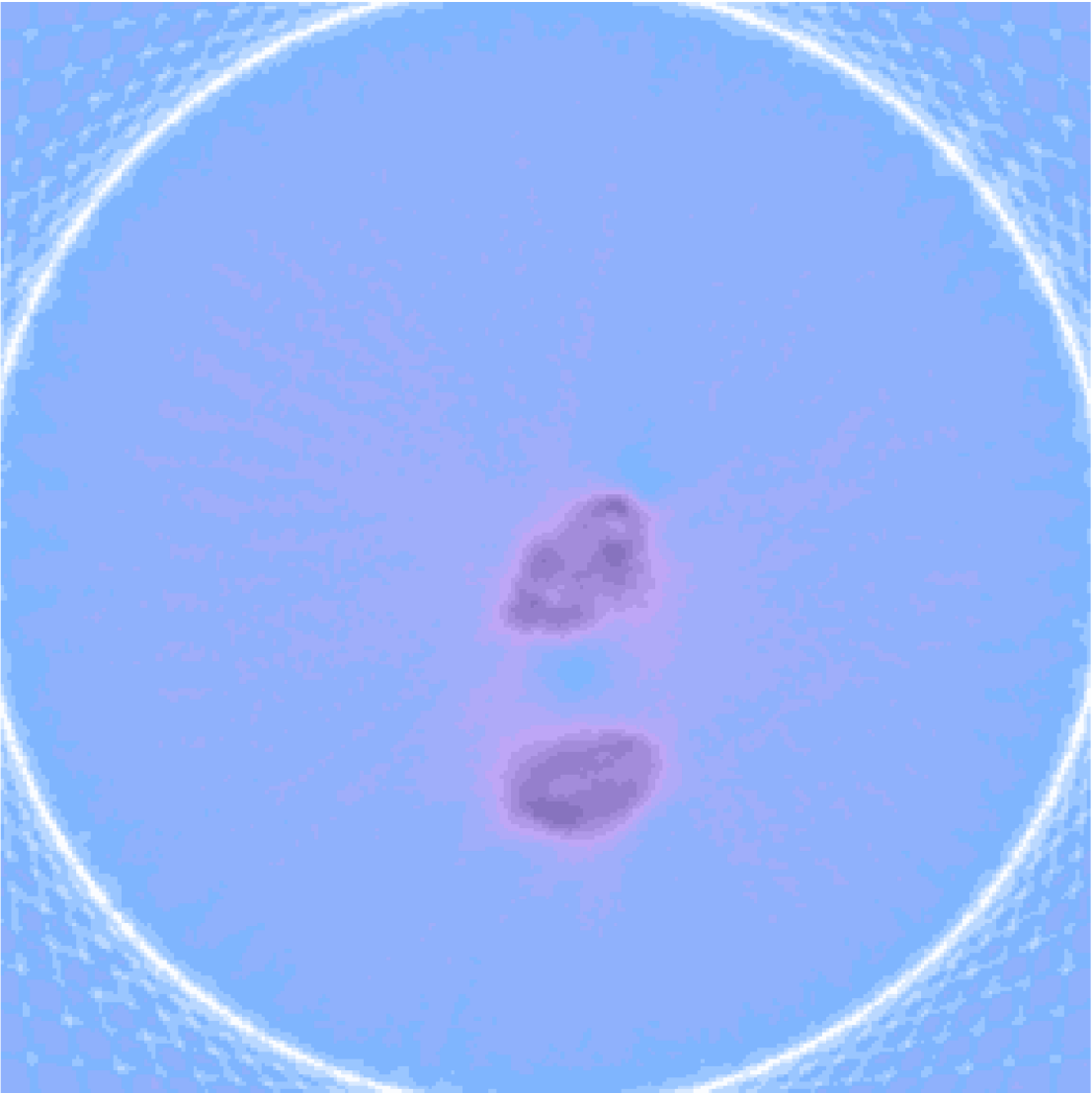}};
    \node[text width = 3cm] at (5.3,-5.6) {Cross-Section \\ (Slice)};

    \draw [-stealth] 
    (7.3,-7.5) -- (8.2,-6.5);
    \node[text width = 3cm] at (7.7,-6.0) {Full \\ Slices};

    \node[xshift=-1.5cm, yshift=-6cm] at (current page.south)
    {\adjincludegraphics[width= 1.6\textwidth, trim={{.2\width} {.18\height} {.3\width} {.15\height}}, clip]{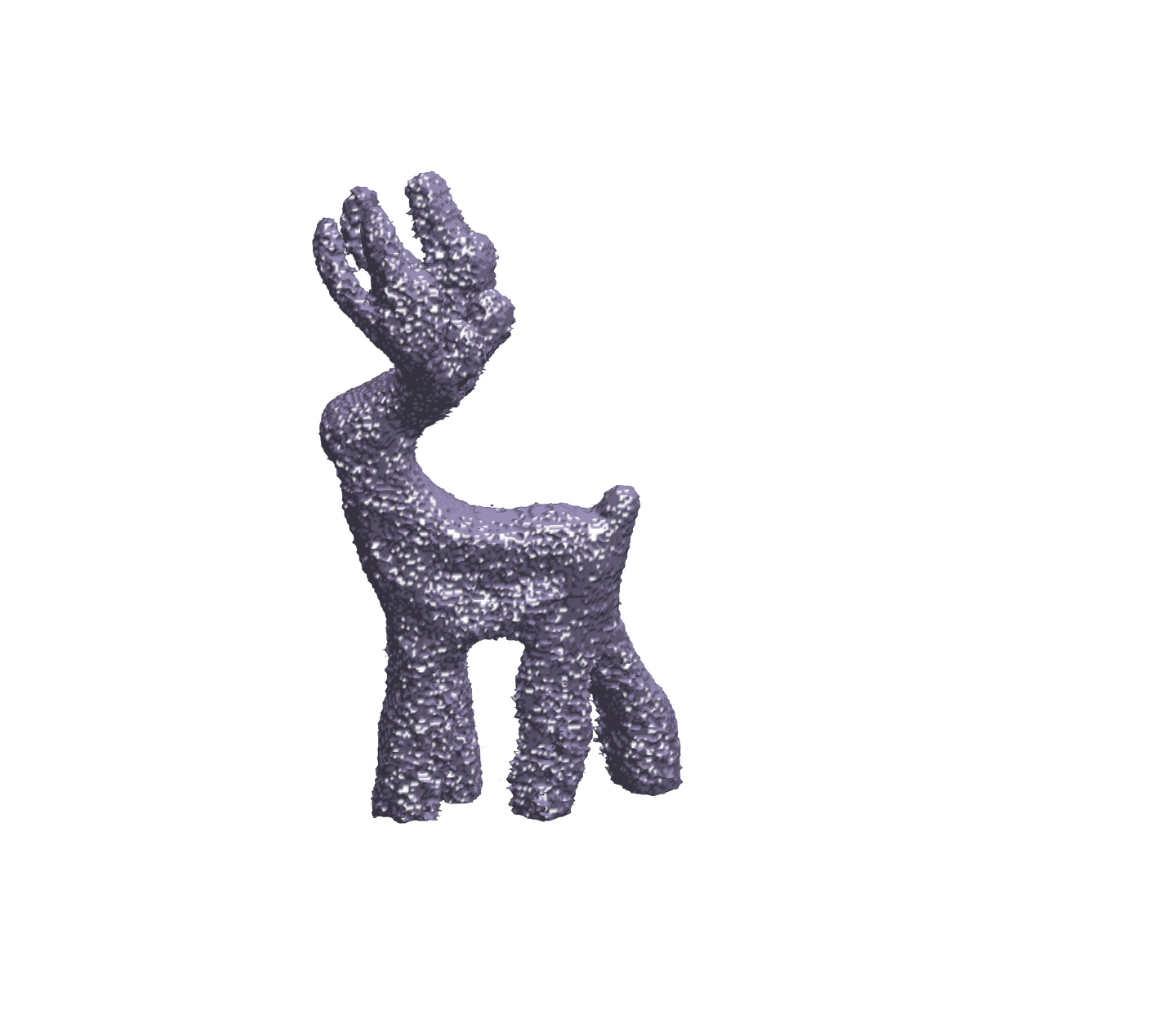}};

    \node[xshift=2.5cm, yshift=-6cm] at (current page.south)
    {\adjincludegraphics[width= 1.6\textwidth, trim={{.2\width} {.15\height} {.3\width} {.18\height}}, clip]{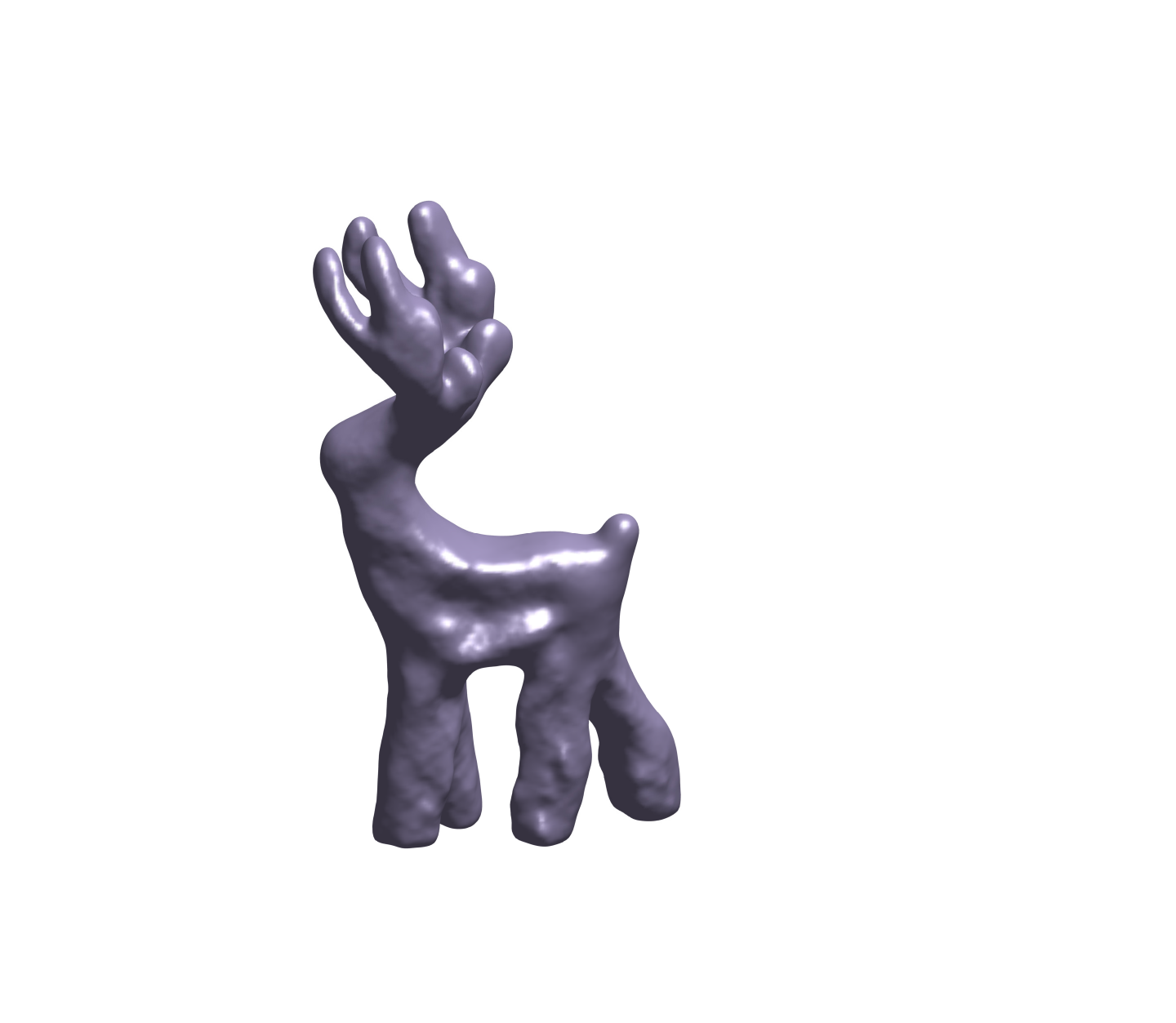}};

    \node[xshift=6.5cm, yshift=-6cm] at (current page.south)
    {\adjincludegraphics[width= 1.6\textwidth, trim={{.2\width} {.15\height} {.3\width} {.18\height}}, clip]{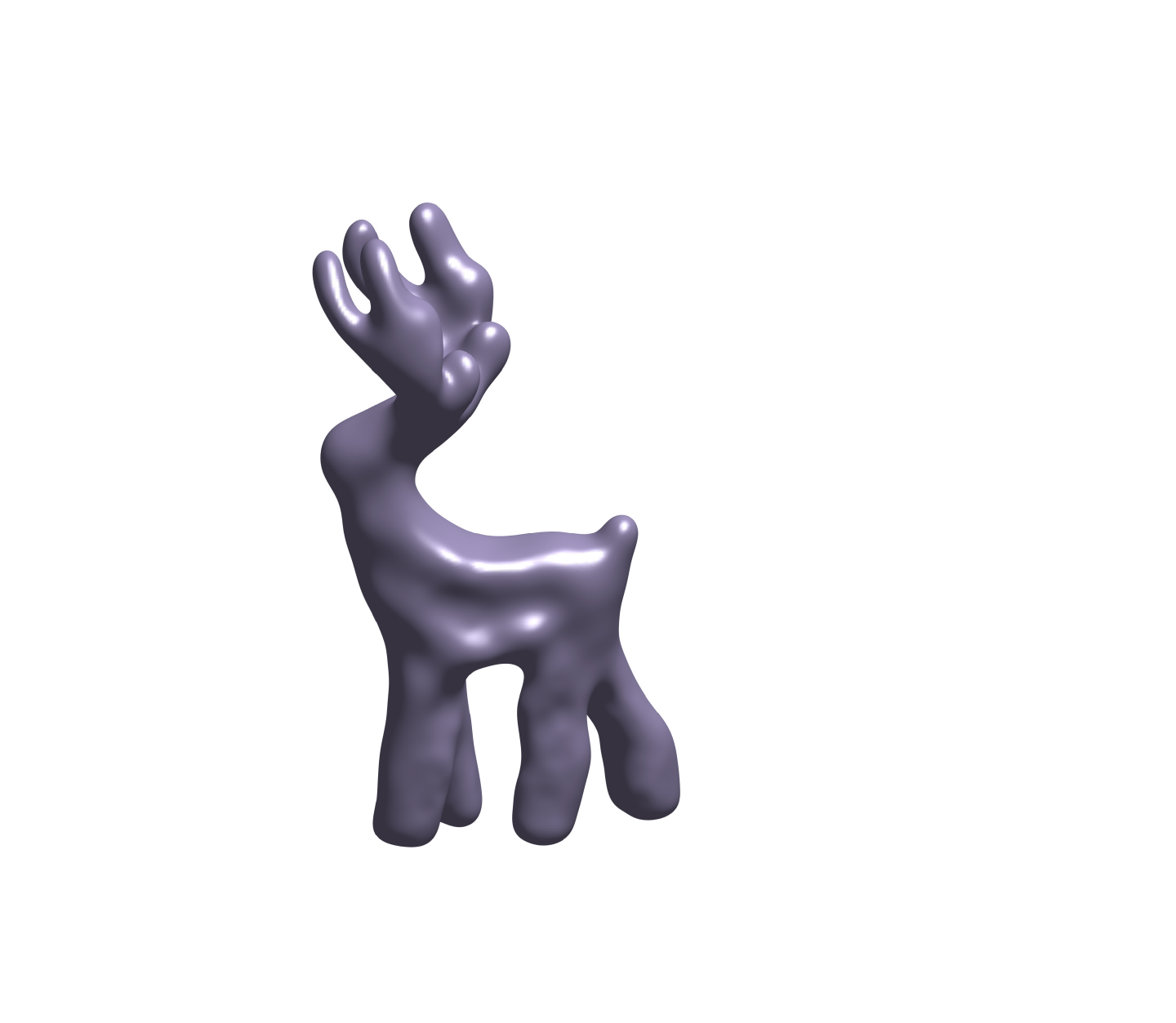}};

    \draw [-stealth] 
    (7.3,-8.8) -- (8.2,-9.8);
    \node[text width = 3cm] at (7.7,-10.3) {Fewer \\ Slices};

    \node[xshift=-1.5cm, yshift=-10.5cm] at (current page.south)
    {\adjincludegraphics[width= 1.6\textwidth, trim={{.2\width} {.18\height} {.3\width} {.15\height}}, clip]{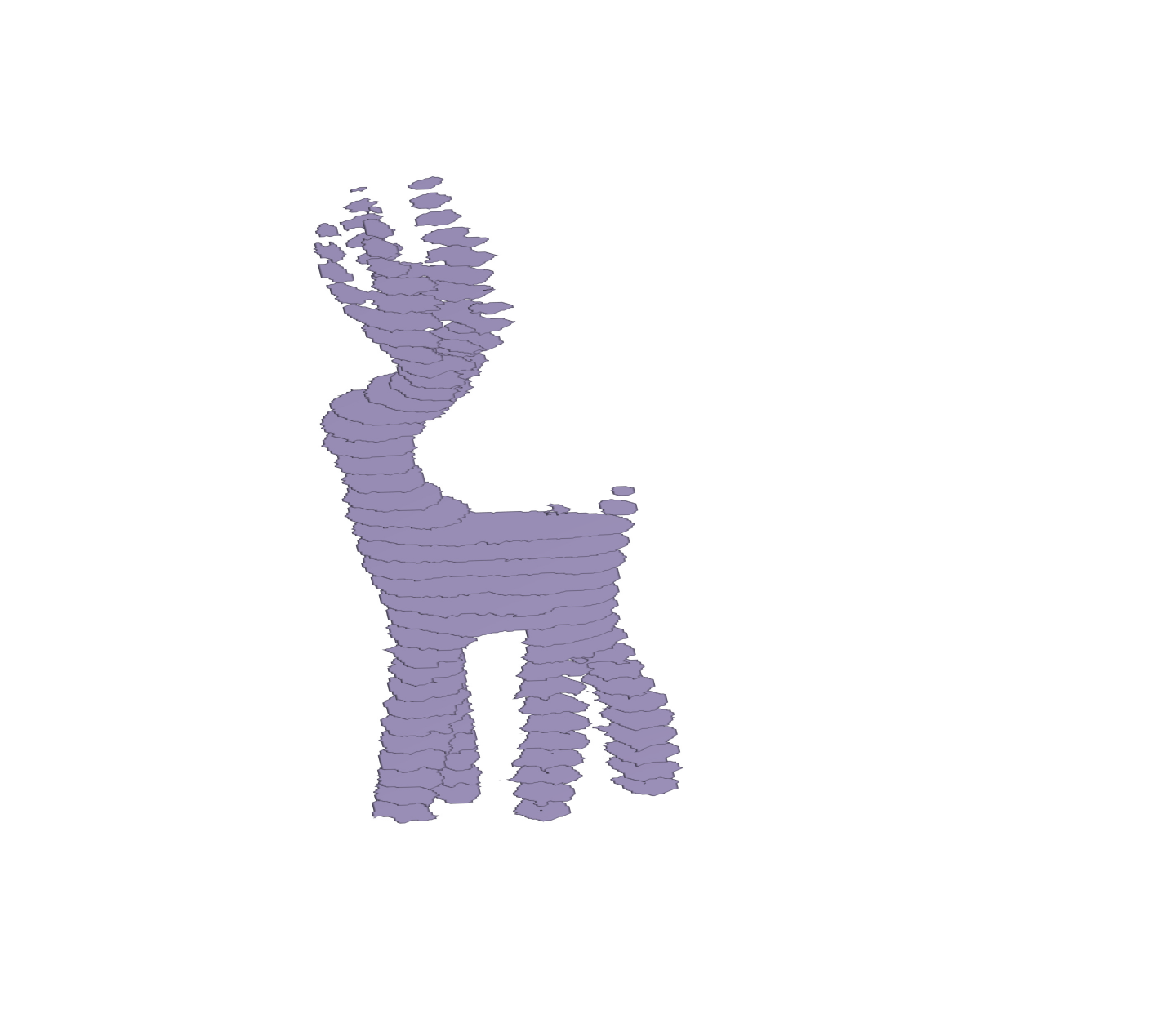}};

    \node[xshift=2.5cm, yshift=-10.5cm] at (current page.south)
    {\adjincludegraphics[width= 1.6\textwidth, trim={{.2\width} {.15\height} {.3\width} {.18\height}}, clip]{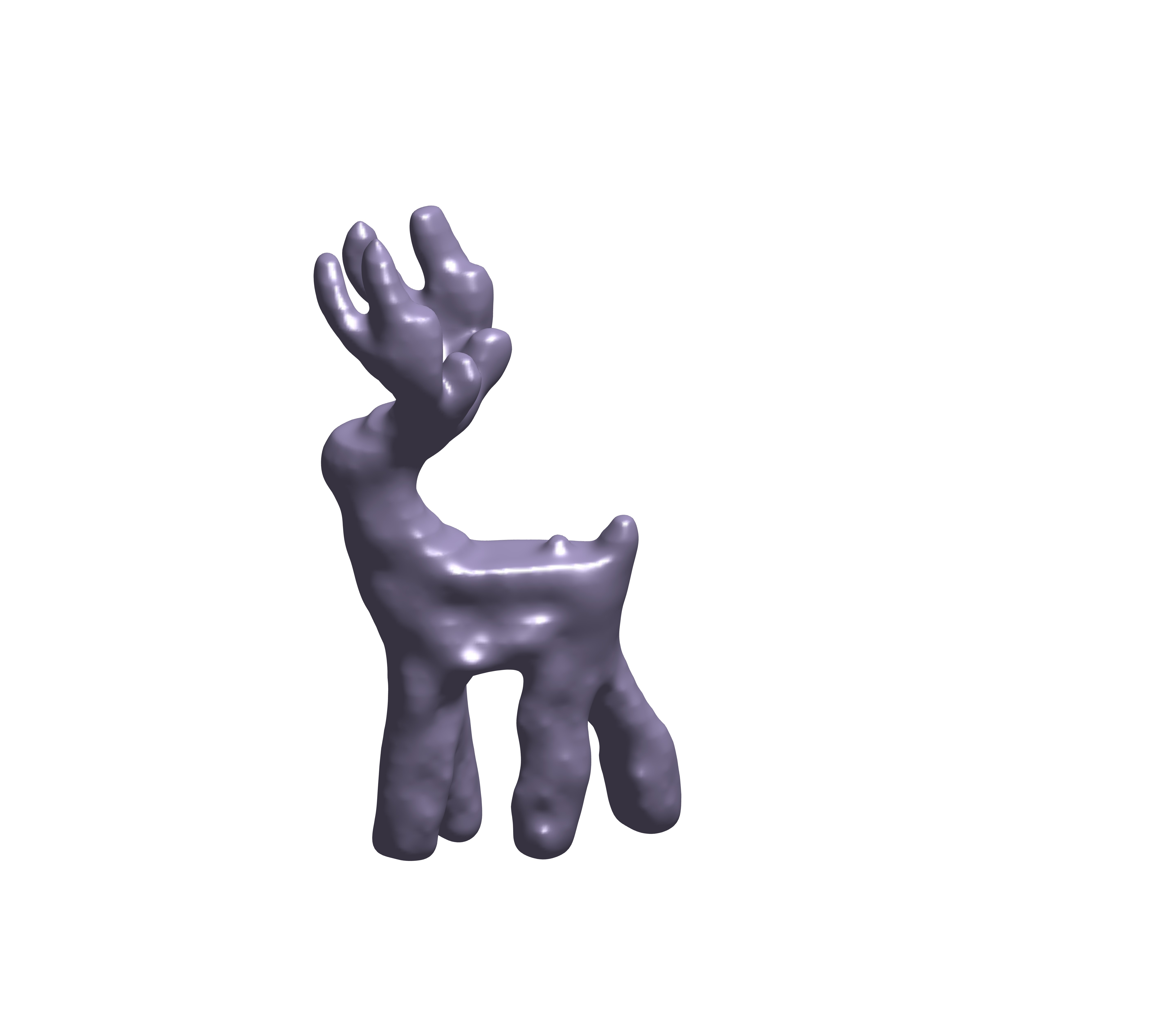}};

    \node[xshift=6.5cm, yshift=-10.5cm] at (current page.south)
    {\adjincludegraphics[width= 1.6\textwidth, trim={{.2\width} {.15\height} {.3\width} {.18\height}}, clip]{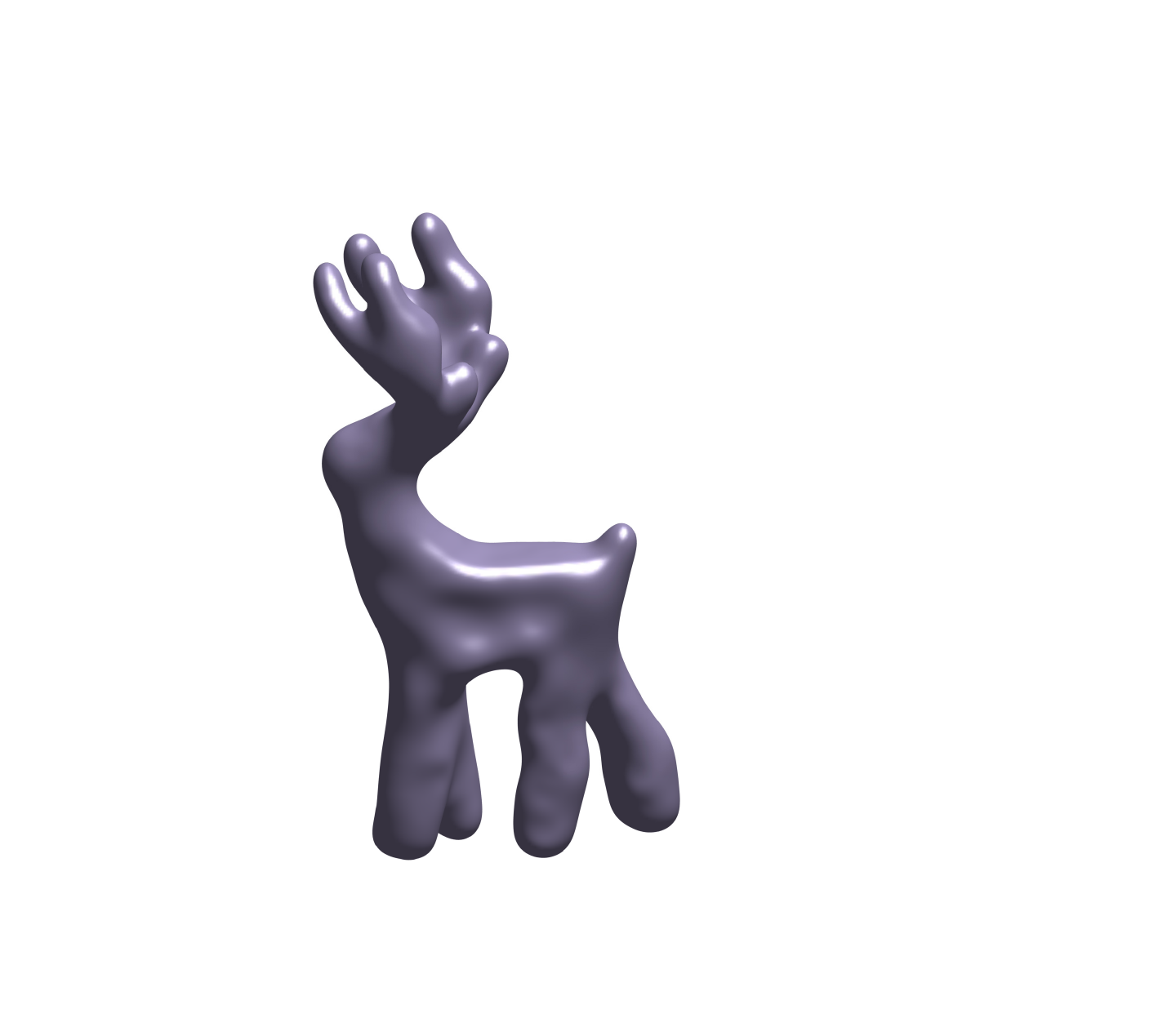}};

    \node[text width = 10cm] at (10,-8.3) {Rough Reconstruction};

    \node[text width = 10cm] at (13.8,-8.3) {Willmore Model};

    \node[text width = 10cm] at (17.5,-8.3) {Euler-Elastica Model};

    \draw[nthupurple!30!black!50, rounded corners] (-1.2,-4.05) rectangle (12.1,-12.5);
	
	\node[nthupurple!30!black!80, text width = 10cm] at (5.8,-12.2) {\bf{Conventional THz CT}};

    \draw[nthupurple!30!black!50, rounded corners] (12.2,-4.05) rectangle (19.45,-12.5);
	
	\node[nthupurple!30!black!80, text width = 10cm] at (15.75,-12.2) {\bf{Our Work}};

    \node[text width = 3cm] at (15.75,-6.0) {Full};

    \node[text width = 3cm] at (15.75,-10.3) {Fewer};
 
\end{tikzpicture}
        \end{adjustbox}
        % \vspace{-0.9cm}
        \caption{Schematic illustration (top) and visualization (bottom) of BLIss in our THz SR 3D reconstruction: conventional methodologies are integrated with proposed variational models to improve quality and resolution (highlighted nodes within the inner color). }
        \label{p20230412_Framework}
    \end{figure}

    \textcolor{revision_black}{In this work, we focus on improving 3D THz reconstruction by integrating variational frameworks with conventional CT methods. 
    Here, we introduce BLIss (as delineated in~\cref{p20230412_Framework}) with the aim of facilitating rapid and smooth SR 3D reconstruction, even when working with a limited pool of LR images. 
    In detail, this framework integrates traditional CT techniques, including signal processing (e.g., projections, sinograms, and slices by IRT $\hat{\mathcal{R}}$), with the underpinning mathematical principles of variational formulations, i.e., Willmore-based formulation and Euler-Elastica-based formulation.}
    To assess the quality and fidelity of the reconstructed results by our BLIss, the conventional evaluation metrics, e.g. mean-squared error (MSE), peak signal-to-noise ratio (PSNR), and structural similarity index (SSIM), which are commonly employed in 2D image assessment, \textcolor{revision_black}{are not sufficient} due to the inconsistency of data type in 3D. 
    To address this, we introduce an approach that quantifies the global smoothness stability of 3D surface quality using concepts from discrete geometry. 
    This approach involves calculating the standard deviation of Gaussian and mean curvatures for the triangular meshes of the reconstructed surfaces. 
    In addition, we evaluate the fidelity of reconstruction when using limited input data. 
    To achieve this, we extend our analysis to incorporate the calculation of the multi-scale structural similarity index measure (MS-SSIM), a 3D structural similarity metric that builds upon the foundation of the 2D SSIM. 
    Our proposed variational framework effectively addresses our concerns (i) and (ii), presenting a pathway to rapid and high-precision 3D THz imaging. 
    This advancement holds significant promise for non-destructive sensing and inspection applications. 

\section{Methods}
\label{sec:methods}

\subsection{Experimental Setup for Data Acquisition}
\label{subsec:setup}

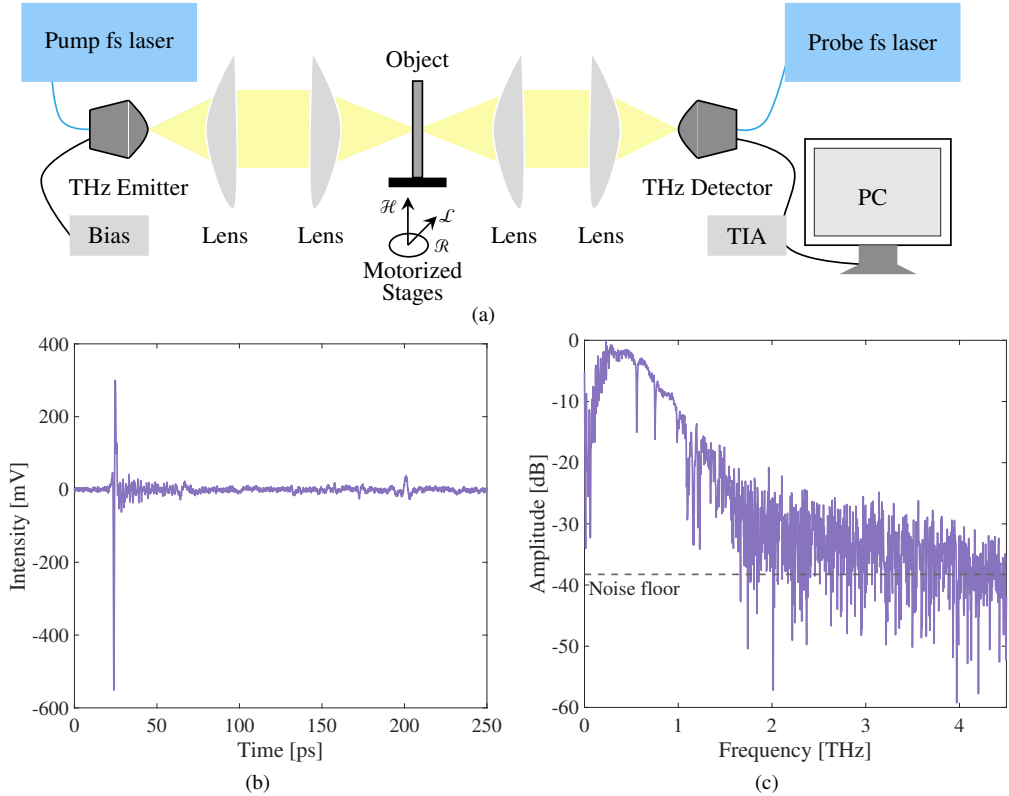
\begin{figure}[htbp]
        \begin{subfigure}[b]{0.95\textwidth}
        \begin{adjustbox}{width=\textwidth}
        \hspace{-0.5cm}
        % \input{0_Figures/p20230725_T_Setup.tex}
        % 20220525_THz Setup for Transmission

\definecolor{nottblue}{RGB}{123, 192, 249} % cyan
\definecolor{oppurple}{RGB}{180,167,214} % optica-purple
\definecolor{limegreen}{RGB}{90, 186, 182} % limegreen
\definecolor{limeblue}{RGB}{0, 145, 222} % limeblue
\definecolor{nthupurple}{RGB}{127,15,133}
\definecolor{lightyellow}{RGB}{249, 248, 113} % lightyellow

\begin{tikzpicture}[scale=.8]
\tikzstyle{every node}=[scale=.8]

% Cable - Detector to PC
\draw  plot[smooth, tension=.7] coordinates {(-3.2091,3.5) (-2.6245,3.2143) (-2.7,2.4) (-1.8,2.2)};

% Fiber - Laser to Emitter
\draw[limeblue!80]  plot[smooth, tension=.7] coordinates {(-10.3,4.2) (-10.2,3.7) (-9.9,3.6)};
% Fiber - Laser to Detector
\draw[limeblue!80]  plot[smooth, tension=.7] coordinates {(-2.6619,4.4091) (-2.9,3.7) (-3.2091,3.6)};

% Beam
\fill[lightyellow!60] (-9.3,3.6) -- (-8.5,3.2) -- (-8.5,4) ;
\fill[lightyellow!60] (-8.5,4) -- (-8.5,3.2) -- (-7.5,3.2) -- (-7.5,4) ;

\fill[lightyellow!60] (-6.5091,3.6) -- (-5.5293,3.2) -- (-5.5293,4) ;
\fill[lightyellow!60] (-6.5091,3.6) -- (-7.5,3.2) -- (-7.5,4) ;

\fill[lightyellow!60] (-5.5293,4) -- (-5.5293,3.2) -- (-4.6091,3.2) -- (-4.6091,4) ;
\fill[lightyellow!60] (-3.8091,3.6) -- (-4.6091,3.2) -- (-4.6091,4) ;

% Emitter
\fill[gray!90] (-9.9,3.8) -- (-9.9,3.4) -- (-9.5,3.3) -- (-9.5,3.9) plot[smooth, tension=.7] coordinates {(-9.5,3.9) (-9.3,3.6) (-9.5,3.3)};
\draw plot[smooth, tension=.7, double] coordinates {(-9.9,3.8) (-9.9,3.4)};
\draw plot[smooth, tension=.7] coordinates {(-9.9,3.8) (-9.5,3.9)};
\draw plot[smooth, tension=.7] coordinates {(-9.9,3.4) (-9.5,3.3)};
\draw plot[smooth, tension=.7] coordinates {(-9.5,3.9) (-9.3,3.6) (-9.5,3.3)};

% Lens 1
\fill[gray!30] plot[smooth, tension=.7] coordinates {(-8.4,4.4) (-8.7,3.6) (-8.4,2.8) (-8.4,4.4)} ;
\draw[white] plot[smooth, tension=.7] coordinates {(-8.4,4.4) (-8.7,3.6) (-8.4,2.8) (-8.4,4.4)} ;

% Lens 2
\fill[gray!30] plot[smooth, tension=.7] coordinates {(-7.6,4.4) (-7.3,3.6) (-7.6,2.8) (-7.6,4.4)};
\draw[white] plot[smooth, tension=.7] coordinates {(-7.6,4.4) (-7.3,3.6) (-7.6,2.8) (-7.6,4.4)} ;

% Lens 4
\fill[gray!30] plot[smooth, tension=.7] coordinates {(-4.7091,4.4) (-4.4091,3.6) (-4.7091,2.8) (-4.7091,4.4)} ;
\draw[white] plot[smooth, tension=.7] coordinates {(-4.7091,4.4) (-4.4091,3.6) (-4.7091,2.8) (-4.7091,4.4)} ;

% Lens 3
\fill[gray!30] plot[smooth, tension=.7] coordinates {(-5.4293,4.4) (-5.7293,3.6) (-5.4293,2.8)  (-5.4293,4.4)};
\draw[white] plot[smooth, tension=.7] coordinates {(-5.4293,4.4) (-5.7293,3.6) (-5.4293,2.8) (-5.4293,4.4)};

% Detector
\fill[gray!90] (-3.2091,3.8) -- (-3.2091,3.4) -- (-3.6091,3.3) -- (-3.6091,3.9) plot[smooth, tension=.7] coordinates {(-3.6091,3.9) (-3.8091,3.6) (-3.6091,3.3)};
\draw plot[smooth, tension=.7] (-3.6091,3.9) -- (-3.2091,3.8) -- (-3.2091,3.4) -- (-3.6091,3.3) ;
\draw plot[smooth, tension=.7] coordinates {(-3.6091,3.9) (-3.8091,3.6) (-3.6091,3.3)};

%  Sample
\fill[gray!70] plot[smooth, tension=.7, thick] (-6.4591,3.1) -- (-6.4591,4.1) -- (-6.5591,4.1) -- (-6.5591,3.1) -- (-6.4591,3.1) ;
\draw plot[smooth, tension=.7, thick] (-6.4591,3.1) -- (-6.4591,4.1) -- (-6.5591,4.1) -- (-6.5591,3.1) -- (-6.4591,3.1) ;

\node at (-9.5,3) {\scriptsize{THz Emitter}};

\node at (-8.5,2.5) {\scriptsize{Lens}};
\node at (-7.5091,2.5) {\scriptsize{Lens}};

\node at (-6.5091,4.3) {\scriptsize{Object}};

\node at (-5.5091,2.5) {\scriptsize{Lens}};
\node at (-4.6091,2.5) {\scriptsize{Lens}};

\node at (-3.5091,3) {\scriptsize{THz Detector}};

% Laser 1
\draw[fill, nottblue!80]  (-10.6,4.9) rectangle (-8.8,4.1);
\node at (-9.6999,4.5) {\scriptsize{Pump fs laser}};

% Bias
\draw  plot[smooth, tension=.7] coordinates {(-9.9,3.5) (-10.4,3.1) (-10.1,2.5)};
\draw[fill, gray!30]  (-10.1,2.7) rectangle (-9.3,2.3);
\node at (-9.7,2.5) {\scriptsize{Bias}};

% Laser 2
\draw[fill, nottblue!80]  (-2.7,4.9) rectangle (-0.9,4.1);
\node at (-1.8,4.5) {\scriptsize{Probe fs laser}};

% TIA - TransImpedance Amplifier
\draw[fill, gray!30]  (-3.5,2.7) rectangle (-2.7,2.3);
\node at (-3.1,2.5) {\scriptsize{TIA}};
% \node at (-2.1,1.9) {\tiny{TIA: TransImpedance Amplifier}};

% PC
\draw  (-2.5,3.5) rectangle (-0.9909,2.4);
\draw (-2.4,3.4) rectangle (-1.1,2.5);
\fill[gray!20] plot  (-2.4,3.4) rectangle (-1.1,2.5);

\fill[gray!90] plot[tension=.7] coordinates {(-1.9444,2.4) (-1.9444,2.2) (-2.1444,2.1) (-1.3444,2.1) (-1.5444,2.2) (-1.5444,2.4) (-1.9444,2.4)};

\node at (-1.8,2.9) {\scriptsize{PC}};

% Rotation Stage
\fill plot  (-6.8091,3.1) rectangle (-6.2091,3);
\node at (-6.5587,2.1402) {\scriptsize{Motorized}};
\node at (-6.5587,1.8902) {\scriptsize{Stages}};
% \draw  (-2.5,1) rectangle (-2.5,1);

\draw[-stealth] (-6.6091,2.4) -- (-6.6091,2.9);
\node at (-6.8091,2.8) {\tiny{$\mathcal{H}$}};
\draw[-stealth] (-6.6091,2.4) -- (-6.3091,2.7);
\node at (-6.2,2.7) {\tiny{$\mathcal{L}$}};

\draw  (-6.6,2.4) ellipse (0.2 and 0.12);
\node at (-6.25,2.4) {\tiny{$\mathcal{R}$}};

\end{tikzpicture}
        \end{adjustbox}
        \vspace{-0.8cm} \caption{}
        \label{p20230725_T_Setup}
        \end{subfigure}
        \\
        \begin{subfigure}[b]{0.5\textwidth}
        \includegraphics[width=\textwidth]{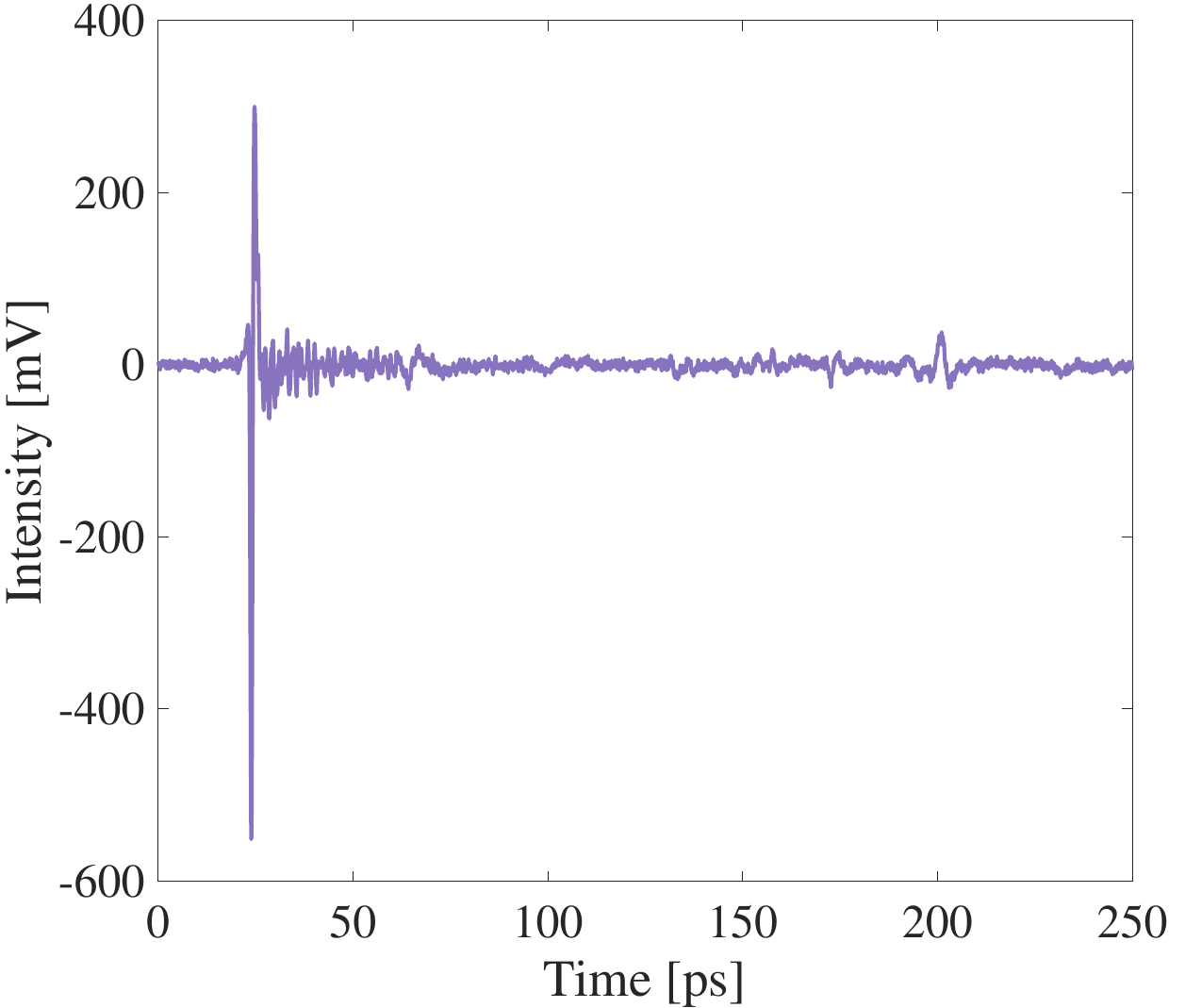}
        \vspace{-0.5cm} \caption{}
        \label{p20230801_THz_Pulse_1}
        \end{subfigure}
        \,
        \begin{subfigure}[b]{0.48\textwidth}
        \includegraphics[width=\textwidth]{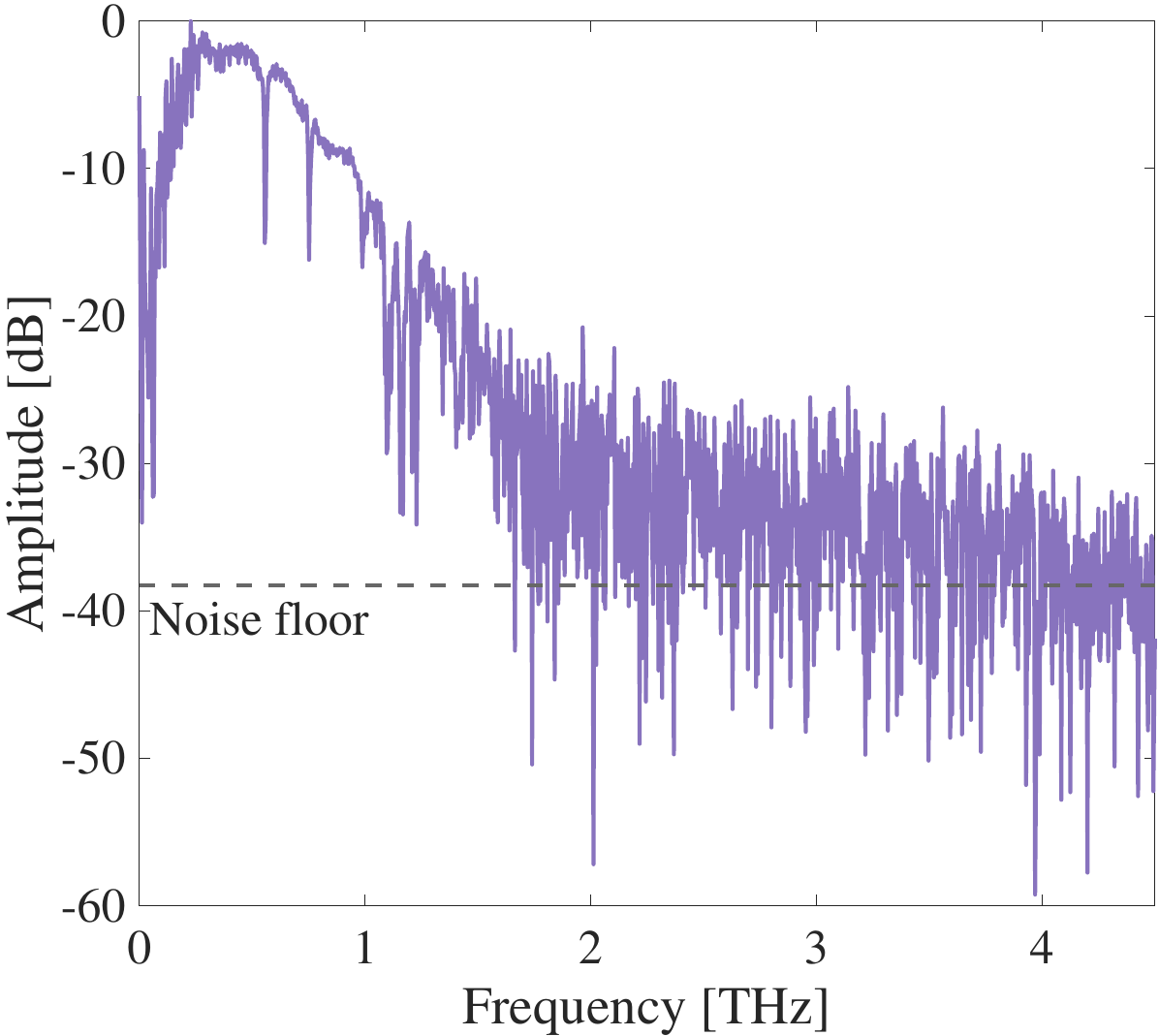}
        \vspace{-0.5cm} \caption{}
        \label{p20230801_THz_Pulse_2}
        \end{subfigure}
        % \vspace{-0.5cm}
        \caption{\subref{p20230725_T_Setup} Schematic diagram of the THz 3D imaging system setup using transmission geometry (TIA: transimpedance amplifier).~\subref{p20230801_THz_Pulse_1} Illustrations of one THz pulse without averaging in the time domain.~\subref{p20230801_THz_Pulse_2} Illustrations of normalized THz power spectrum obtained through fast Fourier transform (FFT) operation from~\subref{p20230801_THz_Pulse_1}. }
        \label{p20230725_T_Setup_p20230801_THz_Pulse}
    \end{figure}

    The experimental setup is shown in~\cref{p20230725_T_Setup}. 
    The THz 3D imaging system records time-resolved THz signals for each voxel using an ASOPS THz-TDS system (TERA ASOPS, MenloSystems). 
    The pulsed THz wave is generated by a THz photoconductive antenna (PCA) emitter and is directed through four plane-convex polypropylene (PP) THz lenses, which focus it onto the sample with a beam size of 1.5 mm before refocusing it onto a THz PCA detector. 
    The photocurrent signal is subsequently amplified by a transimpedance amplifier (TIA) and digitized by an analog-to-digital converter (ADC) processing unit. 
    The data collection process accelerates due to the usage of two asynchronous femtosecond (fs) Er-doped fiber lasers in the ASOPS system. 
    The repetition rate of the pump laser is 100 MHz + 50 Hz ($f_{\mathrm{pump}}$), the probe laser operates at 100 MHz ($f_{\mathrm{probe}}$), and the fixed offset frequency ($\Delta f = f_{\mathrm{pump}} - f_{\mathrm{probe}}$) is 50 Hz. 
    The sampling rate ($n_{\mathrm{sr}}$) is 10 MHz, which allows for a data acquisition time of approximately 20 ms per pulse with a temporal resolution of 50 fs. 
    Each THz signal pulse in the dataset corresponds to 5000 sampling points per voxel and spans 250 ps, as instantiated in~\cref{p20230801_THz_Pulse_1}. 
    This allows for a dynamic range of approximately 38.26 dB from 0.1 to 4 THz, as shown in~\cref{p20230801_THz_Pulse_2}. 
    To facilitate the acquisition of a multi-projection temporal THz dataset for further THz CT purposes, the objects under test are positioned on a rotational stage $\mathcal{R}$ equipped with motorized stages in two directions (length $\mathcal{L}$ and height $\mathcal{H}$). 

\subsection{Conventional THz CT Imaging}
\label{subsec:ConventionalTechniques}

\begin{figure}[htbp]
    \centering
        \begin{subfigure}[b]{0.32\textwidth}
        \includegraphics[width=\textwidth]{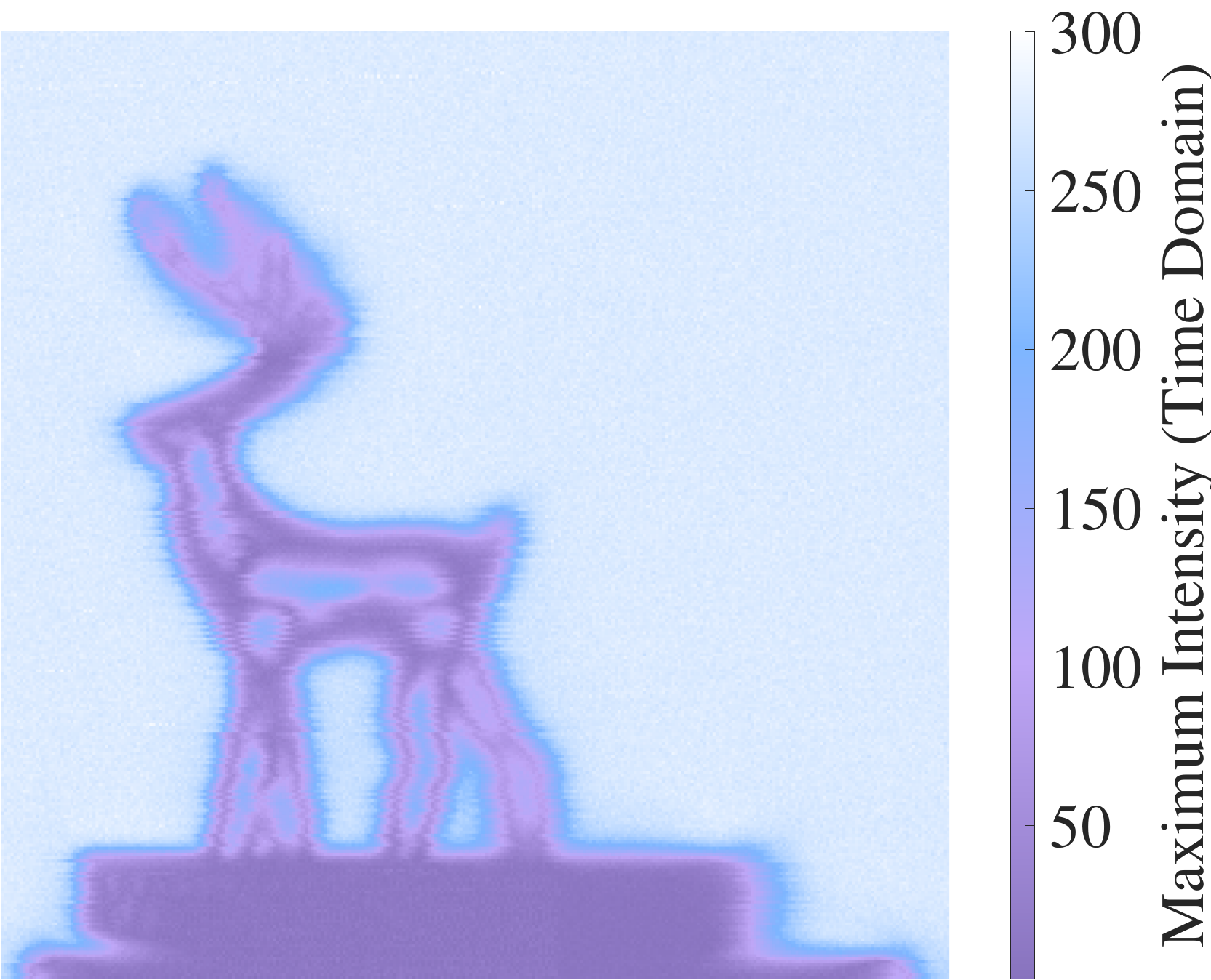}
        \caption{}
        \label{p20230801_THz_Deer_IRT_288_Sinogram_1}
        \end{subfigure}
        \,
        \begin{subfigure}[b]{0.32\textwidth}
        \includegraphics[width=\textwidth]{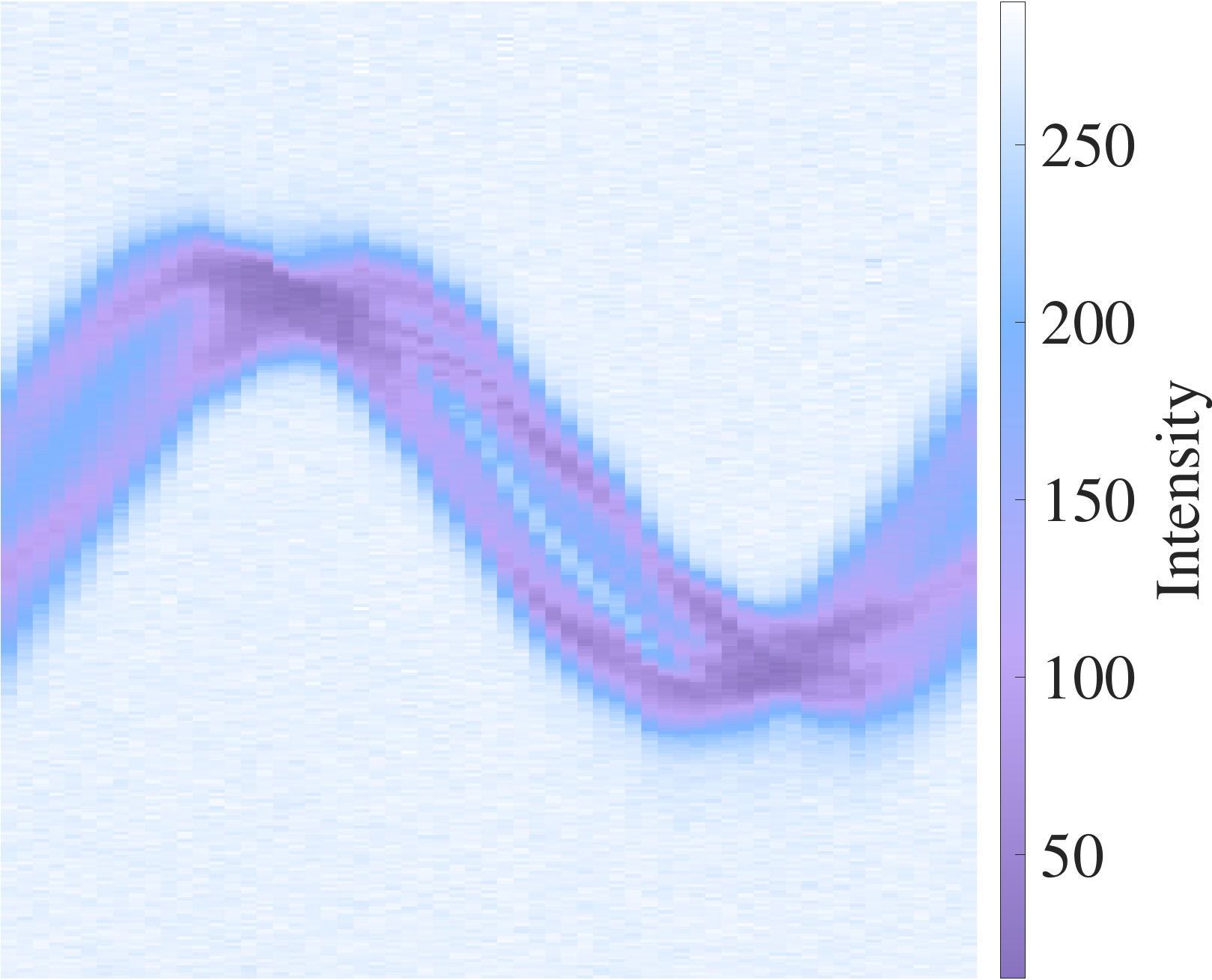}
        \caption{}
        \label{p20230801_THz_Deer_IRT_288_Sinogram_2}
        \end{subfigure}
        \,
        \begin{subfigure}[b]{0.30\textwidth}
        \includegraphics[width=\textwidth]{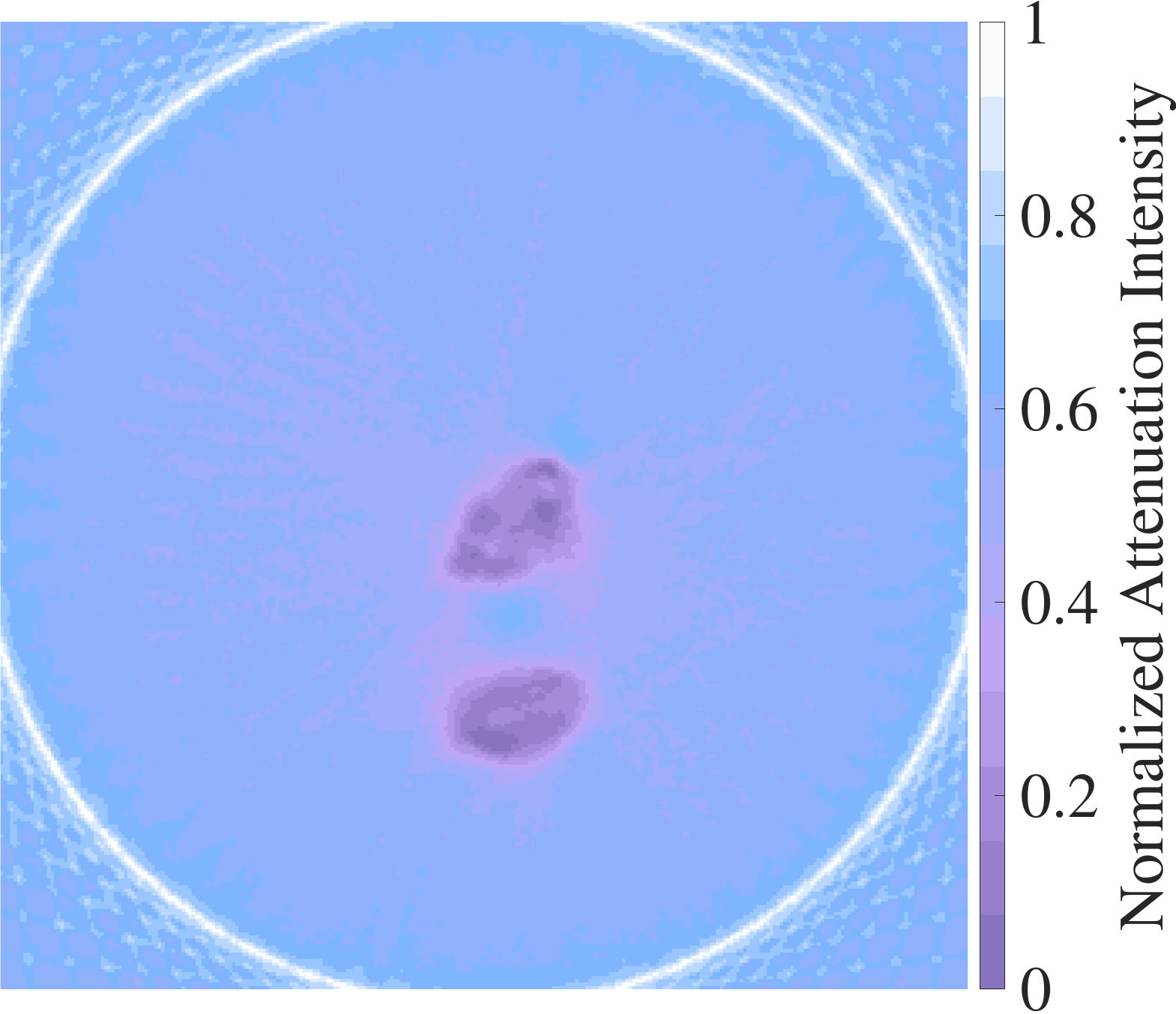}
        \caption{}
        \label{p20230801_THz_Deer_IRT_288_Sinogram_3}
        \end{subfigure}
        \caption{
        Illustrations of the conventional techniques for THz 3D CT Imaging:~\subref{p20230801_THz_Deer_IRT_288_Sinogram_1} projection by Time-MAX;~\subref{p20230801_THz_Deer_IRT_288_Sinogram_2} sinogram;~\subref{p20230801_THz_Deer_IRT_288_Sinogram_3} cross section (slice) by IRT. }
        \label{p20230801_THz_Deer_IRT_288_Sinogram}
        % \vspace{-0.3cm}
    \end{figure}

    Conventional techniques for THz 3D CT imaging usually involve a series of steps to construct 3D images from the collected THz dataset. 
    The initial step in the THz 3D CT imaging process is the acquisition of projection data. 
    In this step, the Time-MAX projection technique is one of the most fundamental methods in THz 3D CT imaging. 
    Time-MAX involves extracting the maximum field strength from each time-resolved THz signal trace, which is then used to represent the projection of the scanned object at a particular projection angle~\cite{Hung_2022_CT_DL, Zhang_2022_CT_VM}. 
    This method effectively reduces the temporal dimensionality of the THz signals to a 2D projection image (see~\cref{p20230801_THz_Deer_IRT_288_Sinogram_1}). 
    The data obtained from Time-MAX is primarily associated with object thickness, surface contour, and material contrast, providing potential information into the spatial distribution of composite materials within the objects. 
    Next, a sinogram (see~\cref{p20230801_THz_Deer_IRT_288_Sinogram_2}) is generated by stacking all the 2D projection images at different rotation angles in a 3D array. 
    The sinogram visualizes the internal structure of the object from multiple angles and forms the basis for image reconstruction. 
    The horizontal axis represents the spatial location along the projection, and the vertical axis corresponds to the projection angle. 
    The value at each pixel in the sinogram indicates the THz signal intensity at the corresponding location and angle. 
    The final step in the conventional THz CT imaging process is the inverse Radon transform (IRT) $\hat{\mathcal{R}}$~\cite{Kak_2001_IRT, Zhang_2022_CT_VM}. 
    This mathematical procedure reconstructs an image $f(\mathbf{x})$ from its projections 
    \begin{eqnarray*}
    f(\mathbf{x})
    \approx
    \sum_{k}
    \left(
    \mathcal{R}f(t, \theta_{k} * h)
    (\langle \mathbf{x}, \mathbf{n}_{\theta_{k}} \rangle)
    \right)
    \end{eqnarray*}
    where the Radon transform $\mathcal{R}f(t, \theta)$ is the integral of $f$ along the line $L_{t, \theta} = \left\{ \mathbf{x}: \langle \mathbf{x}, \mathbf{n}_{\theta} \rangle \right. $ $ \left. = t \right\}$ whose distance from the origin is $t$ and whose direction is orthogonal to the unit vector $\mathbf{n}_{\theta} = (\cos{\theta}, \sin{\theta})$ with the convolution kernel $h$ satisfied $\hat{h}(\omega) = |\omega|$. 
    Applied to the THz field, IRT takes the sinogram as input and reconstructs a 2D cross-section (slice) of the object (see~\cref{p20230801_THz_Deer_IRT_288_Sinogram_3}). 
    The value of each reconstructed pixel represents the normalized attenuated intensity, rescaled to the interval $[0, 1]$ based on the minimum and maximum across all pixels. 
    Then, a 3D volume image of the object can be obtained (see~\cref{p20230712_THz_Deer_S218G0_W_1}) after stacking all 2D slices along the vertical direction and selecting an appropriate threshold. 
    Here, the threshold can be determined based on characteristics of the histogram, such as peaks and valleys. 
    \textcolor{revision_black}{
    These stages are fundamental to conventional THz 3D CT imaging. 
    While} they can yield decent reconstructed THz 3D images, they are still limited by computational cost and complexity, the need for extensive data processing, and potential issues related to artifacts in the reconstructed image. 
    Nevertheless, they continue to serve as the cornerstone methods in THz 3D CT imaging, thanks to their firmly established theoretical foundation and widespread implementation. 

\subsection{Variational Frameworks and Numerical Algorithm with ADMM}
\label{subsec:newproposed}

    In this section, we present a variational approach based on the adapted Euler-Elastica-based formulation and implement it through a numerical algorithm using the alternating direction method of multipliers (ADMM). 
    The variational approach, often referred to as the mathematical approach, and the physical modeling approach are two distinct methodologies used in the field of computational modeling. 
    The physical modeling approach is domain-specific and directly relies on the physical laws governing a particular system, whereas variational frameworks are primarily rooted in mathematical principles and optimization techniques. 
    The variational frameworks operate based on the principle of optimizing a function, often referred to as a functional, which typically represents energy/cost forms associated with a physical or computational system. 
    Through the optimization of this functional, we can infer the optimal parameters or states of the system. 
    This approach is more versatile and applicable to a broad spectrum of problems. 
    
    Our proposed variational framework is particularly designed to address the issues of prolonged data acquisition time and resolution constraints by scanning resolution with diffraction effects (referred to as concerns (i) and (ii) in ~\Cref{sec:introduction}). 
    This framework has the ability to handle LR input slices, even if they are limited in number, and adeptly reconstruct them into a smooth SR surface. 
    Reconstruction from these $s$ slices $\{\Pi_{i}\}_{i = 1, \ldots, s}$ can be framed as a variational model, which is motivated by the works~\cite{Roger_2006_EE_Theory, Bretin_2017_VolumeReconstruction, Zhang_2022_CT_VM, Zhang_2023_CLEO_CT_VM_EE, Zhang_2022-2023_IPI}. 
    Mathematically, the goal is to find the (local) minimum $u^{*}$ that satisfies our new adapted Euler-Elastica-based formulation
    \begin{eqnarray}
    \label{equ:EEad}
    \begin{aligned}
        &
        \arg \min_{u}
        \mathscr{m}
        \int_{\Omega} 
            \left( 
            \frac{\varepsilon}{2} |\grad u|^{2} 
            + 
            \frac{1}{\varepsilon} W(u) 
            \right)
            \, \id \Omega
            +
            \frac{\mathscr{n}}{2\varepsilon} 
            \int_{\Omega} 
            \left( 
            \varepsilon \laplace u 
            - 
            \frac{1}{\varepsilon} W'(u) 
            \right)^{2} 
            \, \id \Omega
            \\
            &
            \mbox{ s.t. }
            u^{in}
            \le 
            u 
            \le 
            u^{ex}. 
    \end{aligned}
    \end{eqnarray}
    Here, 
    $u \in \Omega \subset \R^{3}$ is the phase-field function to provide a 3D representation of the object, governed by the profile function $q$; 
    $q$ is the profile function designed to be piecewise, taking a value of 1 within the object, half on the boundary, and 0 outside the object, to maintain continuity and smoothness in the phase-field function; 
    $\nabla, \laplace$ is the gradient and Laplacian operator (note that, $\laplace = \operatorname{div}\nabla$ with divergence operator $\operatorname{div}$); 
    $\varepsilon$ signifies the phase-field variable and is linked to the thickness parameter; 
    $u^{in} \le u \le u^{ex}$ is the linear obstacle restrictions, ensuring that the shape of the input data is preserved where the phase-field profiles, $u^{in}$ and $u^{ex}$, are derived from the interior $\{\Pi_{s}^{in}\}$ and exterior region $\{\Pi_{s}^{ex}\}$ through the profile function $q$; and 
    $W(u)$ is the double-well potential defined by $W(u) = \frac{1}{2}u^{2}(1-u)^{2}$, which has two minima related to the profile function $q$ and its first derivative is $W'(u) = u(u-1)(2u-1)$. 
    Specifically, each pair of coefficients $(\mathscr{m}, \mathscr{n})$ is used to control the type of formulation applied. 
    If $\mathscr{m} = 1, \mathscr{n} = 0$, the formulation is perimeter-based $\mathscr{P_{\varepsilon}}$ from the perimeter energy, focusing on preserving the edge features of the object. 
    Conversely, if $\mathscr{m} = 0, \mathscr{n} = 1$, the Willmore-based formulation $\mathscr{W_{\varepsilon}}$ is applied, which is related to the mean curvature from Willmore energy and promotes smoothness in the output. 
    Remark that, without certain restrictions, this formulation could lead to an output that takes the shape of a sphere, as the sphere is the configuration that minimizes the Willmore energy. 
    Lastly, when $\mathscr{m} = 1, \mathscr{n} = 1$, the Euler-Elastica-based formulation $\mathscr{E_{\varepsilon}}$ is used, which is an optimal combination of edge preservation and smoothness. 
    For detailed mathematical preliminaries and clarifications, please refer to~\hyperref[supp]{Sec. 1 of Supplement 1}~\cite{Bretin_2017_VolumeReconstruction, Zhang_2022_CT_VM, Zhang_2023_CLEO_CT_VM_EE, Zhang_2022-2023_IPI}. 

    To address our model as represented in~\eqref{equ:EEad}, we employ the alternating direction method of multiplier (ADMM). 
    This approach breaks down complex optimization problems into smaller, more manageable parts for swifter numerical approximation. 
    We further implement the linear obstacle restriction through orthogonal projection, thereby managing the inequality with improved flexibility, which is achieved by
    \begin{eqnarray}
    \label{eqn:linear_obstacle_restriction_apprximate}
        u^{in} 
        \leq 
        u 
        \leq 
        u^{ex}
        \quad
        \leftrightsquigarrow
        \quad
        \max(\min(u, u^{ex}), u^{in}). 
    \end{eqnarray}

    \pagebreak
    \noindent
    With $\mathbf{w} = \grad u$, the augmented Lagrangian functional is then expressed using a penalty parameter $\rho > 0$ and a Lagrange multiplier $\boldsymbol{\lambda}$, as follows
    \begin{eqnarray}
    \begin{aligned}
    \mathcal{L}^{\rho}
    (u, \mathbf{w}; \boldsymbol{\lambda}) 
    % &
    % =
    % \int_{\Omega} 
    %     \left[
    %     \mathscr{m}
    %     \left( 
    %     \frac{\varepsilon}{2} |\mathbf{w}|^{2} 
    %     + 
    %     \frac{1}{\varepsilon} W(u) 
    %     \right)
    %     +
    %     \frac{\mathscr{n}}{2\varepsilon} 
    %     \left( 
    %     \varepsilon \operatorname{div} \mathbf{w} 
    %     - 
    %     \frac{1}{\varepsilon} W'(u) 
    %     \right)^{2} 
    %     \right.
    %     \\
    %     &
    %     \qquad \quad
    %     +
    %     \left \langle 
    %     \boldsymbol{\lambda}, \nabla u - \mathbf{w} 
    %     \right \rangle
    %     +
    %     \left.
    %     \frac{\rho}{2}
    %     \left|
    %     \nabla u - \mathbf{w}
    %     \right|^{2}
    %     \right]
    % \, \id \Omega 
    % \\
    % & 
    =
    \int_{\Omega} 
        &
        \left[
        \mathscr{m}
        \left( 
        \frac{\varepsilon}{2} |\mathbf{w}|^{2} 
        + 
        \frac{1}{\varepsilon} W(u) 
        \right)
        +
        \frac{\mathscr{n}}{2\varepsilon} 
        \left( 
        \varepsilon \operatorname{div} \mathbf{w} 
        - 
        \frac{1}{\varepsilon} W'(u) 
        \right)^{2} 
        \right.
        \\
        &
        % \qquad \quad
        +
        \left.
        \frac{\rho}{2}
        \left|
        \nabla u - \mathbf{w} 
        + \rho^{-1} \boldsymbol{\lambda}
        \right|^{2}
        -
        \frac{\boldsymbol{\lambda}^{2}}{2 \rho}
        \right]
    \, \id \Omega. 
    \end{aligned}
    \end{eqnarray}
    Hence, the application of the ADMM involves a three-step process: solving two subproblems $u, \mathbf{w}$ and updating one multiplier $\boldsymbol{\lambda}$
    \begin{eqnarray}
    \label{eqn:ADMMpro}
    \left\{
    \begin{array}{l}
    u_{k+1}
    =
    \arg \min \limits_{u} 
    \int_{\Omega}
    \left[
    \frac{\mathscr{m}}{\varepsilon} W(u)
    -
    \frac{\mathscr{n}}{\varepsilon} (\operatorname{div} \mathbf{w}_{k})  W'(u)
    +
    \frac{\mathscr{n}}{2\varepsilon^{3}}
    (W'(u))^{2}
    \right. 
    \\
    \left.
    \qquad \qquad \qquad \qquad
    % +
    % \boldsymbol{\lambda} \nabla u
    % +
    % \frac{\rho}{2}
    %     \left(
    %     \nabla u - \mathbf{w}_{k}
    %     \right)^{2}
    +
    \frac{\rho}{2}
    \left|
    \nabla u - \mathbf{w}_{k} 
    + \rho^{-1} \boldsymbol{\lambda}_{k}
    \right|^{2}
    \right] 
    \, \id \Omega 
    \\
    \mathbf{w}_{k+1}
    =
    \arg \min \limits_{\mathbf{w}} 
    \int_{\Omega}
    \left[
    \frac{\mathscr{m}\varepsilon}{2} |\mathbf{w}|^{2}
    +
    \frac{\mathscr{n}\varepsilon}{2}
    (\operatorname{div} \mathbf{w})^{2}
    -
    \frac{\mathscr{n}}{\varepsilon} (\operatorname{div} \mathbf{w})  W'(u_{k+1})
    \right. 
    \\
    \left.
    \qquad \qquad \qquad \qquad
    % -
    % \boldsymbol{\lambda} \mathbf{w}
    % +
    % \frac{\rho}{2}
    %     \left(
    %     \nabla u_{k+1} - \mathbf{w}
    %     \right)^{2}
    +
    \frac{\rho}{2}
    \left|
    \mathbf{w} 
    - \nabla u_{k+1} 
    - \rho^{-1} \boldsymbol{\lambda}_{k}
    \right|^{2}
    \right] 
    \, \id \Omega 
    \\
    \boldsymbol{\lambda}_{k+1}
    =
    \boldsymbol{\lambda}_{k}
    +
    \rho
    \left(
    \nabla u_{k+1}-\mathbf{w}_{k+1}
    \right) 
    \end{array}
    \right..
    \end{eqnarray}
    
    Utilizing the numerical Euler semi-implicit discretization scheme with a time step $\tau$,  the optimal solutions for the $u$- and $\mathbf{w}$-subproblems in~\eqref{eqn:ADMMpro} can be updated as follows: 
    \begin{eqnarray}
    \begin{aligned}
        u_{k+1}
        =
        (I - \tau \rho \laplace)^{-1}
        &
        \left[
        u_{k}
        +
        \tau
        \left(
        - \rho \nabla \cdot \mathbf{w}_{k} 
        + \nabla \cdot \boldsymbol{\lambda}_{k}
        -
        \frac{\mathscr{m}}{\varepsilon} W'(u_{k}) 
        \right. 
        \right.
        \\
        & 
        \left.
        \left.
        +
        \frac{\mathscr{n}}{\varepsilon} 
        \operatorname{div} \mathbf{w}_{k} W''(u_{k})
        +
        \frac{\mathscr{n}}{\varepsilon^{3}} 
        W'(u_{k}) W''(u_{k})
        \right)
        \right], 
    \end{aligned}
    \end{eqnarray}
    \begin{eqnarray}
    \begin{aligned}
        \mathbf{w}_{k+1}
        =
        (I + \mathscr{m} \varepsilon \tau - \mathscr{n} \varepsilon \tau \laplace)^{-1}
        \left[
        \mathbf{w}_{k}
        +
        \frac{\mathscr{n} \tau}{\varepsilon} 
        W'(u_{k+1})
        \laplace \mathbf{w}_{k}
        +
        \tau
        \rho 
        \left(
        \mathbf{w}_{k}
        -
        \nabla u_{k+1} 
        - 
        \rho^{-1} \boldsymbol{\lambda}_{k}
        \right)
        \right]. 
    \end{aligned}
    \end{eqnarray}
    Lastly, recall that the multiplier $\boldsymbol{\lambda}$ will be updated by
    % \begin{eqnarray}
    % \begin{aligned}
    $
        \boldsymbol{\lambda}_{k+1}
        =
        \boldsymbol{\lambda}_{k}
        +
        \rho
        \left(
        \nabla u_{k+1}-\mathbf{w}_{k+1}
        \right). 
    $
    % \end{aligned}
    % \end{eqnarray}
    This marks the completion of one iteration of the ADMM method where the comprehensive algorithm is presented in~\Cref{alg:Alternating_Direction_Method_of_Multipliers} (more details of derivations in~\hyperref[supp]{Sec. 2 of Supplement 1}). 

    %% ADMM
    \begin{algorithm}[htbp]
    \caption{Alternating Direction Method of Multipliers (ADMM)}
    \label{alg:Alternating_Direction_Method_of_Multipliers}
    \begin{algorithmic}[1]
    \Require Input slices $\{\Pi_{s}\}$; Parameters $\mathscr{m}, \mathscr{n}, \tau, \varepsilon, \rho$. 
    \Ensure Numerical solution $u_{k+1}$. 
    \State Initial input: $u_{0} = q(\{\Pi_{s}\})$; $\mathbf{w}_{0} = \grad u_{0}$; $\boldsymbol{\lambda}_{0} = \mathbf{w}_{0}$; 
    \State Interior restriction: $u^{in} = q(\{\Pi_{s}^{in}\})$; Exterior restriction: $u^{ex} = 1 - q(\{\Pi_{s}^{ex}\})$; 
    \For{$k = 0, 1, \ldots$}
    \State $u_{k+\frac{1}{2}} 
        = 
        \max(\min(u_{k}, u^{ex}), u^{in})$; 
    \State $
    \begin{aligned}
    u_{k+1}
        =
        (I - \tau \rho \laplace)^{-1}
        &
        \left[
        u_{k}
        +
        \tau
        \left(
        - \rho \nabla \cdot \mathbf{w}_{k} 
        + \nabla \cdot \boldsymbol{\lambda}_{k}
        -
        \frac{\mathscr{m}}{\varepsilon} W'(u_{k}) 
        \right. 
        \right.
        \\
        & 
        \left.
        \left.
        +
        \frac{\mathscr{n}}{\varepsilon} 
        \operatorname{div} \mathbf{w}_{k} W''(u_{k})
        +
        \frac{\mathscr{n}}{\varepsilon^{3}} 
        W'(u_{k}) W''(u_{k})
        \right)
        \right]; 
    \end{aligned}
    $
    \State $
    \begin{aligned}
    \mathbf{w}_{k+1}
        =
        (I + \mathscr{m} \varepsilon \tau - \mathscr{n} \varepsilon \tau \laplace)^{-1}
        &
        \left[
        \mathbf{w}_{k}
        +
        \frac{\mathscr{n} \tau}{\varepsilon} 
        W'(u_{k+1})
        \laplace \mathbf{w}_{k}
        +
        \tau
        \rho 
        \left(
        \mathbf{w}_{k}
        -
        \nabla u_{k+1} 
        - 
        \rho^{-1} \boldsymbol{\lambda}_{k}
        \right)
        \right]; 
    \end{aligned}
    $
    \State $
        \boldsymbol{\lambda}_{k+1}
        =
        \boldsymbol{\lambda}_{k}
        +
        \rho
        \left(
        \nabla u_{k+1}-\mathbf{w}_{k+1}
        \right)
        $; 
    \EndFor{ when specific stopping criteria are met. }
    \end{algorithmic}
    \end{algorithm}

\section{Experimental Results with Discussion and Application}

    In this section, we showcase experimental results obtained using our BLIss. 
    To facilitate our experiments, these objects are fabricated economically using a fused deposition modeling (FDM) 3D printer and high impact polystyrene (HIPS) material (see optical images in~\hyperref[supp]{Sec. 3 of Supplement 1}). 
    The choice of HIPS material is for its low absorption in the THz spectrum~\cite{Siemion_2021_Material_HIPS}. 
    This selection helps prevent considerable degradation in retrieved time-resolved THz signals. 
    
\begin{figure}[H]
    % \vspace{-1.5cm}
    \centering
        \begin{subfigure}[b]{0.16\textwidth} 
        \begin{tikzpicture}
        \node
        {
        \adjincludegraphics[width=\textwidth, trim={{.2\width} {.18\height} {.3\width} {.15\height}}, clip]{0_Figures/1_Deer/p20230712_THz_Deer_S218G0_W_1-eps-converted-to.pdf}};
        
        \node at (-0.2,1.6) {{\scriptsize{$s = 218, \mbox{gap} = 0$}}};

        \node at (-1.5,0) {{\scriptsize{Conv.}}};
        \end{tikzpicture}
        \vspace{-0.6cm} \caption{}
        \label{p20230712_THz_Deer_S218G0_W_1}
        \end{subfigure}
        \,
        \begin{subfigure}[b]{0.16\textwidth}
        \begin{tikzpicture}
        \node
        {
        \adjincludegraphics[width=\textwidth, trim={{.2\width} {.18\height} {.3\width} {.15\height}}, clip]{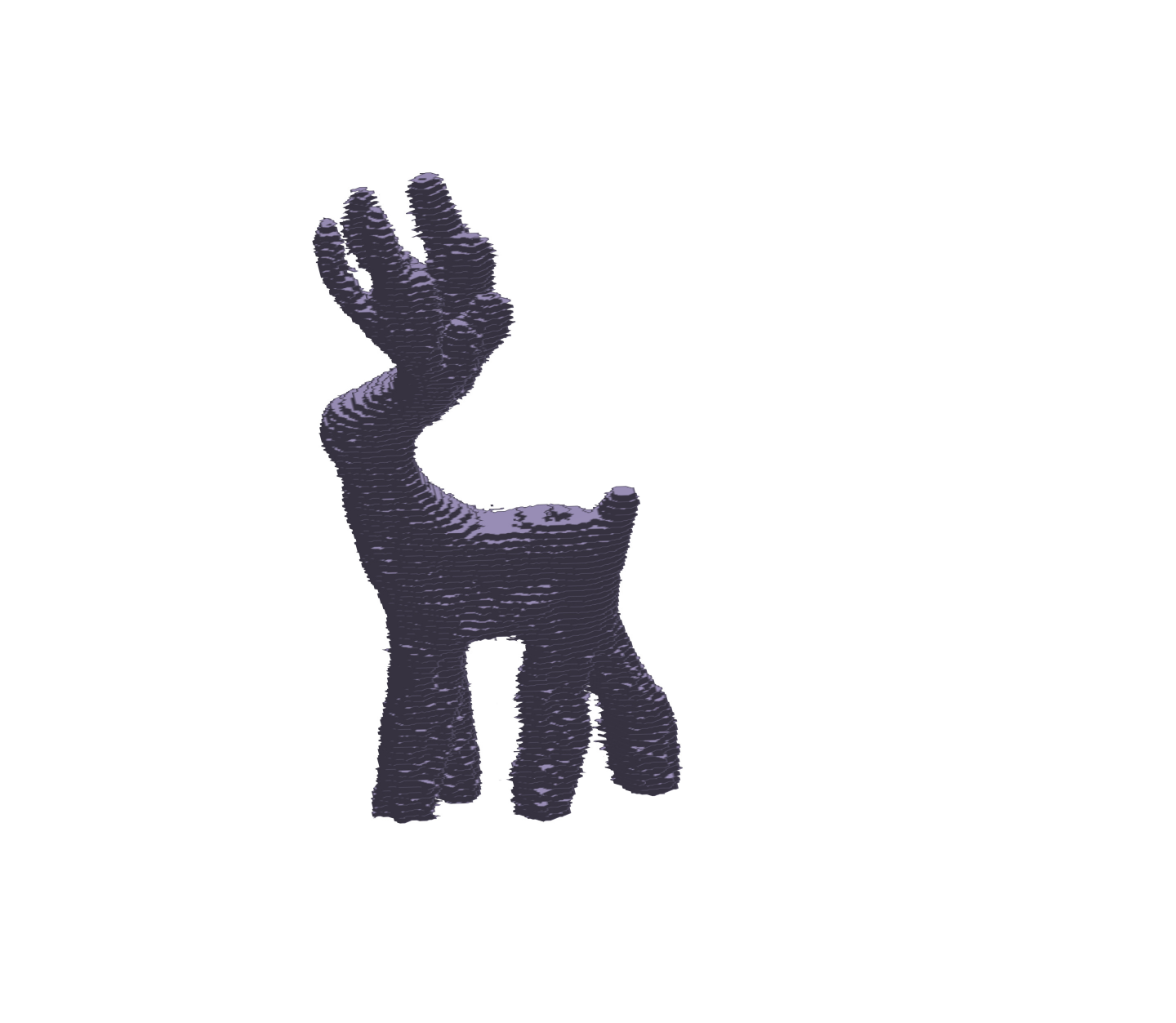}
	};
        
        \node at (-0.2,1.6) {{\scriptsize{$s = 109, \mbox{gap} = 1$}}};

        \node at (-1.5,0) {{\scriptsize{}}};
        \end{tikzpicture}
        \vspace{-0.6cm} \caption{}
        \label{p20230713_THz_Deer_S109G1_W_1}
        \end{subfigure}
        \,
        \begin{subfigure}[b]{0.16\textwidth}
        \begin{tikzpicture}
        \node
        {
        \adjincludegraphics[width=\textwidth, trim={{.2\width} {.18\height} {.3\width} {.15\height}}, clip]{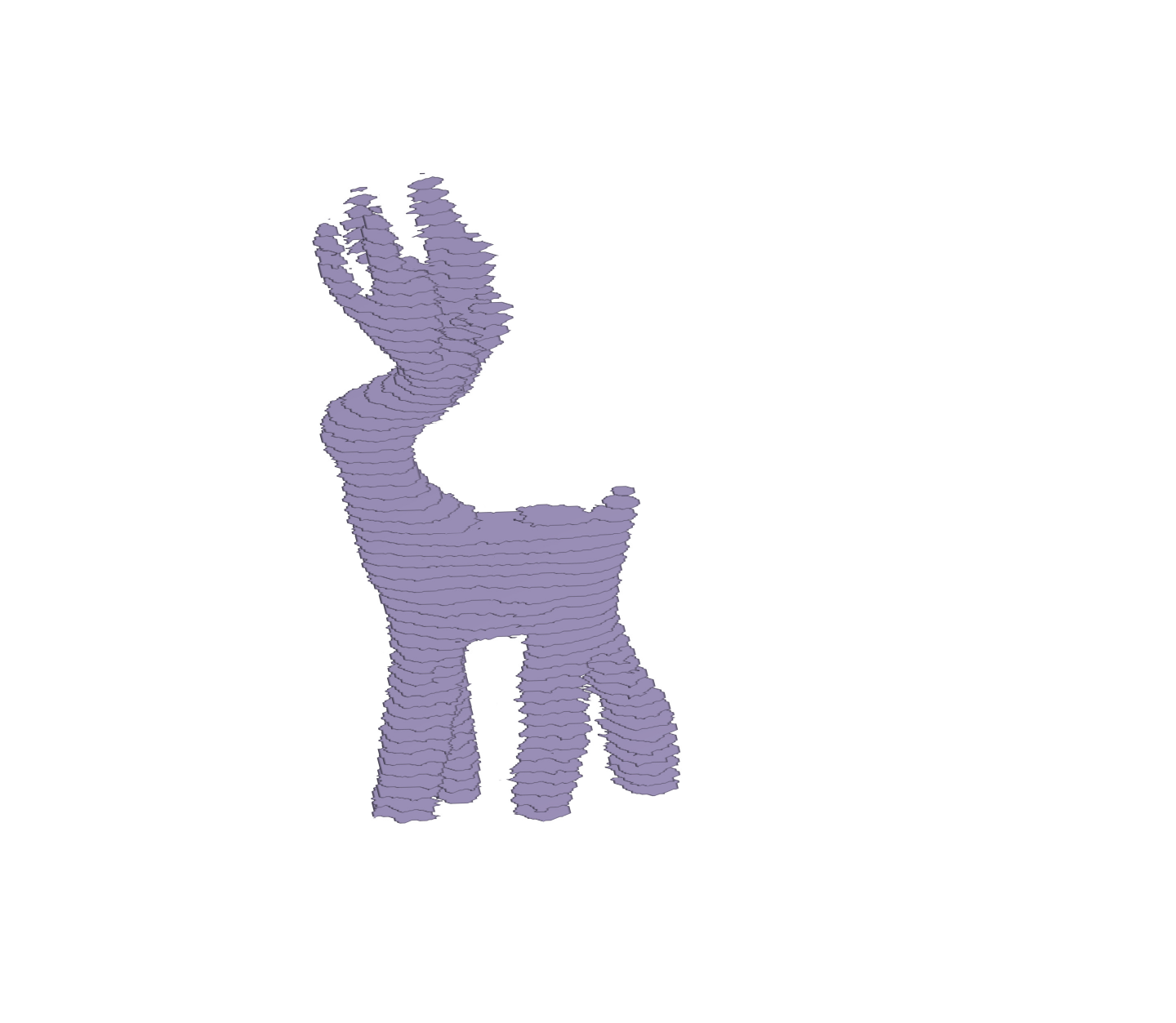}
	};
        
        \node at (-0.2,1.6) {{\scriptsize{$s = 55, \mbox{gap} = 3$}}};

        \node at (-1.5,0) {{\scriptsize{}}};
        \end{tikzpicture}
        \vspace{-0.6cm} \caption{}
        \label{DATE20230713_THz_Deer_S55G3_W_1}
        \end{subfigure}
        \,
        \begin{subfigure}[b]{0.16\textwidth}
        \begin{tikzpicture}
        \node
        {
        \adjincludegraphics[width=\textwidth, trim={{.2\width} {.18\height} {.3\width} {.15\height}}, clip]{0_Figures/1_Deer/p20230713_THz_Deer_S37G5_W_1-eps-converted-to.pdf}
	};
        
        \node at (-0.2,1.6) {{\scriptsize{$s = 37, \mbox{gap} = 5$}}};

        \node at (-1.5,0) {{\scriptsize{}}};
        \end{tikzpicture}
        \vspace{-0.6cm} \caption{}
        \label{p20230713_THz_Deer_S37G5_W_1}
        \end{subfigure}
        \\
        \vspace{-0.5cm}
        \begin{subfigure}[b]{0.16\textwidth}
        \begin{tikzpicture}
        \node
        {
        \adjincludegraphics[width=\textwidth, trim={{.2\width} {.15\height} {.3\width} {.18\height}}, clip]{0_Figures/1_Deer/p20230712_THz_Deer_S218G0_W_5-eps-converted-to.pdf}
	};
        
        \node at (-0.2,1.6) {{\scriptsize{}}};

        \node at (-1.5,0) {{\scriptsize{$\mathscr{W}_{\varepsilon}$}}};
        \end{tikzpicture}
        \vspace{-0.6cm} \caption{}
        \label{p20230712_THz_Deer_S218G0_W_5}
        \end{subfigure}
        \,
        \begin{subfigure}[b]{0.16\textwidth}
        \begin{tikzpicture}
        \node
        {
        \adjincludegraphics[width=\textwidth, trim={{.2\width} {.15\height} {.3\width} {.18\height}}, clip]{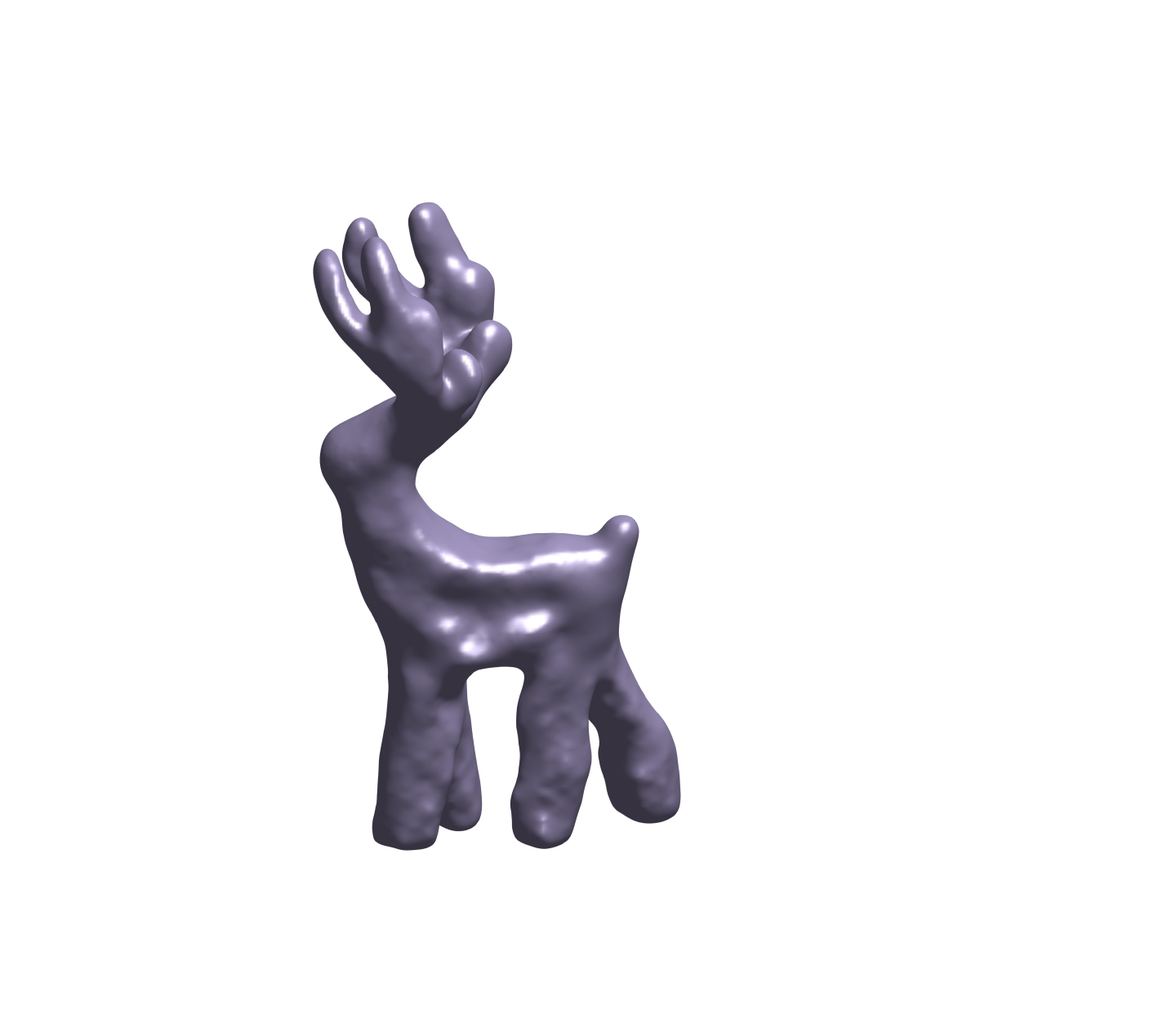}
	};
        
        \node at (-0.2,1.6) {{\scriptsize{}}};

        \node at (-1.5,0) {{\scriptsize{}}};
        \end{tikzpicture}
        \vspace{-0.6cm} \caption{}
        \label{p20230713_THz_Deer_S109G1_W_5}
        \end{subfigure}
        \,
        \begin{subfigure}[b]{0.16\textwidth}
        \begin{tikzpicture}
        \node
        {
        \adjincludegraphics[width=\textwidth, trim={{.2\width} {.15\height} {.3\width} {.18\height}}, clip]{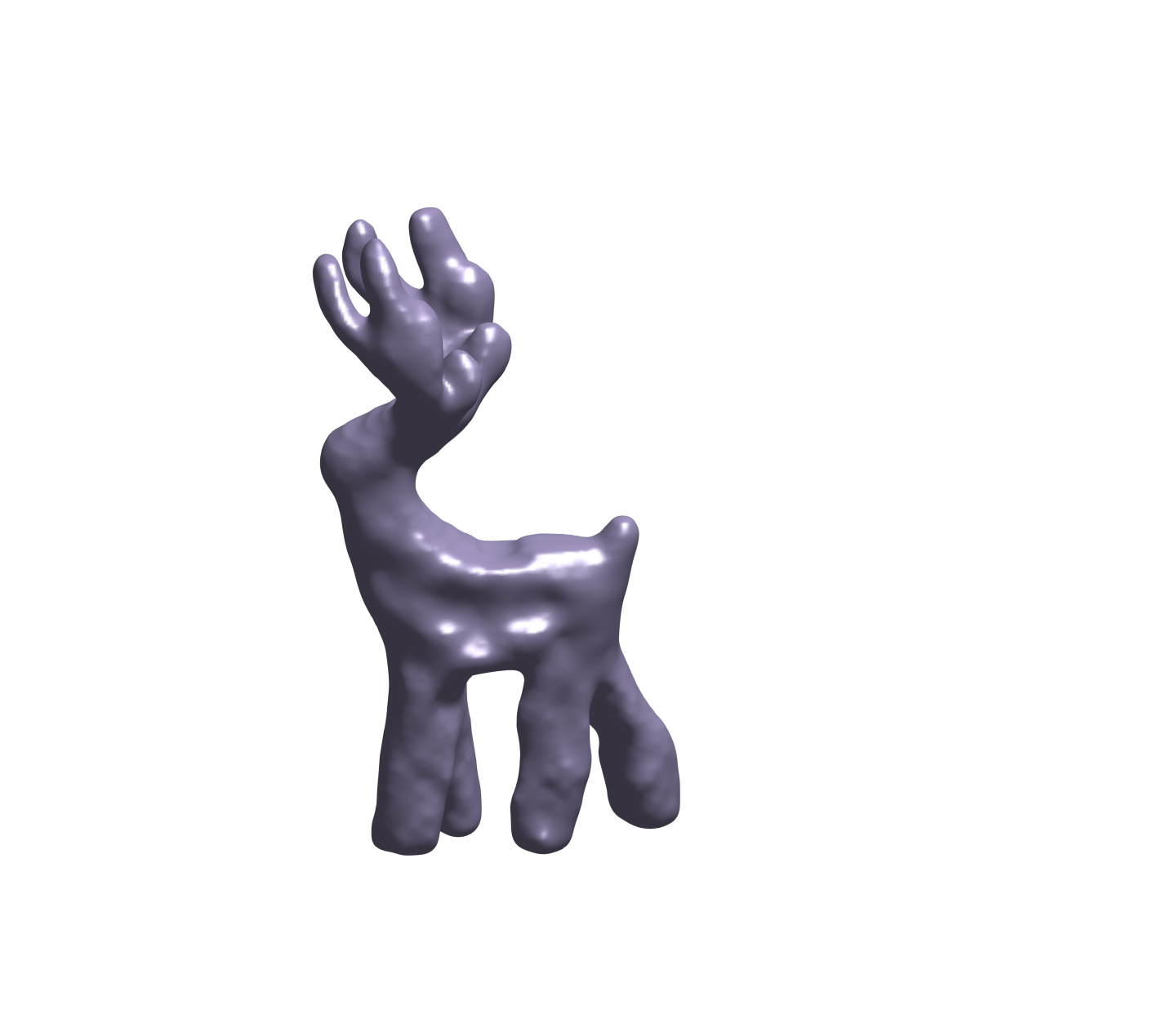}
	};
        
        \node at (-0.2,1.6) {{\scriptsize{}}};

        \node at (-1.5,0) {{\scriptsize{}}};
        \end{tikzpicture}
        \vspace{-0.6cm} \caption{}
        \label{DATE20230713_THz_Deer_S55G3_W_5}
        \end{subfigure}
        \,
        \begin{subfigure}[b]{0.16\textwidth}
        \begin{tikzpicture}
        \node
        {
        \adjincludegraphics[width=\textwidth, trim={{.2\width} {.15\height} {.3\width} {.18\height}}, clip]{0_Figures/1_Deer/p20230713_THz_Deer_S37G5_W_5.jpg}
	};
        
        \node at (-0.2,1.6) {{\scriptsize{}}};

        \node at (-1.5,0) {{\scriptsize{}}};
        \end{tikzpicture}
        \vspace{-0.6cm} \caption{}
        \label{p20230713_THz_Deer_S37G5_W_5}
        \end{subfigure}
        \\
        \vspace{-0.5cm}
        \begin{subfigure}[b]{0.16\textwidth}
        \begin{tikzpicture}
        \node
        {
        \adjincludegraphics[width=\textwidth, trim={{.2\width} {.15\height} {.3\width} {.18\height}}, clip]{0_Figures/1_Deer/p20230713_THz_Deer_S218G0_EE_5-eps-converted-to.pdf}
	};
        
        \node at (-0.2,1.6) {{\scriptsize{}}};

        \node at (-1.5,0) {{\scriptsize{$\mathscr{E}_{\varepsilon}$}}};
        \end{tikzpicture}
        \vspace{-0.6cm} \caption{}
        \label{p20230713_THz_Deer_S218G0_EE_5}
        \end{subfigure}
        \,
        \begin{subfigure}[b]{0.16\textwidth}
        \begin{tikzpicture}
        \node
        {
        \adjincludegraphics[width=\textwidth, trim={{.2\width} {.15\height} {.3\width} {.18\height}}, clip]{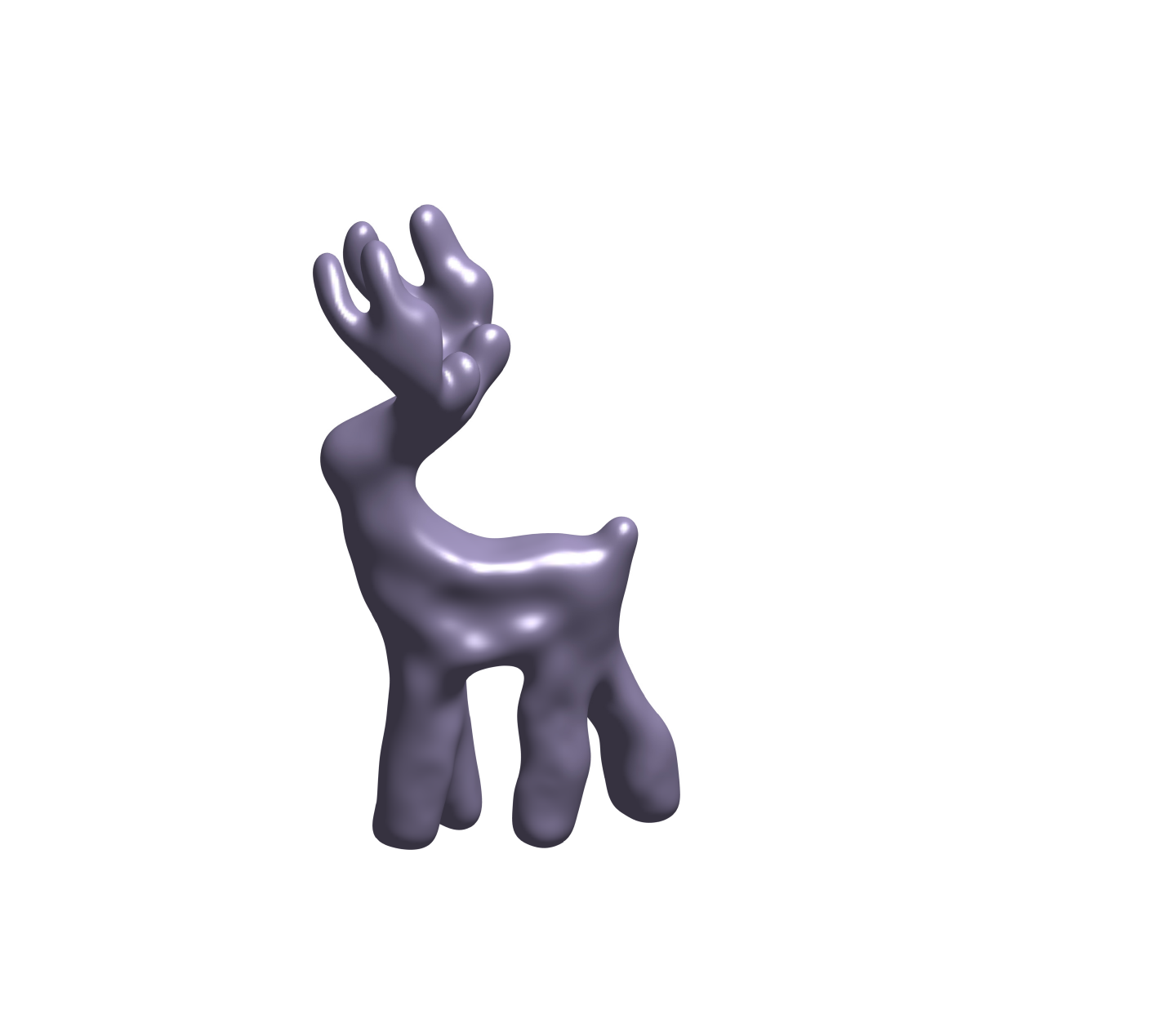}
	};
        
        \node at (-0.2,1.6) {{\scriptsize{}}};

        \node at (-1.5,0) {{\scriptsize{}}};
        \end{tikzpicture}
        \vspace{-0.6cm} \caption{}
        \label{p20230713_THz_Deer_S109G1_EE_11}
        \end{subfigure}
        \,
        \begin{subfigure}[b]{0.16\textwidth}
        \begin{tikzpicture}
        \node
        {
        \adjincludegraphics[width=\textwidth, trim={{.2\width} {.15\height} {.3\width} {.18\height}}, clip]{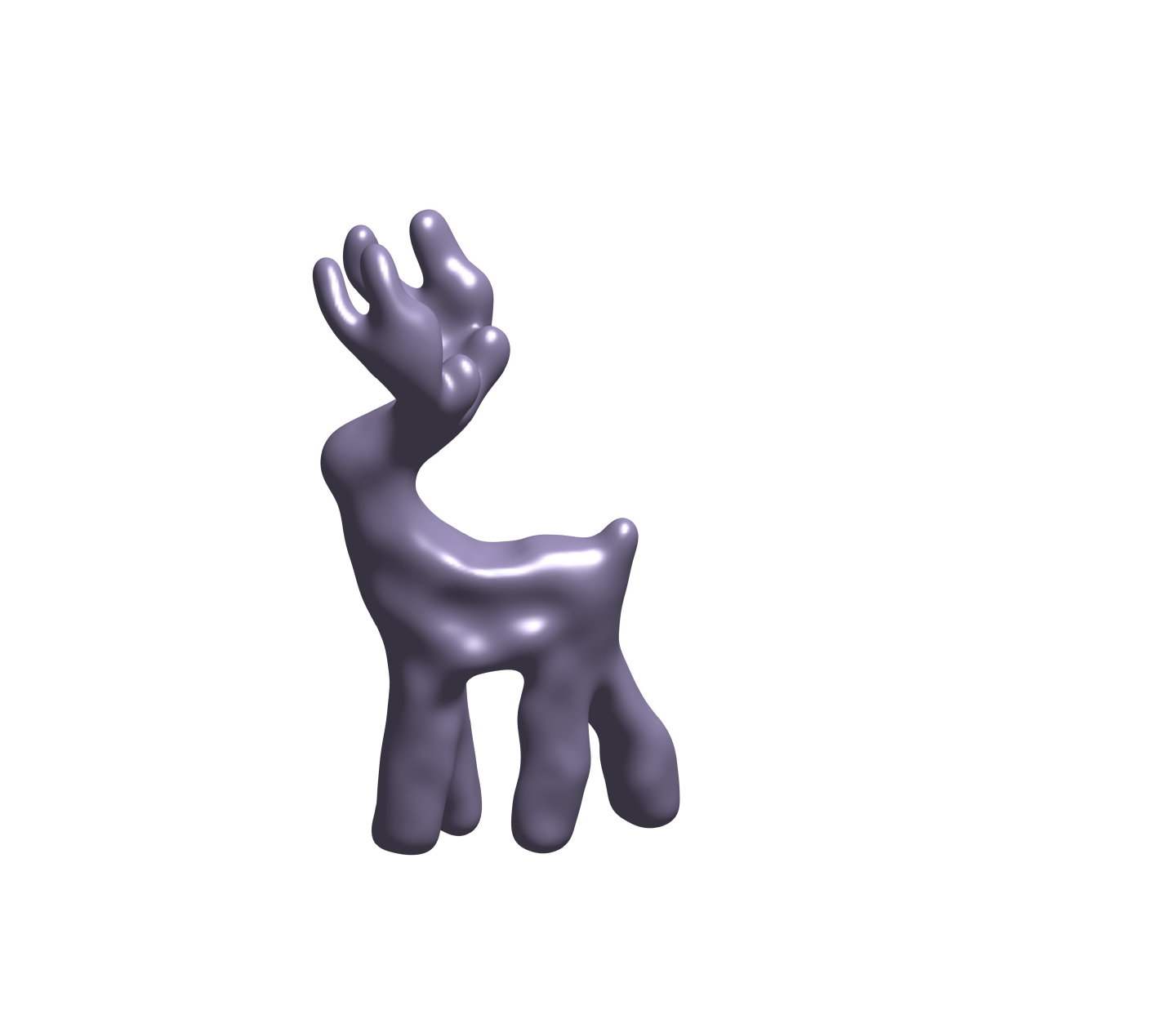}
	};
        
        \node at (-0.2,1.6) {{\scriptsize{}}};

        \node at (-1.5,0) {{\scriptsize{}}};
        \end{tikzpicture}
        \vspace{-0.6cm} \caption{}
        \label{p20230713_THz_Deer_S55G3_EE_23}
        \end{subfigure}
        \,
        \begin{subfigure}[b]{0.16\textwidth}
        \begin{tikzpicture}
        \node
        {
        \adjincludegraphics[width=\textwidth, trim={{.2\width} {.15\height} {.3\width} {.18\height}}, clip]{0_Figures/1_Deer/p20230713_THz_Deer_S37G5_EE_35-eps-converted-to.pdf}
	};
        
        \node at (-0.2,1.6) {{\scriptsize{}}};

        \node at (-1.5,0) {{\scriptsize{}}};
        \end{tikzpicture}
        \vspace{-0.6cm} \caption{}
        \label{p20230713_THz_Deer_S37G5_EE_35}
        \end{subfigure}
        \vspace{-0.3cm}
        \caption{Illustrations of Deer object:~\subref{p20230712_THz_Deer_S218G0_W_1} conventional LR reconstruction with 218 slices where smooth SR reconstruction~\subref{p20230712_THz_Deer_S218G0_W_5} by Willmore-based $\mathscr{W_{\varepsilon}}$ and~\subref{p20230713_THz_Deer_S218G0_EE_5} by Euler-Elastica-based model $\mathscr{E_{\varepsilon}}$;~\subref{p20230713_THz_Deer_S109G1_W_1} input 109 slices with 1 gap where reconstruction~\subref{p20230713_THz_Deer_S109G1_W_5} by $\mathscr{W_{\varepsilon}}$ and~\subref{p20230713_THz_Deer_S109G1_EE_11} by $\mathscr{E_{\varepsilon}}$;~\subref{DATE20230713_THz_Deer_S55G3_W_1} input 55 slices with 3 gaps where reconstruction~\subref{DATE20230713_THz_Deer_S55G3_W_5} by $\mathscr{W_{\varepsilon}}$ and~\subref{p20230713_THz_Deer_S55G3_EE_23} by $\mathscr{E_{\varepsilon}}$;~\subref{p20230713_THz_Deer_S37G5_W_1} input 37 slices with 5 gaps where reconstruction~\subref{p20230713_THz_Deer_S37G5_W_5} by $\mathscr{W_{\varepsilon}}$ and~\subref{p20230713_THz_Deer_S37G5_EE_35} by $\mathscr{E_{\varepsilon}}$. }
        \label{1_Deer}
    \end{figure}

\begin{figure}[htbp]
    \centering
        \begin{subfigure}[b]{0.16\textwidth}
        \begin{tikzpicture}
        \node
        {
        \adjincludegraphics[width=\textwidth, trim={{.2\width} {.12\height} {.3\width} {.15\height}}, clip]{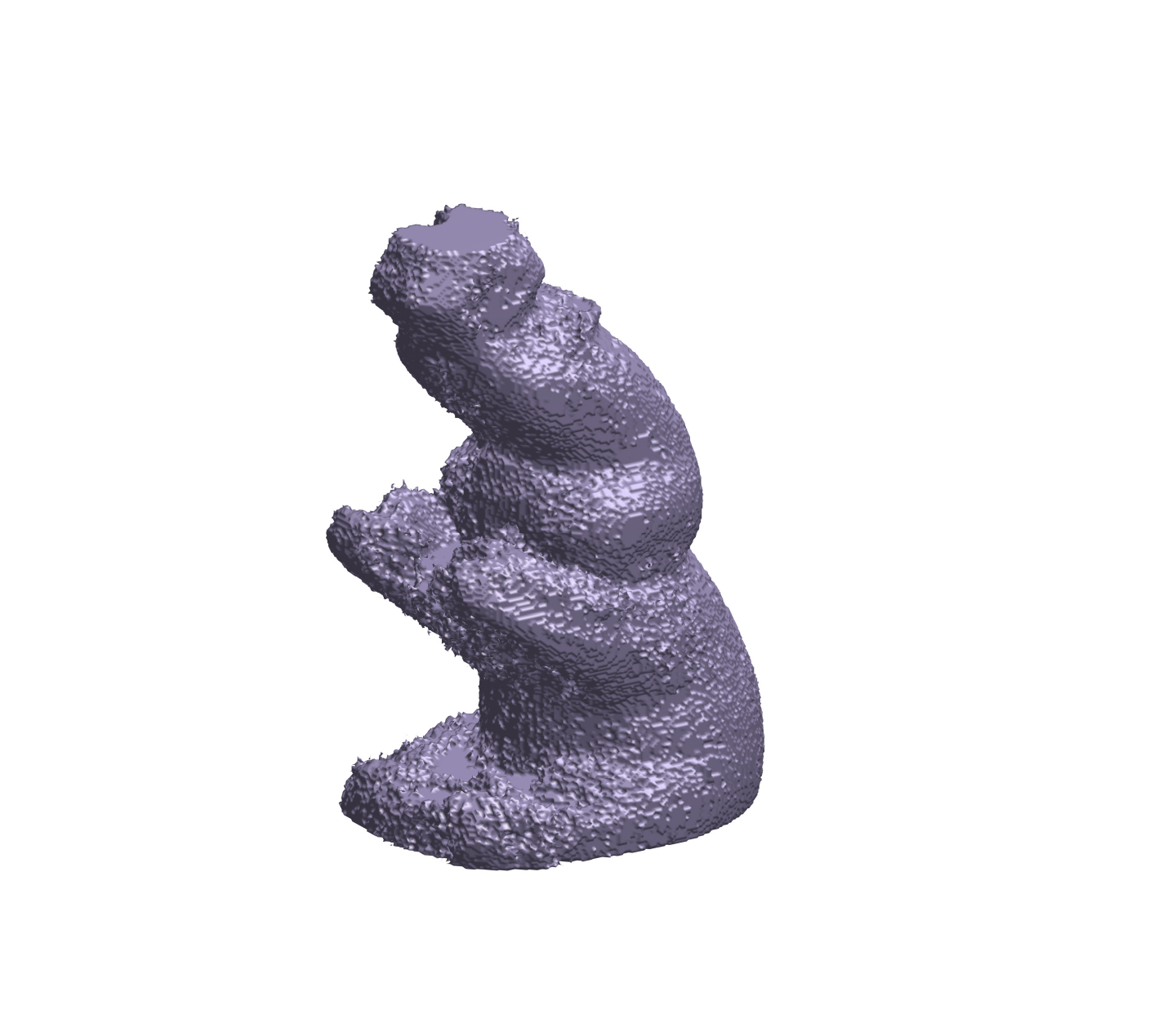}
        };
        
        \node at (-0.2,1.6) {{\scriptsize{$s = 258, \mbox{gap} = 0$}}};

        \node at (-1.5,0) {{\scriptsize{Conv.}}};
        \end{tikzpicture}
        \vspace{-0.6cm} \caption{}
        \label{p20230718_THz_Polarbear_S258G0_W_1}
        \end{subfigure}
        \,
        \begin{subfigure}[b]{0.16\textwidth}
        \begin{tikzpicture}
        \node
        {
        \adjincludegraphics[width=\textwidth, trim={{.2\width} {.12\height} {.3\width} {.15\height}}, clip]{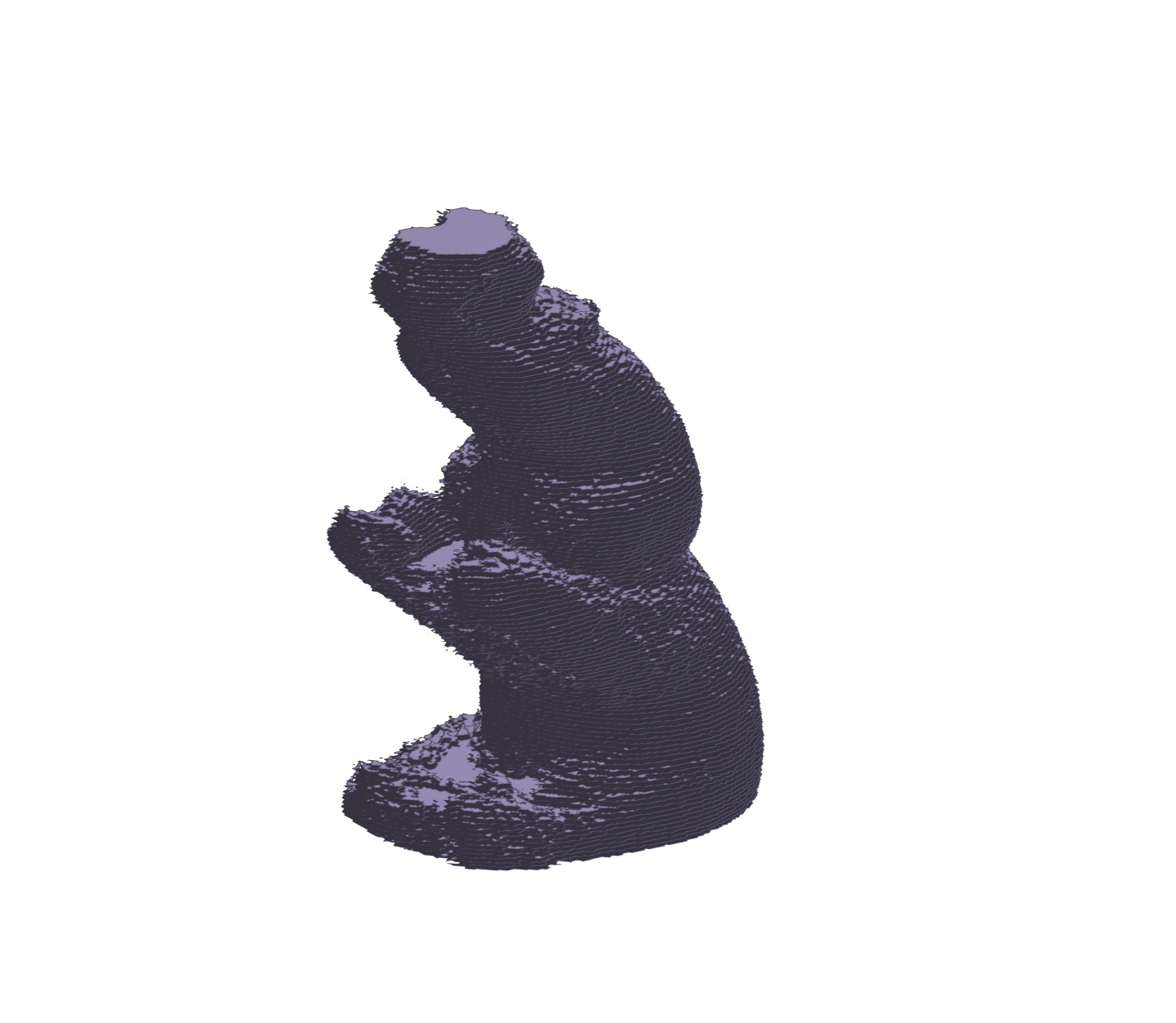}
	};
        
        \node at (-0.2,1.6) {{\scriptsize{$s = 129, \mbox{gap} = 1$}}};

        \node at (-1.5,0) {{\scriptsize{}}};
        \end{tikzpicture}
        \vspace{-0.6cm} \caption{}
        \label{p20230718_THz_Polarbear_S129G1_W_7}
        \end{subfigure}
        \,
        \begin{subfigure}[b]{0.16\textwidth}
        \begin{tikzpicture}
        \node
        {
        \adjincludegraphics[width=\textwidth, trim={{.2\width} {.12\height} {.3\width} {.15\height}}, clip]{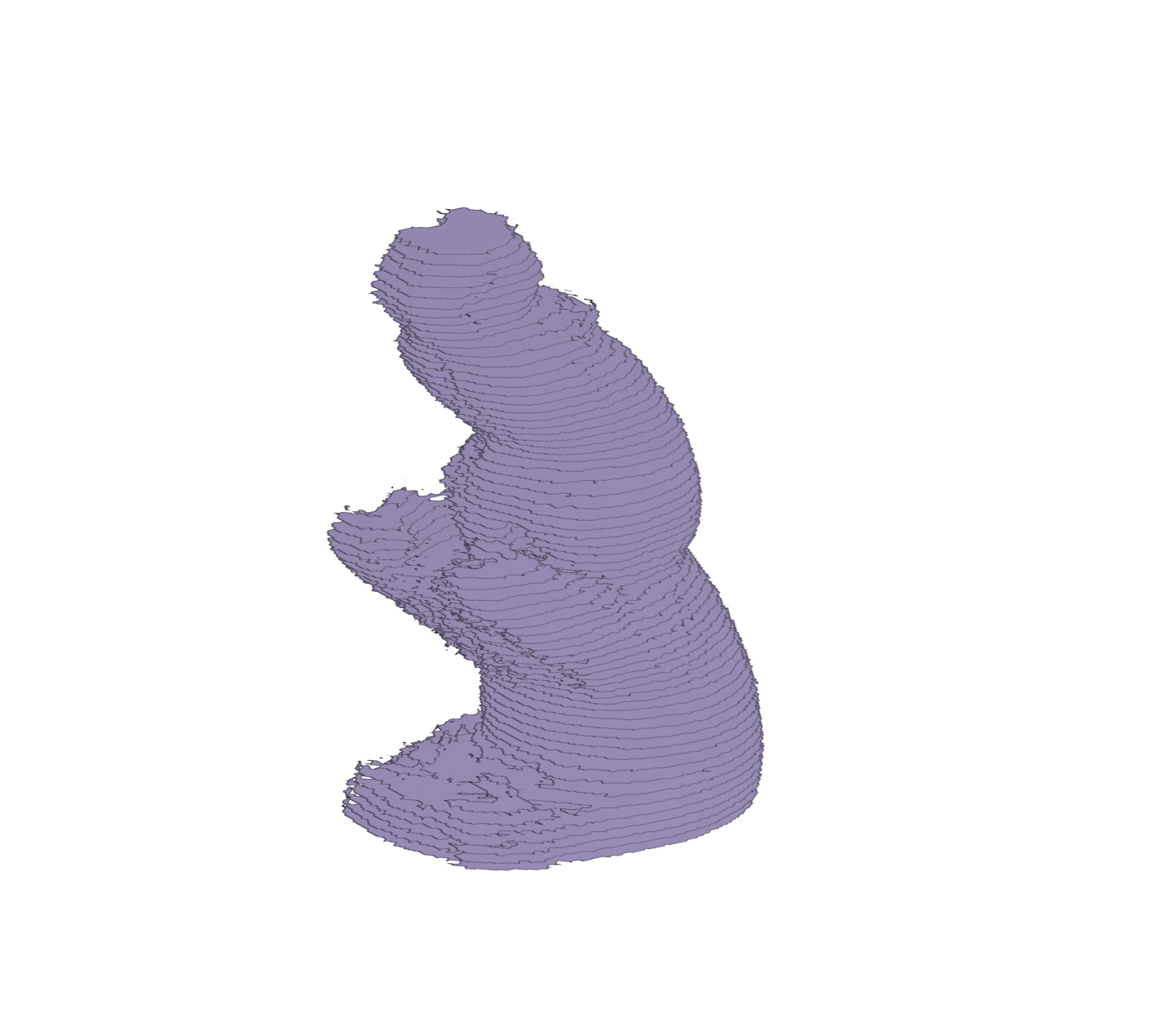}
	};
        
        \node at (-0.2,1.6) {{\scriptsize{$s = 65, \mbox{gap} = 3$}}};

        \node at (-1.5,0) {{\scriptsize{}}};
        \end{tikzpicture}
        \vspace{-0.6cm} \caption{}
        \label{p20230718_THz_Polarbear_S65G3_W_19}
        \end{subfigure}
        \,
        \begin{subfigure}[b]{0.16\textwidth}
        \begin{tikzpicture}
        \node
        {
        \adjincludegraphics[width=\textwidth, trim={{.2\width} {.12\height} {.3\width} {.15\height}}, clip]{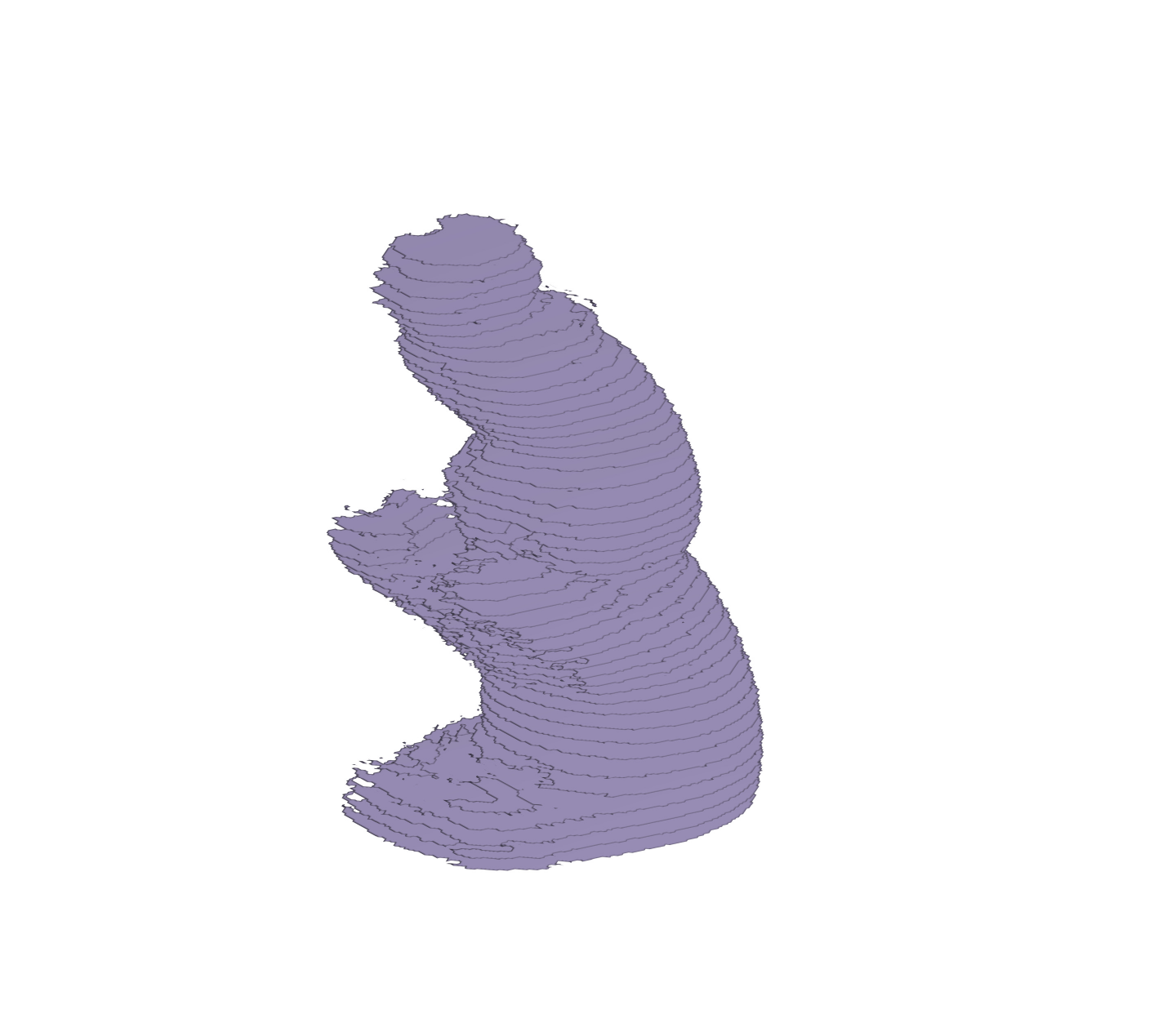}
	};
        
        \node at (-0.2,1.6) {{\scriptsize{$s = 43, \mbox{gap} = 5$}}};

        \node at (-1.5,0) {{\scriptsize{}}};
        \end{tikzpicture}
        \vspace{-0.6cm} \caption{}
        \label{p20230718_THz_Polarbear_S43G5_W_31}
        \end{subfigure}
        \\
        \vspace{-0.4cm}
        \begin{subfigure}[b]{0.16\textwidth}
        \begin{tikzpicture}
        \node
        {
        \adjincludegraphics[width=\textwidth, trim={{.2\width} {.12\height} {.3\width} {.15\height}}, clip]{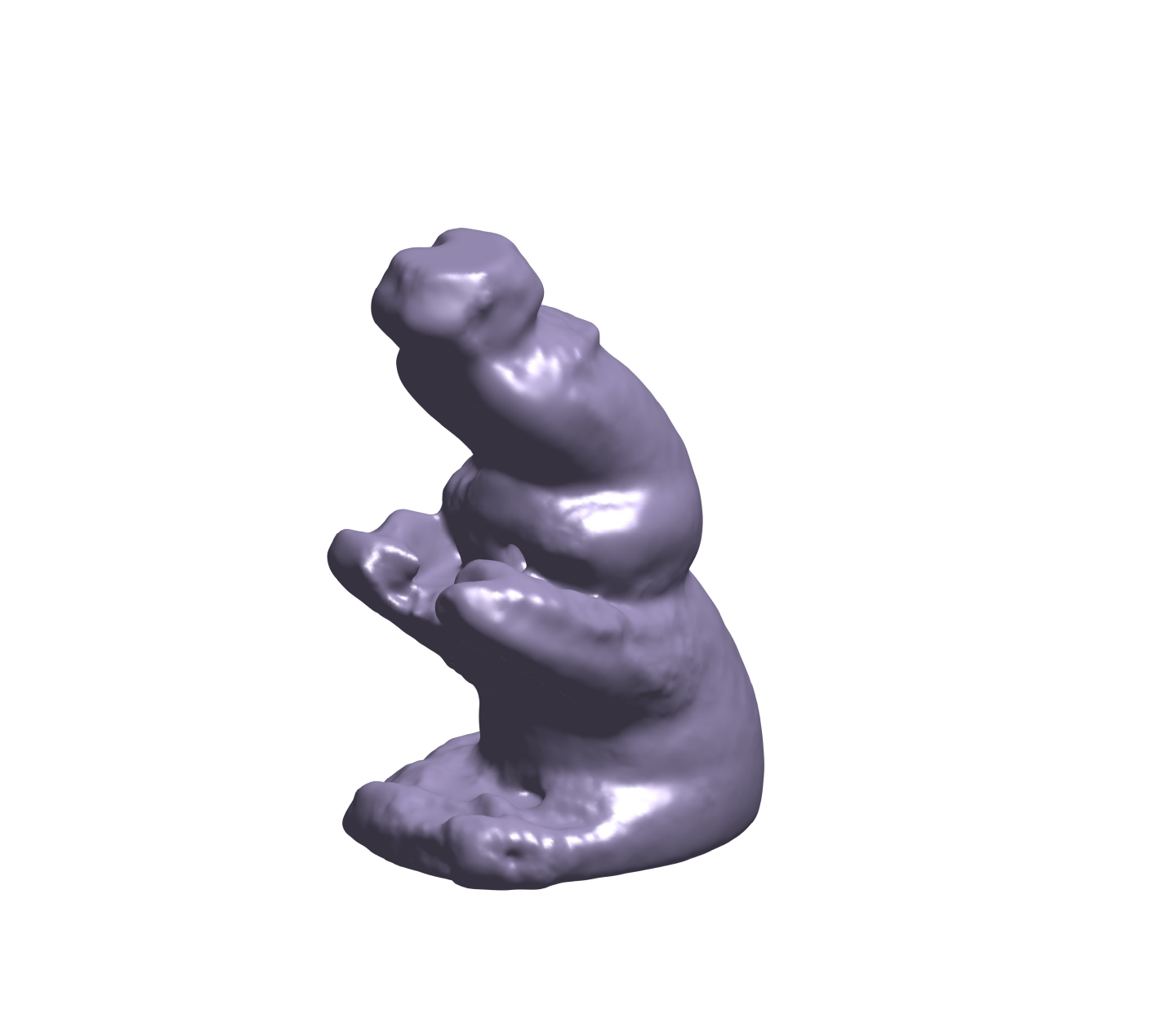}
	};
        
        \node at (-0.2,1.6) {{\scriptsize{}}};

        \node at (-1.5,0) {{\scriptsize{$\mathscr{W}_{\varepsilon}$}}};
        \end{tikzpicture}
        \vspace{-0.6cm} \caption{}
        \label{p20230718_THz_Polarbear_S258G0_W_5}
        \end{subfigure}
        \,
        \begin{subfigure}[b]{0.16\textwidth}
        \begin{tikzpicture}
        \node
        {
        \adjincludegraphics[width=\textwidth, trim={{.2\width} {.12\height} {.3\width} {.15\height}}, clip]{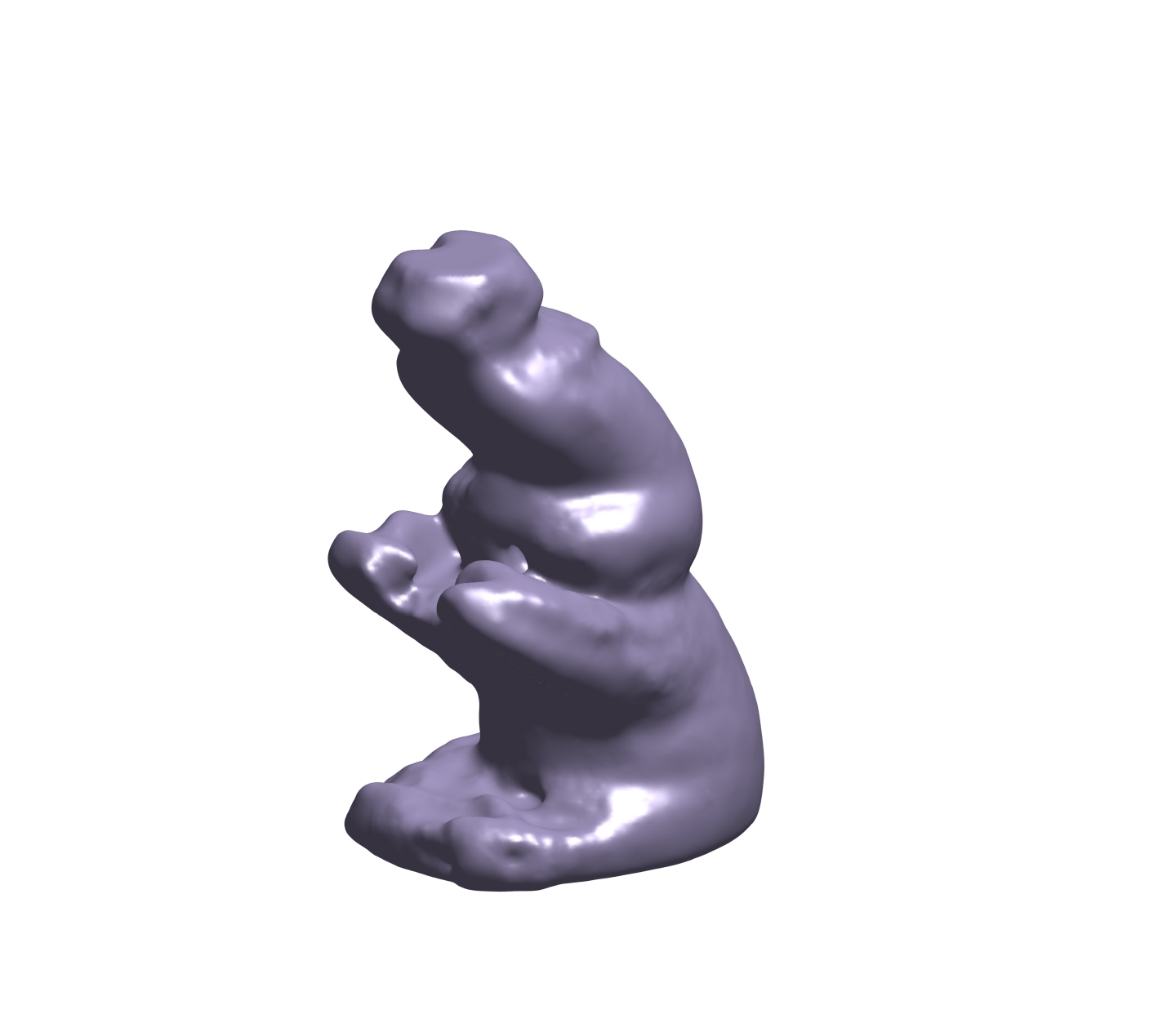}
	};
        
        \node at (-0.2,1.6) {{\scriptsize{}}};

        \node at (-1.5,0) {{\scriptsize{}}};
        \end{tikzpicture}
        \vspace{-0.6cm} \caption{}
        \label{p20230718_THz_Polarbear_S129G1_W_11}
        \end{subfigure}
        \,
        \begin{subfigure}[b]{0.16\textwidth}
        \begin{tikzpicture}
        \node
        {
        \adjincludegraphics[width=\textwidth, trim={{.2\width} {.12\height} {.3\width} {.15\height}}, clip]{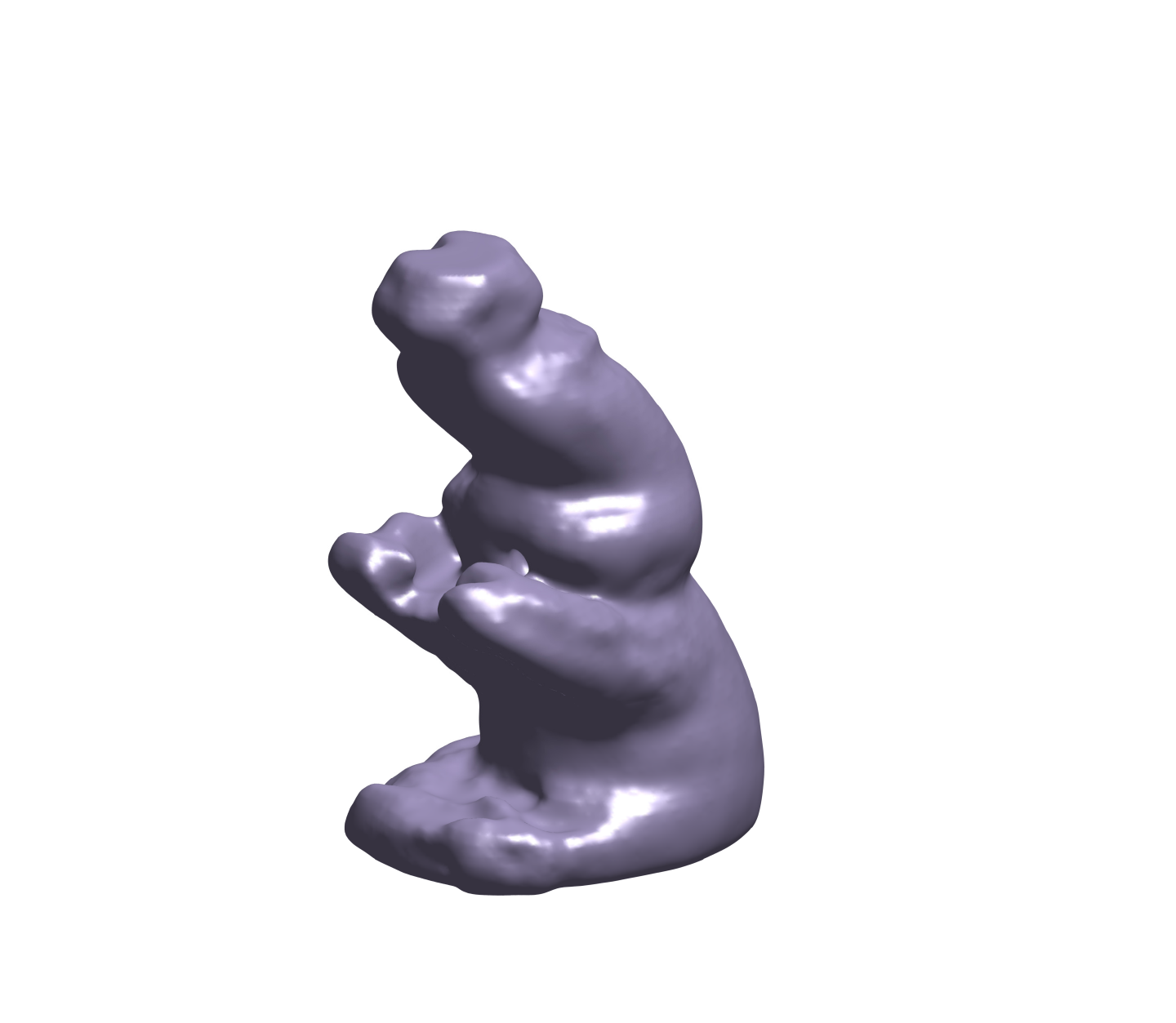}
	};
        
        \node at (-0.2,1.6) {{\scriptsize{}}};

        \node at (-1.5,0) {{\scriptsize{}}};
        \end{tikzpicture}
        \vspace{-0.6cm} \caption{}
        \label{p20230718_THz_Polarbear_S65G3_W_23}
        \end{subfigure}
        \,
        \begin{subfigure}[b]{0.16\textwidth}
        \begin{tikzpicture}
        \node
        {
        \adjincludegraphics[width=\textwidth, trim={{.2\width} {.12\height} {.3\width} {.15\height}}, clip]{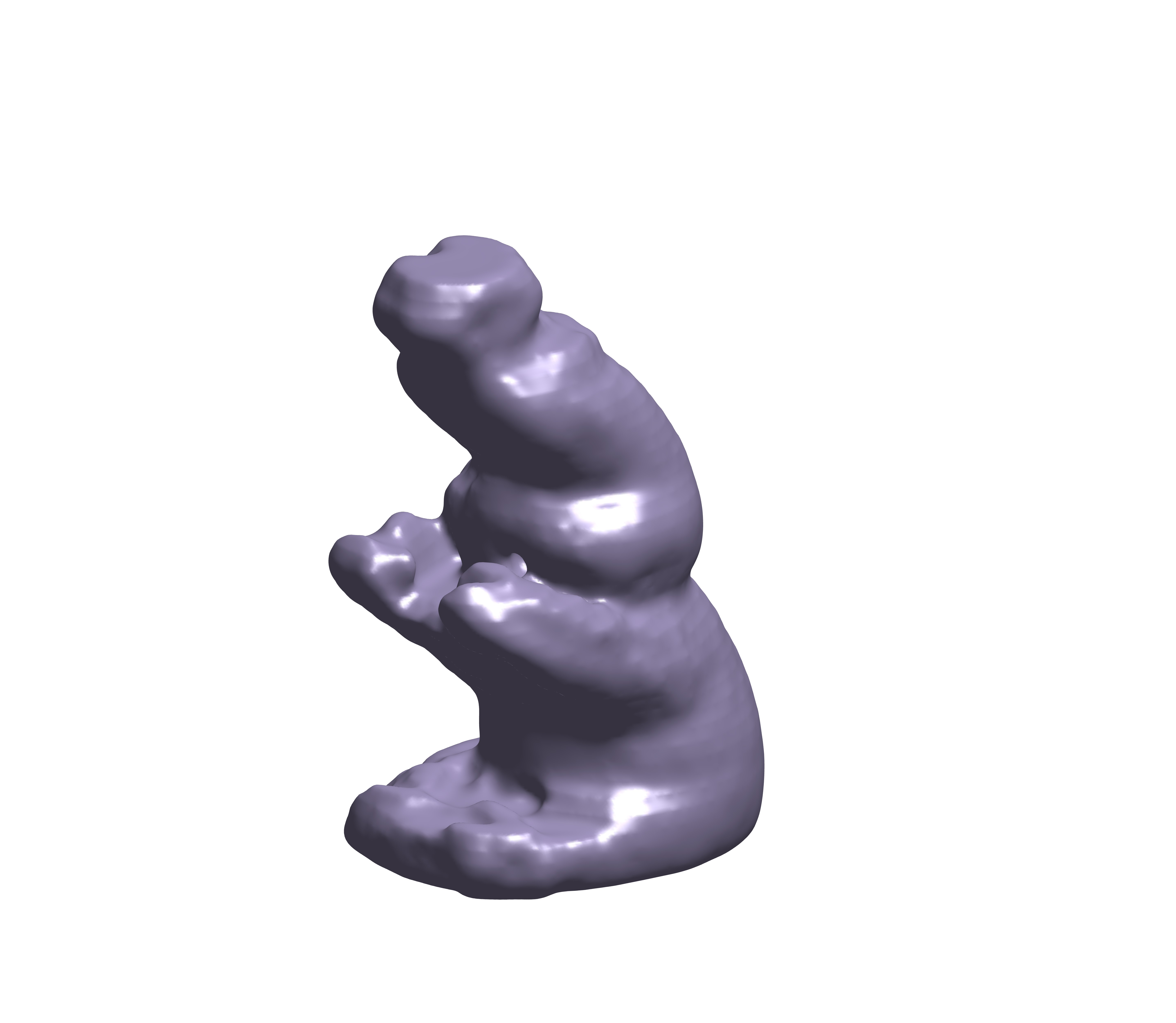}
	};
        
        \node at (-0.2,1.6) {{\scriptsize{}}};

        \node at (-1.5,0) {{\scriptsize{}}};
        \end{tikzpicture}
        \vspace{-0.6cm} \caption{}
        \label{p20230718_THz_Polarbear_S43G5_W_35}
        \end{subfigure}
        \\
        \vspace{-0.4cm}
        \begin{subfigure}[b]{0.16\textwidth}
        \begin{tikzpicture}
        \node
        {
        \adjincludegraphics[width=\textwidth, trim={{.2\width} {.12\height} {.3\width} {.15\height}}, clip]{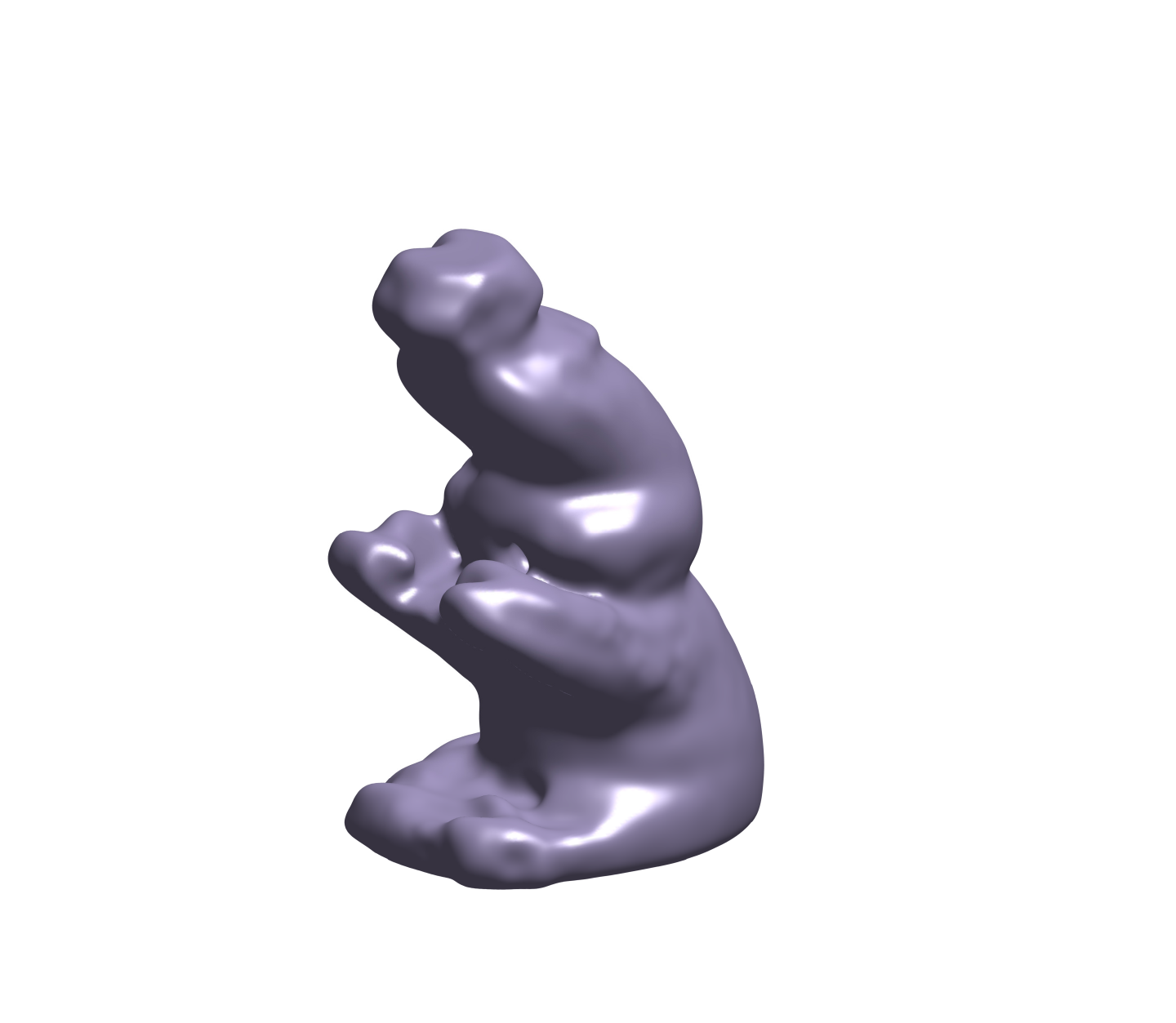}
	};
        
        \node at (-0.2,1.6) {{\scriptsize{}}};

        \node at (-1.5,0) {{\scriptsize{$\mathscr{E}_{\varepsilon}$}}};
        \end{tikzpicture}
        \vspace{-0.6cm} \caption{}
        \label{p20230717_THz_Polarbear_S258G0_EE_5}
        \end{subfigure}
        \,
        \begin{subfigure}[b]{0.16\textwidth}
        \begin{tikzpicture}
        \node
        {
        \adjincludegraphics[width=\textwidth, trim={{.2\width} {.12\height} {.3\width} {.15\height}}, clip]{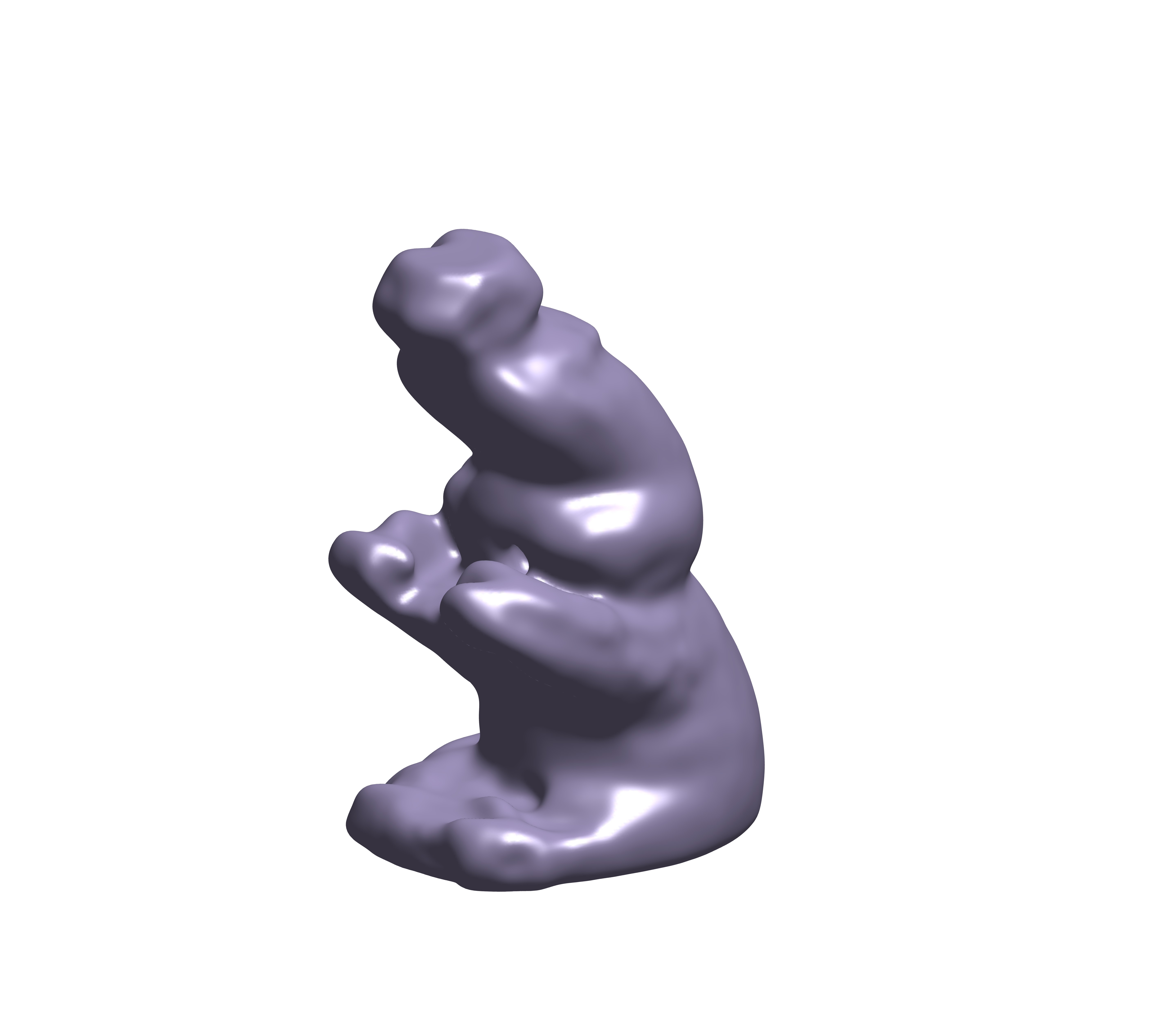}
	};
        
        \node at (-0.2,1.6) {{\scriptsize{}}};

        \node at (-1.5,0) {{\scriptsize{}}};
        \end{tikzpicture}
        \vspace{-0.6cm} \caption{}
        \label{p20230717_THz_Polarbear_S129G1_EE_11}
        \end{subfigure}
        \,
        \begin{subfigure}[b]{0.16\textwidth}
        \begin{tikzpicture}
        \node
        {
        \adjincludegraphics[width=\textwidth, trim={{.2\width} {.12\height} {.3\width} {.15\height}}, clip]{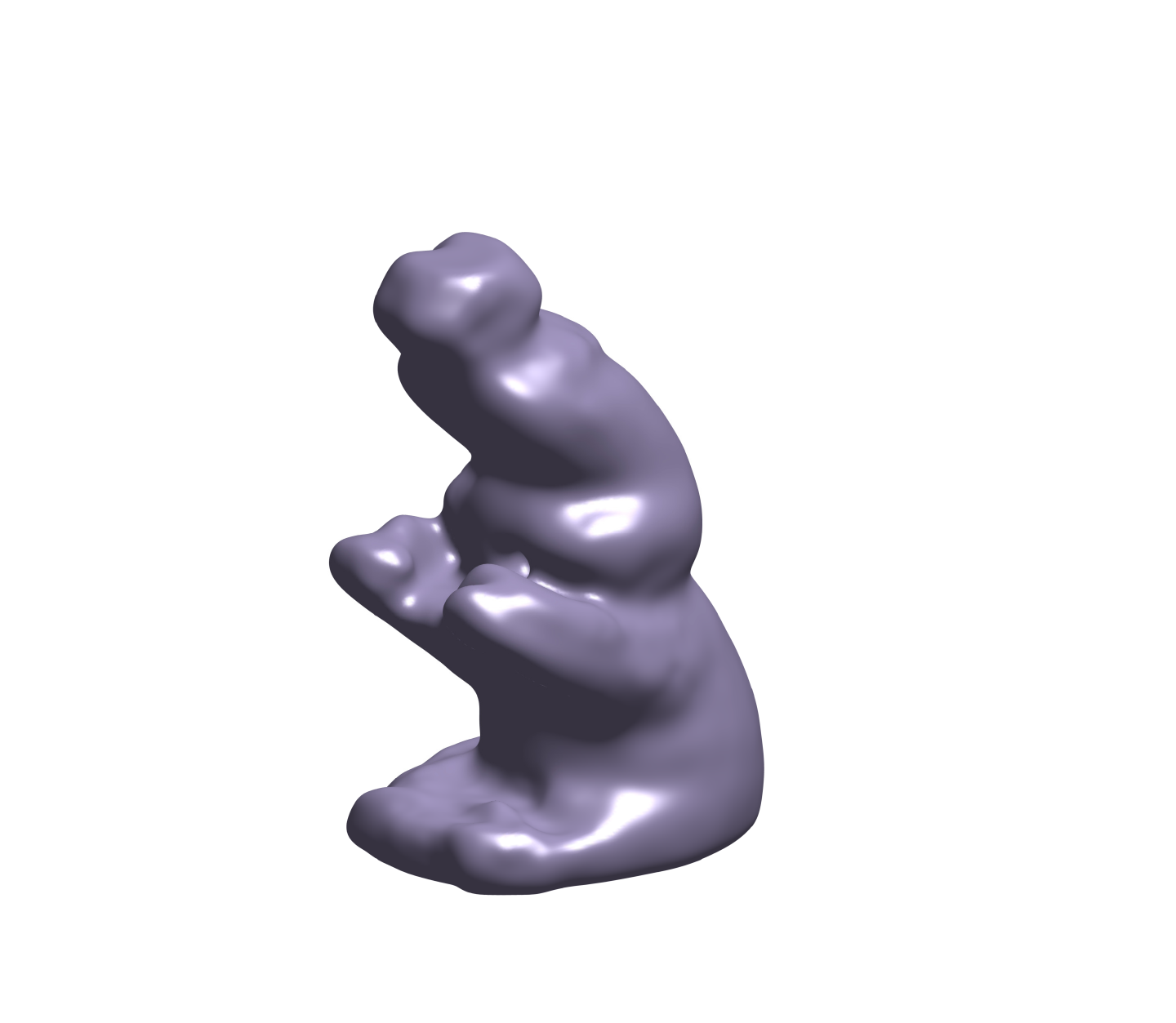}
	};
        
        \node at (-0.2,1.6) {{\scriptsize{}}};

        \node at (-1.5,0) {{\scriptsize{}}};
        \end{tikzpicture}
        \vspace{-0.6cm} \caption{}
        \label{p20230716_THz_Polarbear_S65G3_EE_23}
        \end{subfigure}
        \,
        \begin{subfigure}[b]{0.16\textwidth}
        \begin{tikzpicture}
        \node
        {
        \adjincludegraphics[width=\textwidth, trim={{.2\width} {.12\height} {.3\width} {.15\height}}, clip]{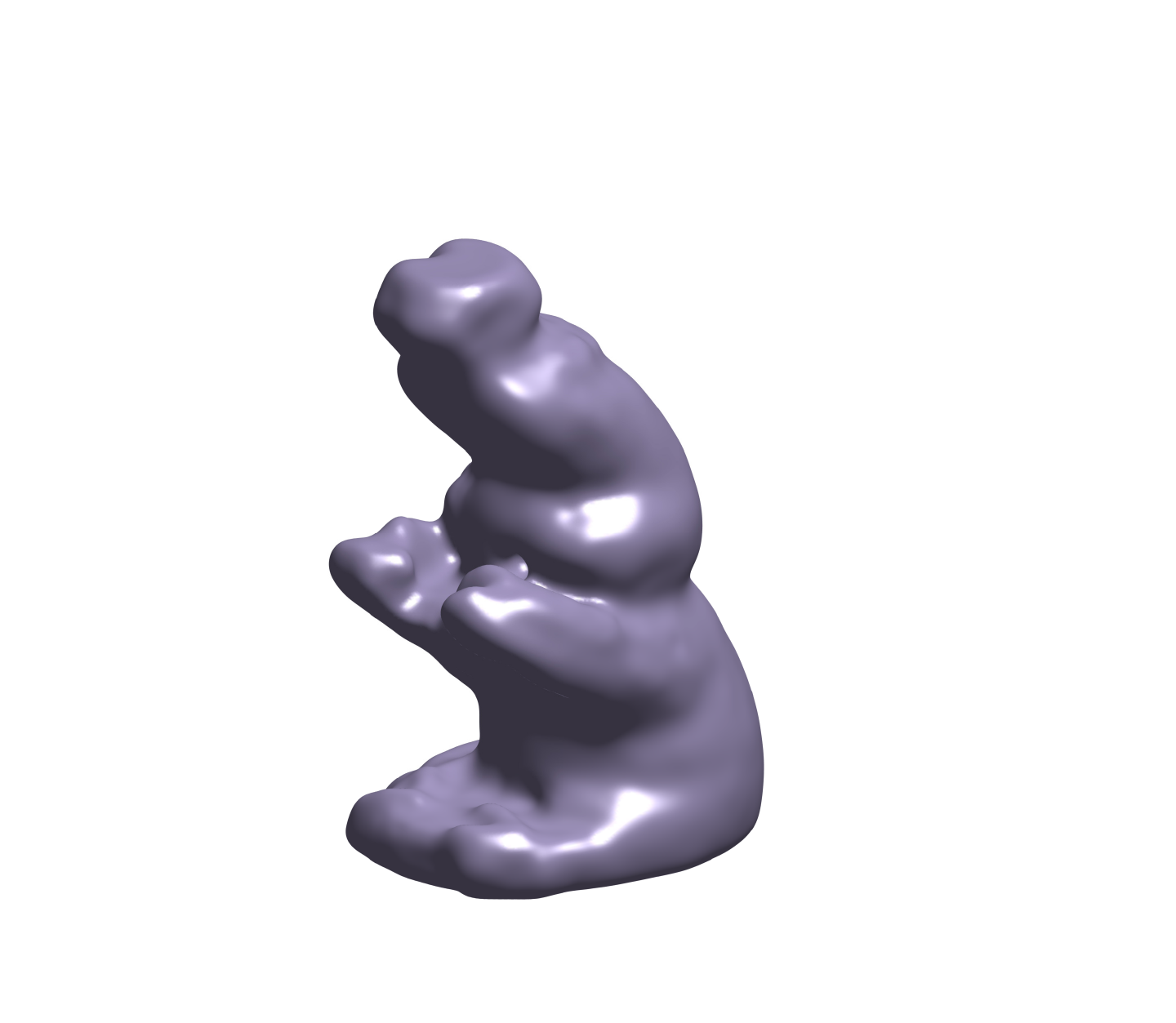}
	};
        
        \node at (-0.2,1.6) {{\scriptsize{}}};

        \node at (-1.5,0) {{\scriptsize{}}};
        \end{tikzpicture}
        \vspace{-0.6cm} \caption{}
        \label{p20230716_THz_Polarbear_S43G5_EE_35}
        \end{subfigure}
        \vspace{-0.3cm}
        \caption{Illustrations of Polarbear object:~\subref{p20230718_THz_Polarbear_S258G0_W_1} conventional LR reconstruction with 258 slices where smooth reconstruction~\subref{p20230718_THz_Polarbear_S258G0_W_5} by Willmore-based $\mathscr{W_{\varepsilon}}$ and~\subref{p20230717_THz_Polarbear_S258G0_EE_5} by Euler-Elastica-based formulation $\mathscr{E_{\varepsilon}}$;~\subref{p20230718_THz_Polarbear_S129G1_W_7} input 129 slices with 1 gap where reconstruction~\subref{p20230718_THz_Polarbear_S129G1_W_11} by $\mathscr{W_{\varepsilon}}$ and~\subref{p20230717_THz_Polarbear_S129G1_EE_11} by $\mathscr{E_{\varepsilon}}$;~\subref{p20230718_THz_Polarbear_S65G3_W_19} input 65 slices with 3 gaps where reconstruction~\subref{p20230718_THz_Polarbear_S65G3_W_23} by $\mathscr{W_{\varepsilon}}$ and~\subref{p20230716_THz_Polarbear_S65G3_EE_23} by $\mathscr{E_{\varepsilon}}$;~\subref{p20230718_THz_Polarbear_S43G5_W_31} input 43 slices with 5 gaps where reconstruction~\subref{p20230718_THz_Polarbear_S43G5_W_35} by $\mathscr{W_{\varepsilon}}$ and~\subref{p20230716_THz_Polarbear_S43G5_EE_35} by $\mathscr{E_{\varepsilon}}$. }
        \label{2_Polarbear}
    \end{figure}

\begin{figure}[htbp]
    \centering
        \begin{subfigure}[b]{0.16\textwidth}
        \begin{tikzpicture}
        \node
        {
        \adjincludegraphics[width=\textwidth, trim={{.2\width} {.18\height} {.3\width} {.15\height}}, clip]{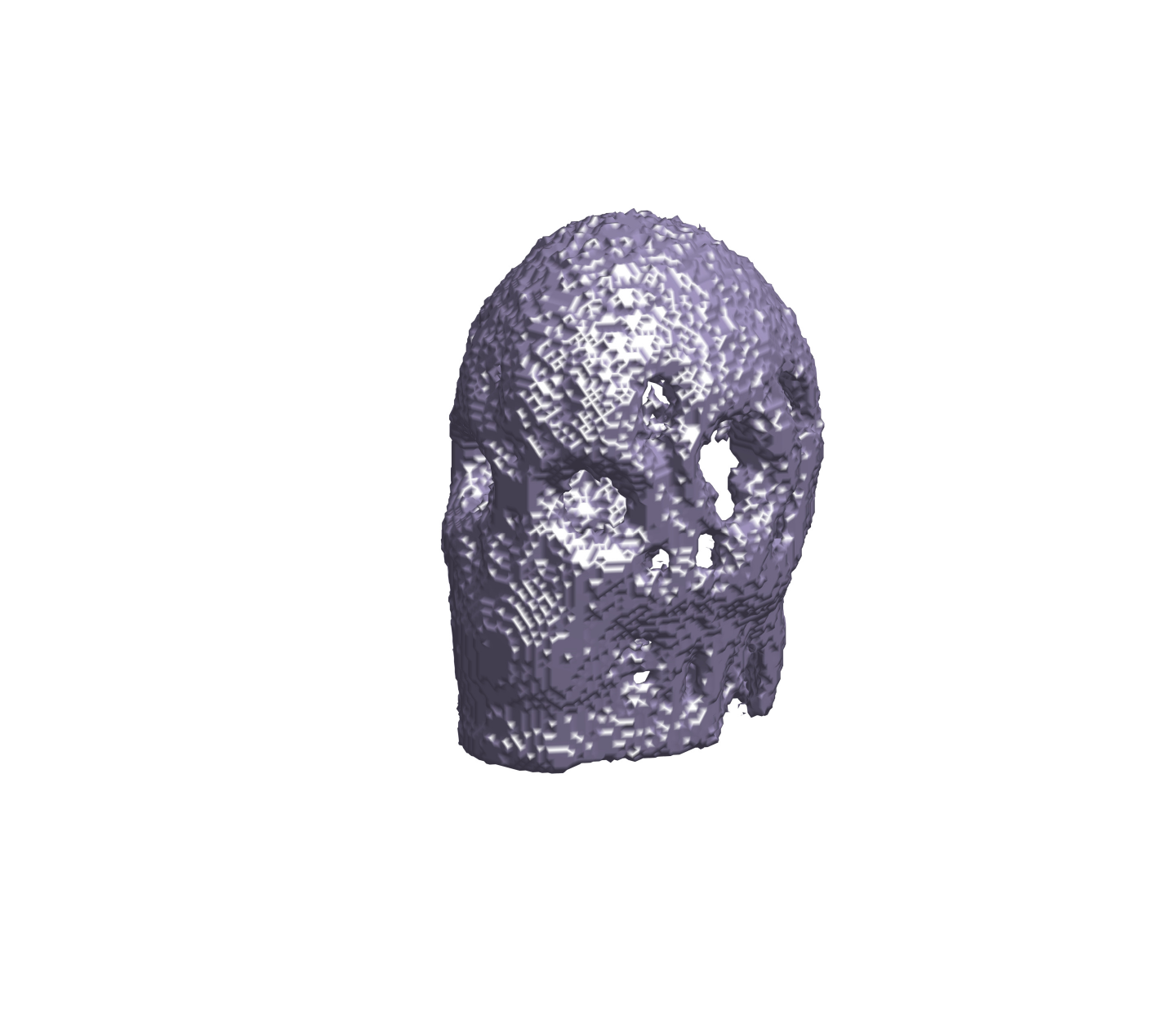}
        };
        
        \node at (0.3,1.4) {{\scriptsize{$s = 138, \mbox{gap} = 0$}}};

        \node at (-1.,0) {{\scriptsize{Conv.}}};
        \end{tikzpicture}
        \vspace{-0.6cm} \caption{}
        \label{p20230718_THz_Skull_S138G0_W_1}
        \end{subfigure}
        \,
        \begin{subfigure}[b]{0.16\textwidth}
        \begin{tikzpicture}
        \node
        {
        \adjincludegraphics[width=\textwidth, trim={{.2\width} {.18\height} {.3\width} {.15\height}}, clip]{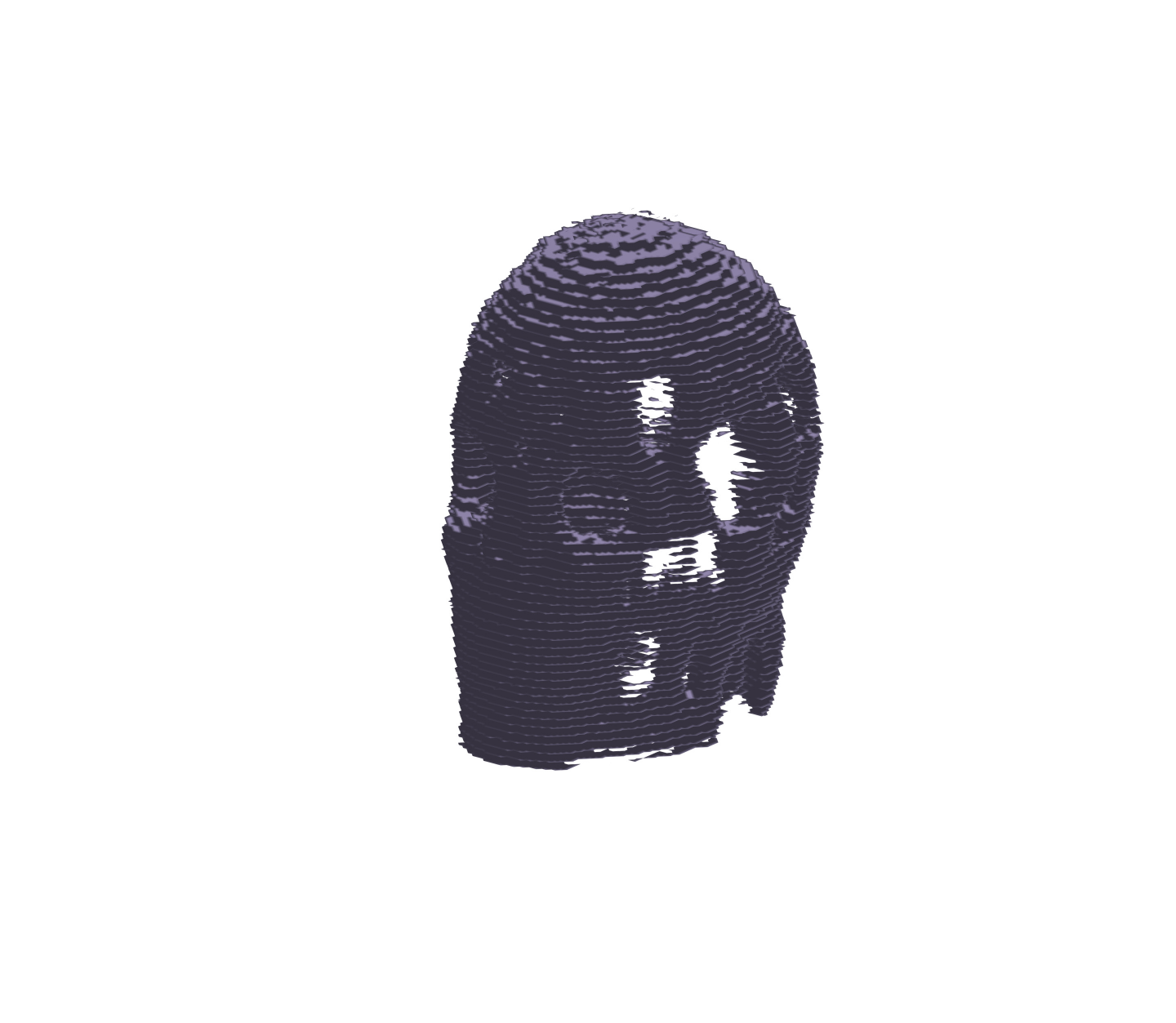}
	};
        
        \node at (0.3,1.4) {{\scriptsize{$s = 69, \mbox{gap} = 1$}}};

       \node at (-1.,0) {{\scriptsize{}}};
        \end{tikzpicture}
        \vspace{-0.6cm} \caption{}
        \label{p20230718_THz_Skull_S69G1_W_7}
        \end{subfigure}
        \,
        \begin{subfigure}[b]{0.16\textwidth}
        \begin{tikzpicture}
        \node
        {
        \adjincludegraphics[width=\textwidth, trim={{.2\width} {.18\height} {.3\width} {.15\height}}, clip]{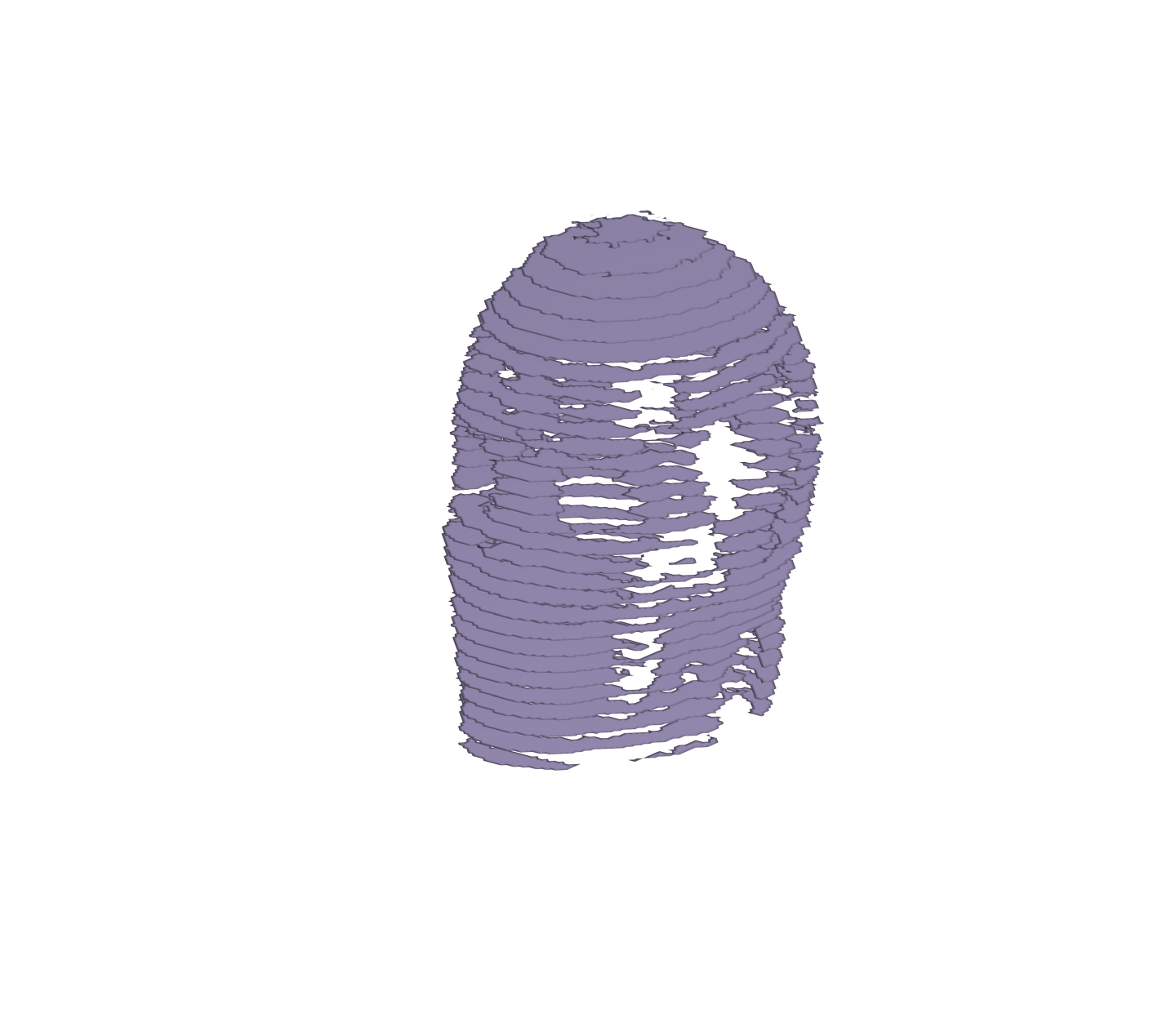}
	};
        
        \node at (0.3,1.4) {{\scriptsize{$s = 35, \mbox{gap} = 3$}}};

       \node at (-1.,0) {{\scriptsize{}}};
        \end{tikzpicture}
        \vspace{-0.6cm} \caption{}
        \label{p20230718_THz_Skull_S35G3_W_19}
        \end{subfigure}
        \,
        \begin{subfigure}[b]{0.16\textwidth}
        \begin{tikzpicture}
        \node
        {
        \adjincludegraphics[width=\textwidth, trim={{.2\width} {.18\height} {.3\width} {.15\height}}, clip]{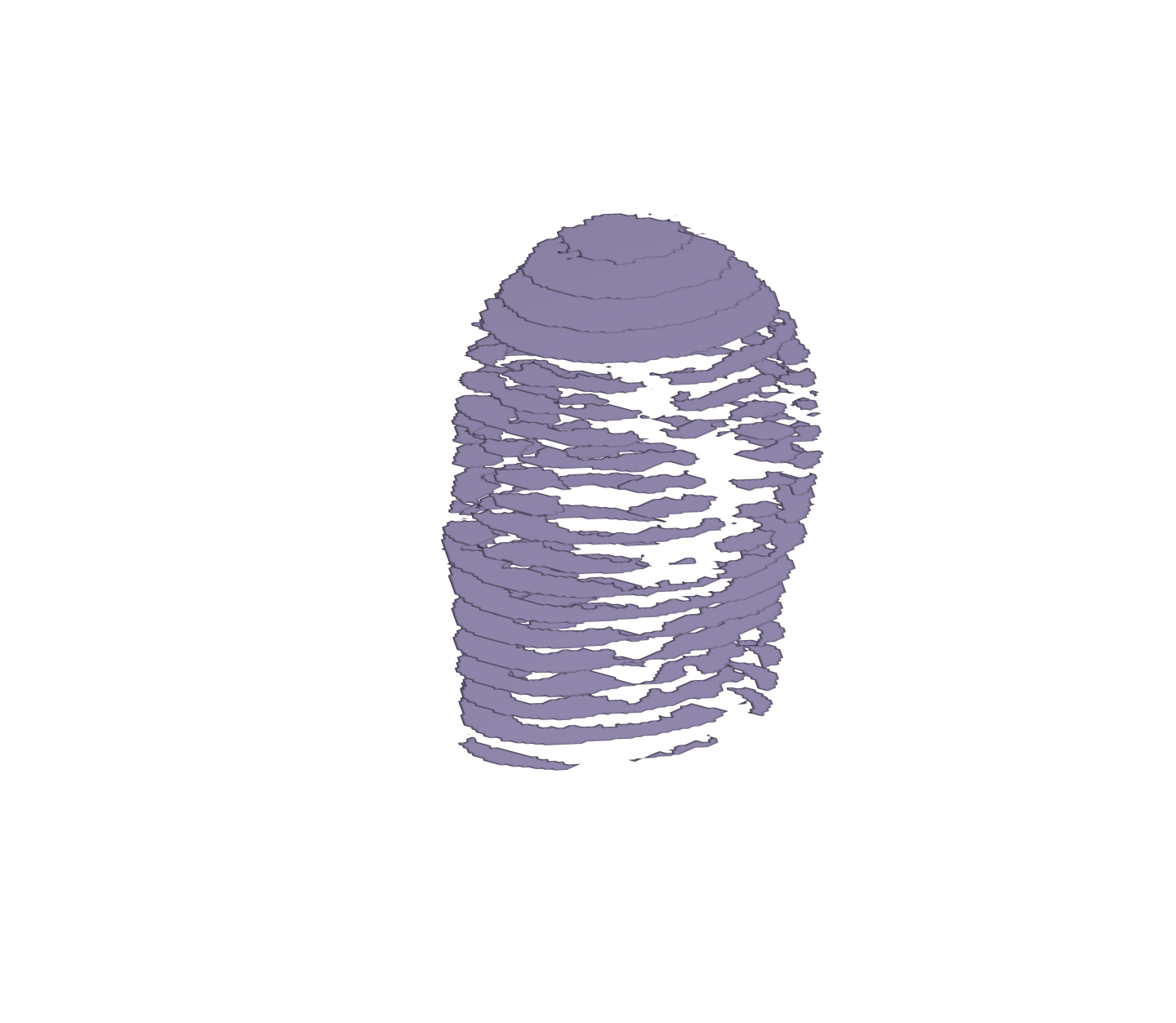}
	};
        
        \node at (0.3,1.4) {{\scriptsize{$s = 23, \mbox{gap} = 5$}}};

       \node at (-1.,0) {{\scriptsize{}}};
        \end{tikzpicture}
        \vspace{-0.6cm} \caption{}
        \label{p20230718_THz_Skull_S23G5_W_31}
        \end{subfigure}
        \\
        \vspace{-0.3cm}
        \begin{subfigure}[b]{0.16\textwidth}
        \begin{tikzpicture}
        \node
        {
        \adjincludegraphics[width=\textwidth, trim={{.2\width} {.18\height} {.3\width} {.15\height}}, clip]{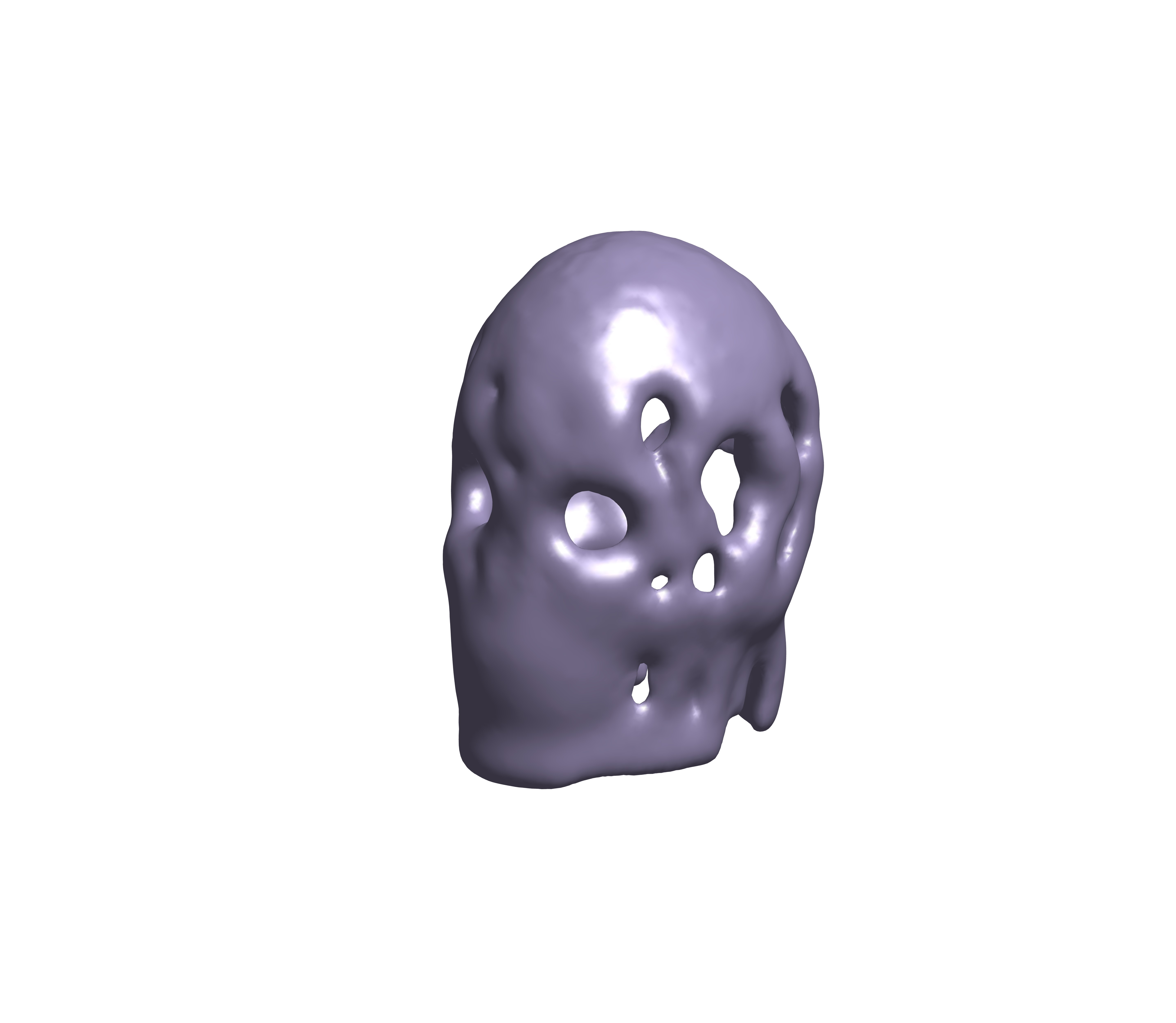}
	};
        
        \node at (0.3,1.4) {{\scriptsize{}}};

       \node at (-1.,0) {{\scriptsize{$\mathscr{W}_{\varepsilon}$}}};
        \end{tikzpicture}
        \vspace{-0.6cm} \caption{}
        \label{p20230718_THz_Skull_S138G0_W_5}
        \end{subfigure}
        \,
        \begin{subfigure}[b]{0.16\textwidth}
        \begin{tikzpicture}
        \node
        {
        \adjincludegraphics[width=\textwidth, trim={{.2\width} {.18\height} {.3\width} {.15\height}}, clip]{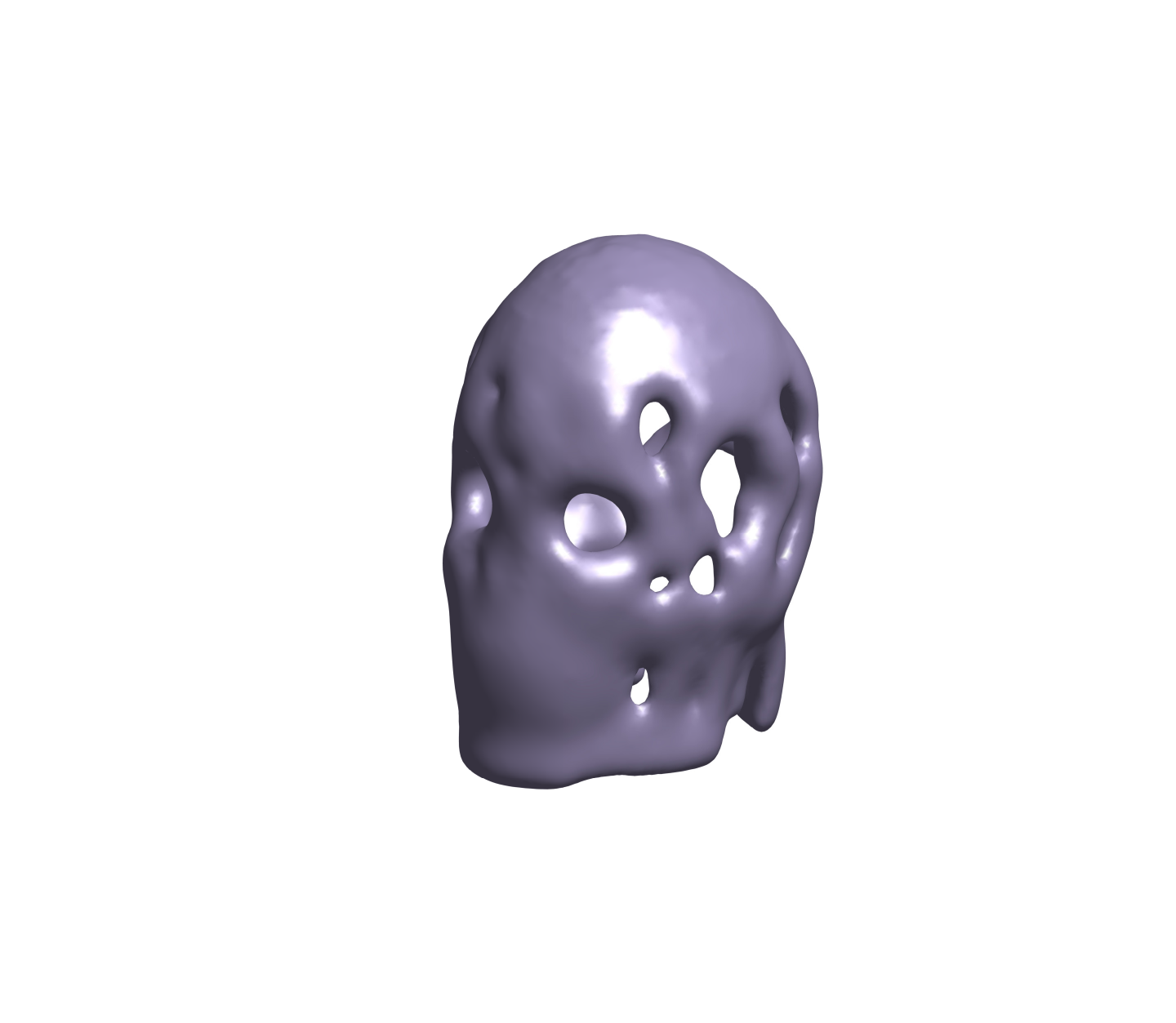}
	};
        
        \node at (0.3,1.4) {{\scriptsize{}}};

       \node at (-1.,0) {{\scriptsize{}}};
        \end{tikzpicture}
        \vspace{-0.6cm} \caption{}
        \label{p20230718_THz_Skull_S69G1_W_11}
        \end{subfigure}
        \,
        \begin{subfigure}[b]{0.16\textwidth}
        \begin{tikzpicture}
        \node
        {
        \adjincludegraphics[width=\textwidth, trim={{.2\width} {.18\height} {.3\width} {.15\height}}, clip]{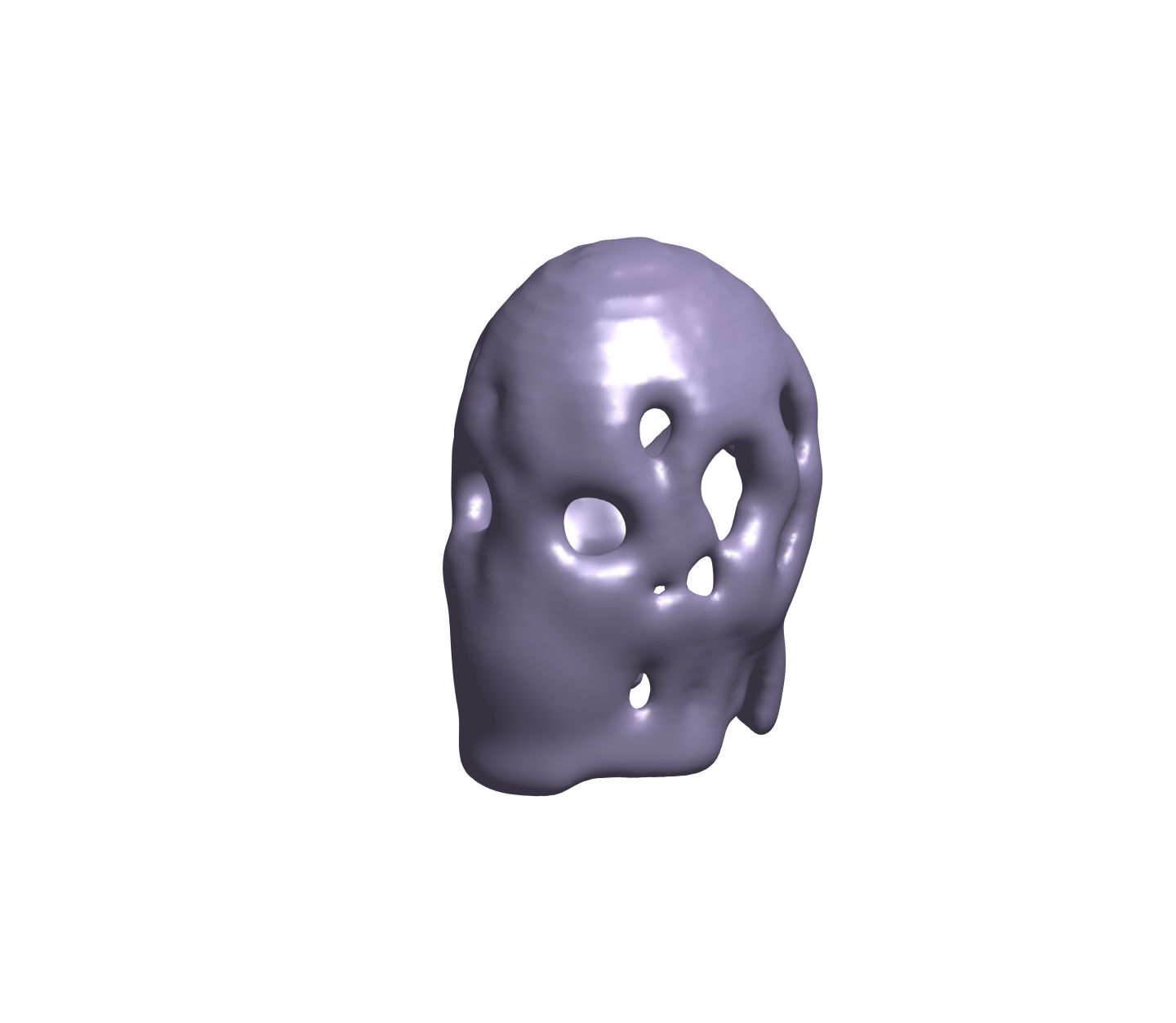}
	};
        
        \node at (0.3,1.4) {{\scriptsize{}}};

       \node at (-1.,0) {{\scriptsize{}}};
        \end{tikzpicture}
        \vspace{-0.6cm} \caption{}
        \label{p20230718_THz_Skull_S35G3_W_23}
        \end{subfigure}
        \,
        \begin{subfigure}[b]{0.16\textwidth}
        \begin{tikzpicture}
        \node
        {
        \adjincludegraphics[width=\textwidth, trim={{.2\width} {.18\height} {.3\width} {.15\height}}, clip]{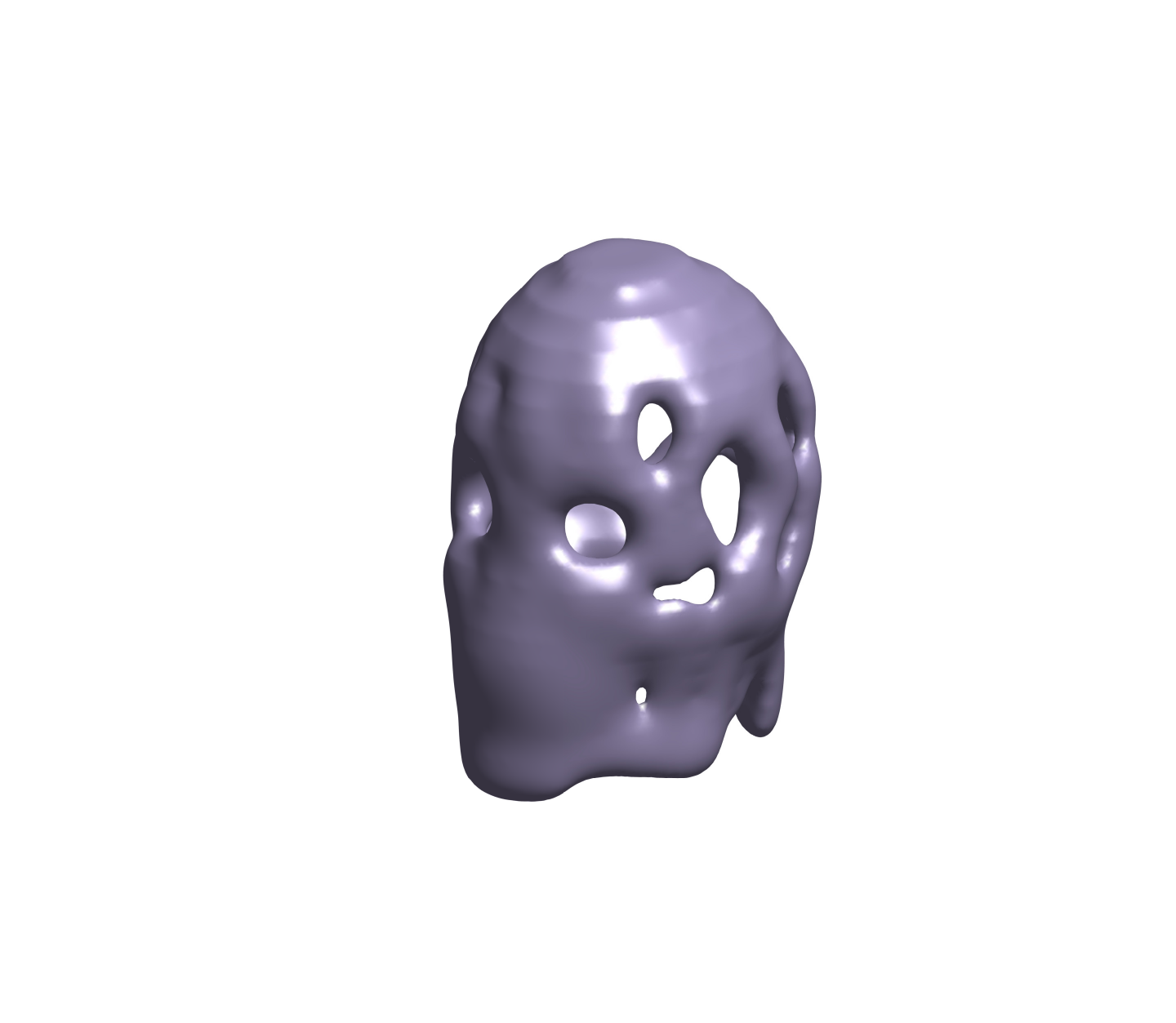}
	};
        
        \node at (0.3,1.4) {{\scriptsize{}}};

       \node at (-1.,0) {{\scriptsize{}}};
        \end{tikzpicture}
        \vspace{-0.6cm} \caption{}
        \label{p20230718_THz_Skull_S23G5_W_35}
        \end{subfigure}
        \\
        \vspace{-0.3cm}
        \begin{subfigure}[b]{0.16\textwidth}
        \begin{tikzpicture}
        \node
        {
        \adjincludegraphics[width=\textwidth, trim={{.2\width} {.18\height} {.3\width} {.15\height}}, clip]{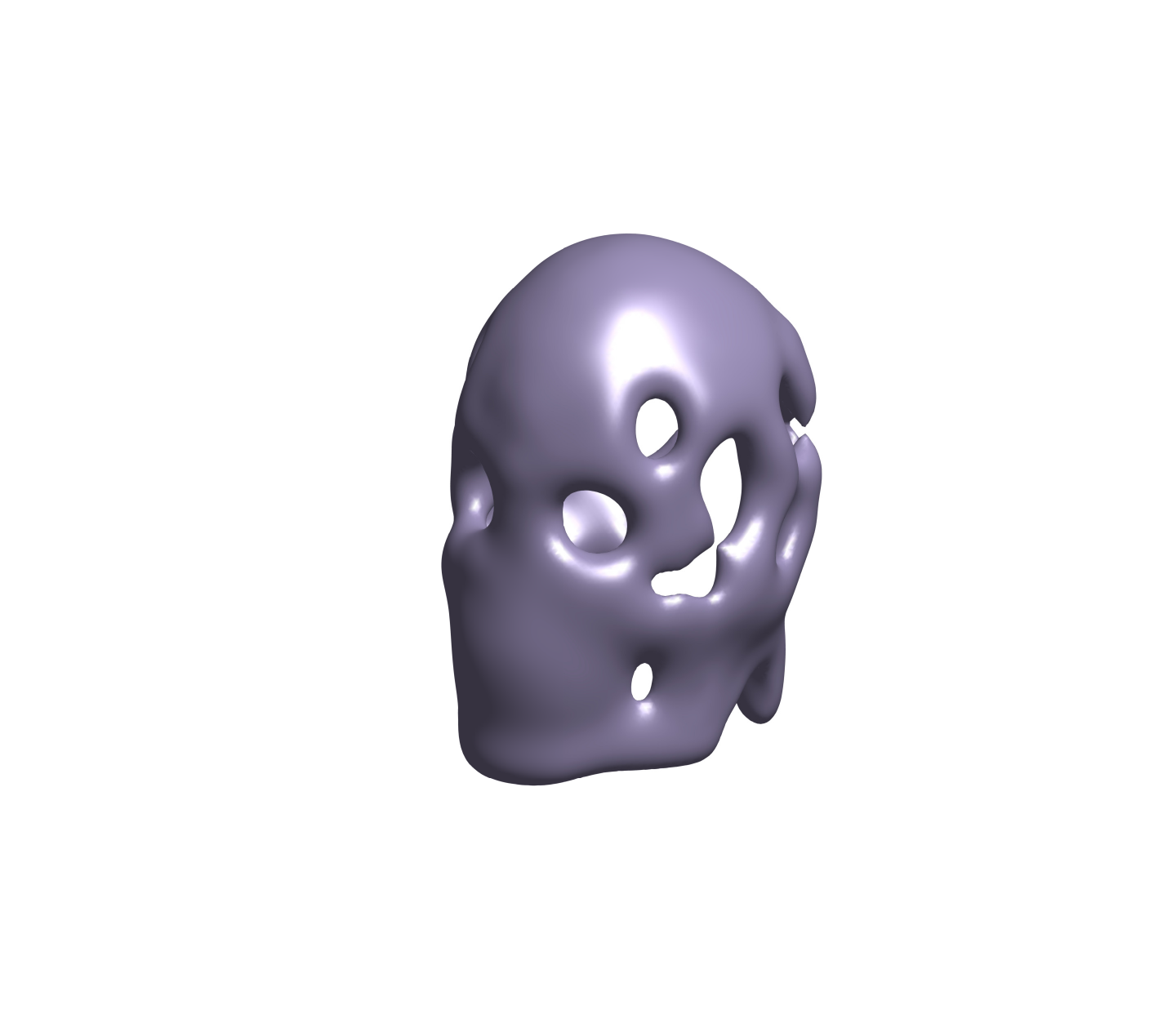}
	};
        
        \node at (0.3,1.4) {{\scriptsize{}}};

       \node at (-1.,0) {{\scriptsize{$\mathscr{E}_{\varepsilon}$}}};
        \end{tikzpicture}
        \vspace{-0.6cm} \caption{}
        \label{p20230716_THz_Skull_S138G0_EE_5}
        \end{subfigure}
        \,
        \begin{subfigure}[b]{0.16\textwidth}
        \begin{tikzpicture}
        \node
        {
        \adjincludegraphics[width=\textwidth, trim={{.2\width} {.18\height} {.3\width} {.15\height}}, clip]{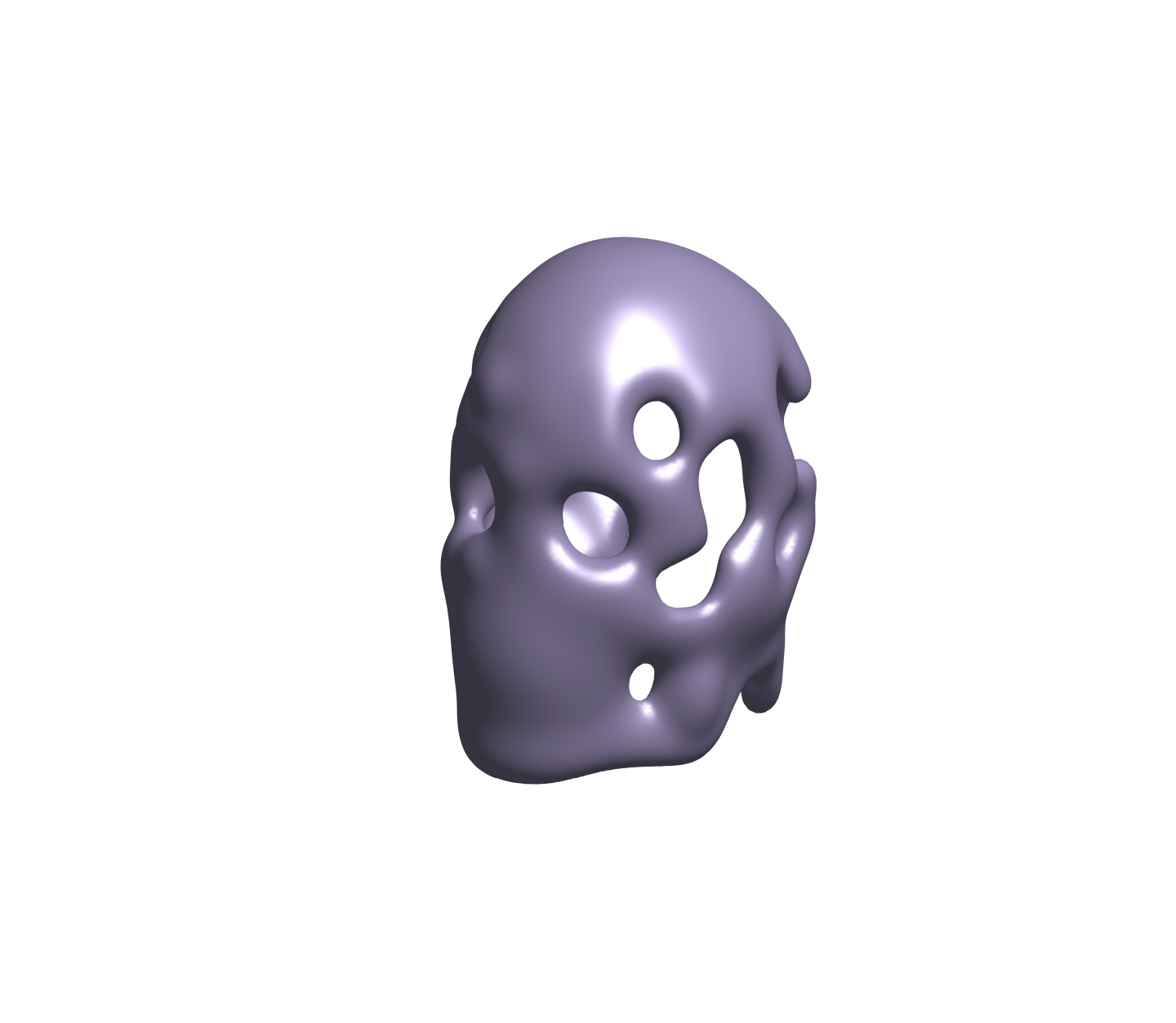}
	};
        
        \node at (0.3,1.4) {{\scriptsize{}}};

       \node at (-1.,0) {{\scriptsize{}}};
        \end{tikzpicture}
        \vspace{-0.6cm} \caption{}
        \label{p20230716_THz_Skull_S69G1_EE_11}
        \end{subfigure}
        \,
        \begin{subfigure}[b]{0.16\textwidth}
        \begin{tikzpicture}
        \node
        {
        \adjincludegraphics[width=\textwidth, trim={{.2\width} {.18\height} {.3\width} {.15\height}}, clip]{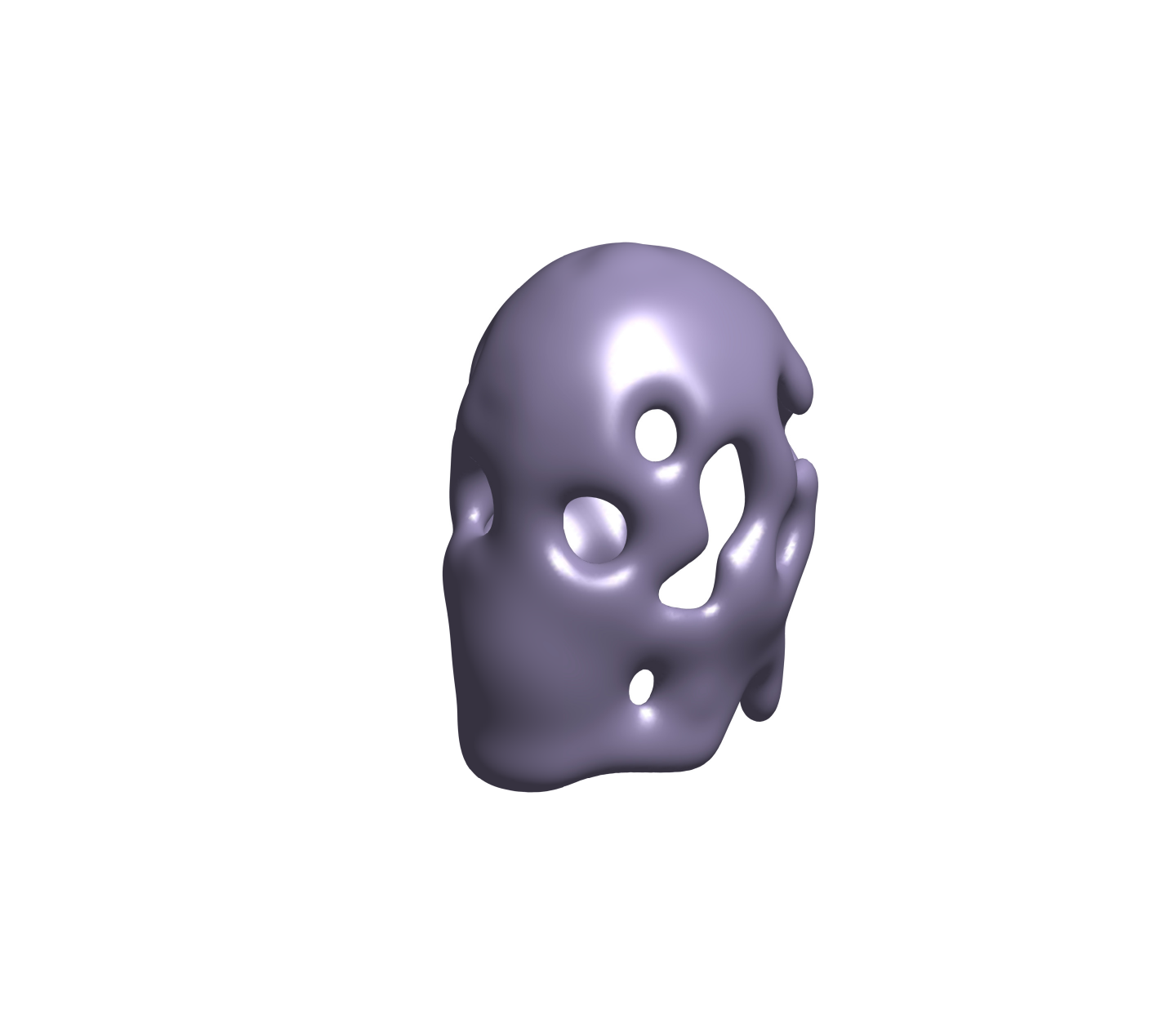}
	};
        
        \node at (0.3,1.4) {{\scriptsize{}}};

       \node at (-1.,0) {{\scriptsize{}}};
        \end{tikzpicture}
        \vspace{-0.6cm} \caption{}
        \label{p20230716_THz_Skull_S35G3_EE_23}
        \end{subfigure}
        \,
        \begin{subfigure}[b]{0.16\textwidth}
        \begin{tikzpicture}
        \node
        {
        \adjincludegraphics[width=\textwidth, trim={{.2\width} {.18\height} {.3\width} {.15\height}}, clip]{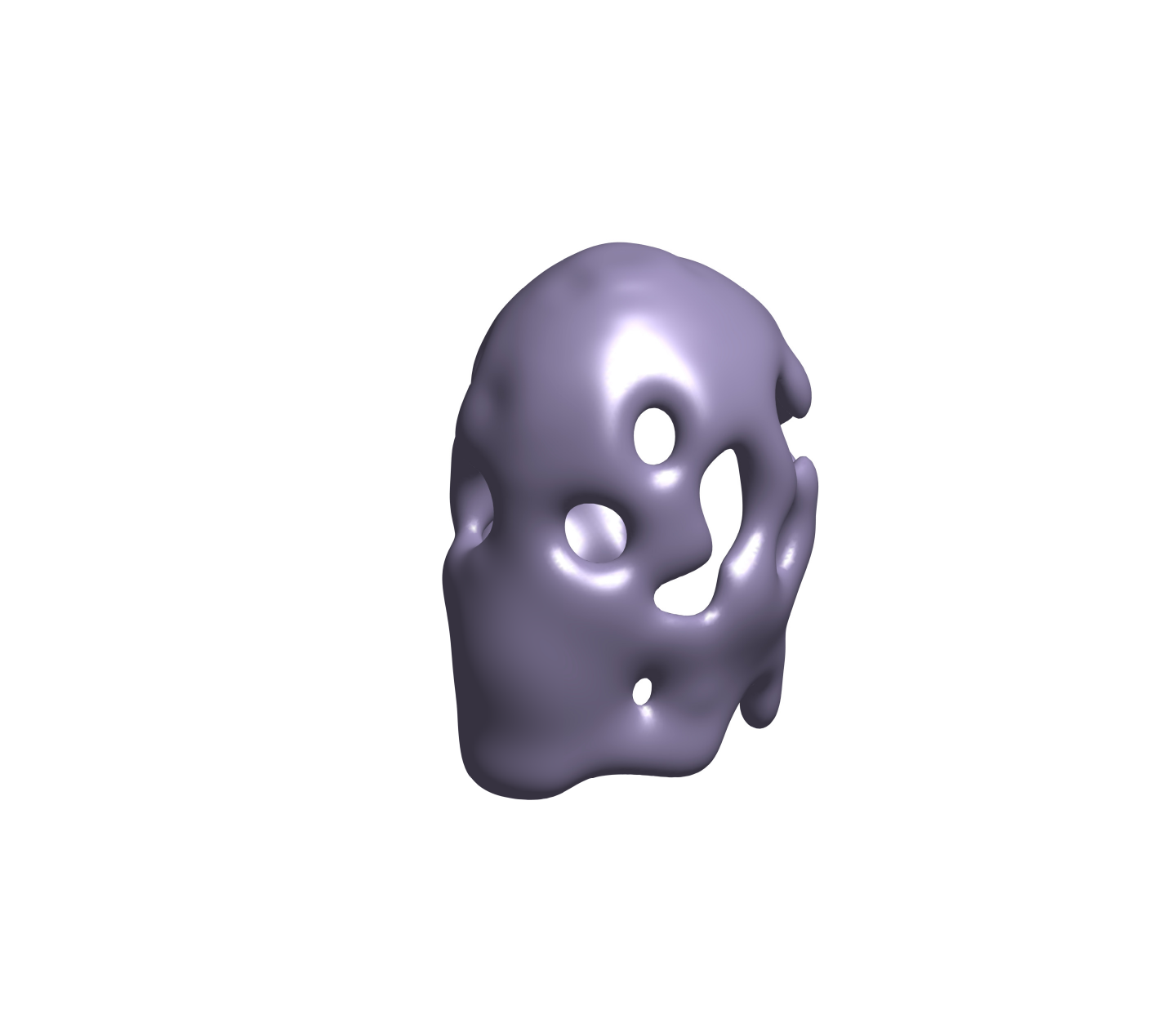}
	};
        
        \node at (0.3,1.4) {{\scriptsize{}}};

       \node at (-1.,0) {{\scriptsize{}}};
        \end{tikzpicture}
        \vspace{-0.6cm} \caption{}
        \label{p20230716_THz_Skull_S23G5_EE_35}
        \end{subfigure}
        % \vspace{-0.3cm}
        \caption{Illustrations of Skull object:~\subref{p20230718_THz_Skull_S138G0_W_1} conventional LR reconstruction with 138 slices where smooth SR reconstruction~\subref{p20230718_THz_Skull_S138G0_W_5} by Willmore-based $\mathscr{W_{\varepsilon}}$ and~\subref{p20230716_THz_Skull_S138G0_EE_5} by Euler-Elastica-based formulation $\mathscr{E_{\varepsilon}}$;~\subref{p20230718_THz_Skull_S69G1_W_7} input 69 slices with 1 gap where reconstruction~\subref{p20230718_THz_Skull_S69G1_W_11} by $\mathscr{W_{\varepsilon}}$ and~\subref{p20230716_THz_Skull_S69G1_EE_11} by $\mathscr{E_{\varepsilon}}$;~\subref{p20230718_THz_Skull_S35G3_W_19} input 35 slices with 3 gaps where reconstruction~\subref{p20230718_THz_Skull_S35G3_W_23} by $\mathscr{W_{\varepsilon}}$ and~\subref{p20230716_THz_Skull_S35G3_EE_23} by $\mathscr{E_{\varepsilon}}$;~\subref{p20230718_THz_Skull_S23G5_W_31} input 23 slices with 5 gaps where reconstruction~\subref{p20230718_THz_Skull_S23G5_W_35} by $\mathscr{W_{\varepsilon}}$ and~\subref{p20230716_THz_Skull_S23G5_EE_35} by $\mathscr{E_{\varepsilon}}$. }
        \label{3_Skull}
        \vspace{-0.1cm}
    \end{figure}

    Building on the conventional techniques for THz 3D CT imaging outlined in~\Cref{subsec:ConventionalTechniques}, we generate preliminary LR reconstructions by stacking the designated $s$ slices for three objects, as illustrated in~\cref{p20230712_THz_Deer_S218G0_W_1} ($s = 218$ for Deer),~\ref{p20230718_THz_Polarbear_S258G0_W_1} ($s = 258$ for Polarbear), and~\ref{p20230718_THz_Skull_S138G0_W_1} ($s = 138$ for Skull). 
    For the technical specifics, each slice is obtained via IRT from $\mathcal{R} = 60$ projections, adopting a rotational angle of 6 degrees. 
    Both $\mathcal{L}$ and $\mathcal{H}$ directions in each projection utilized the raster scanning with a step size of $\Delta \mathcal{L} = \Delta \mathcal{H} = 0.25$ mm. 
    The scanning ranges are set as: $\mathcal{L} = 50$ mm and $\mathcal{H} = 70$ mm for Deer; $\mathcal{L} = 60$ mm and $\mathcal{H} = 70$ mm for Polarbear; and both $\mathcal{L} = 40$ mm and $\mathcal{H} = 40$ mm for Skull. 
    Regarding our experiment using~\Cref{alg:Alternating_Direction_Method_of_Multipliers}, parameters are set to $\tau = \varepsilon^{3.5}$, $\rho = 1$, and $\varepsilon = 3/\max{(N_{x}, N_{y}, N_{z})}$ aligning with the maximum pixel number from the scanning resolution across the three axes of initial input $u_{0}$ where $N_{x} = N_{y} = \mathcal{L}/\Delta \mathcal{L}$ and $N_{z} = \mathcal{H}/\Delta \mathcal{H} \geq s$. 
    The algorithm terminates when the disparity between successive iterative outcomes falls below a designated threshold, with most processes concluding in fewer than ten iterations. 
    Subsequently, the smooth SR results can be reconstructed by the Willmore-based formulation $\mathscr{W}_{\varepsilon}$ ($\mathscr{m} = 0, \mathscr{n} = 1$) and by the Euler-Elastica-based formulation $\mathscr{E}_{\varepsilon}$ ($\mathscr{m} = 1, \mathscr{n} = 1$), as illustrated objects in~\cref{p20230712_THz_Deer_S218G0_W_5},\subref{p20230713_THz_Deer_S218G0_EE_5} for Deer,~\cref{p20230718_THz_Polarbear_S258G0_W_5},\subref{p20230717_THz_Polarbear_S258G0_EE_5} for Polarbear, and~\cref{p20230718_THz_Skull_S138G0_W_5},\subref{p20230716_THz_Skull_S138G0_EE_5} for Skull. 
    
    %% Deer

    %% Polarbear
    
    \noindent
    Furthermore, intuitively, fewer data are required for reconstruction, translating to reduced acquisition time. 
    To underscore the flexibility of our framework, we reconstruct using $\mathscr{W}_{\varepsilon}$ and $\mathscr{E}_{\varepsilon}$ with fewer input slices for reducing the data acquisition time. 
    By setting gaps at intervals of 1, 3, and 5, the results are illustrated in~\cref{p20230713_THz_Deer_S109G1_W_5}-\subref{p20230713_THz_Deer_S37G5_W_5},~\subref{p20230713_THz_Deer_S109G1_EE_11}-\subref{p20230713_THz_Deer_S37G5_EE_35} for Deer,~\cref{p20230718_THz_Polarbear_S129G1_W_11}-\subref{p20230718_THz_Polarbear_S43G5_W_35},~\subref{p20230717_THz_Polarbear_S129G1_EE_11}-\subref{p20230716_THz_Polarbear_S43G5_EE_35} for Polarbear, and~\cref{p20230718_THz_Skull_S69G1_W_11}-\subref{p20230718_THz_Skull_S23G5_W_35},~\subref{p20230716_THz_Skull_S69G1_EE_11}-\subref{p20230716_THz_Skull_S23G5_EE_35} for Skull. 
    All computational tasks are executed on  MATLAB\_R2023b\textsuperscript{\tiny\textregistered} running on macOS Sonoma 14.0 with an Apple M1 Max Chip and 64 GB Memory. 
    
    %% Skull

    \textcolor{revision_black}{
    In our implementation, we preserve the dimensional consistency between inputs and outputs for each object, maintaining isotropic resolution even when using fewer lateral samples. 
    Specifically, our algorithm records and processes all data within the LR 3D matrix space $\R^{3}_{N_{x} \times N_{y} \times N_{z}}$, ensuring a consistent dimensional framework throughout. 
    All results are then reconstructed using isosurface, a 3D triangular mesh representation well-suited for our framework. 
    This method allows for the precise reconstruction of details, effectively demonstrating the isotropic resolution capabilities of our SR algorithm.}
    \textcolor{revision_black}{
    However, due to the inconsistency of triangular mesh surfaces reconstructed by two formulations $\mathscr{W}_{\varepsilon}$ and $\mathscr{E}_{\varepsilon}$ in our BLIss (i.e. variations in the number of vertices and faces), a fair quantitative comparison is challenging. 
    To address this, we introduce a metric to evaluate the performance of surfaces reconstructed by both formulations, $\mathscr{W}_{\varepsilon}$ and $\mathscr{E}_{\varepsilon}$, serving as the indicator of the global smoothness of the surfaces. 
    This metric is to calculate the standard deviation of Gaussian curvatures (GC) $\sigma_{\mbox{\tiny{GC}}}$ and of mean curvatures (MC) $\sigma_{\mbox{\tiny{MC}}}$ based on the theory from discrete geometry (details in~\hyperref[supp]{Sec. 4A of Supplement 1}). 
    In~\Cref{tab:std_GC_MC}, we provide the values of $\sigma_{\mbox{\tiny{GC}}}$ and $\sigma_{\mbox{\tiny{MC}}}$ for various input slices and gap(s). 
    For identical input slices, both $\sigma_{\mbox{\tiny{GC}}}$ and $\sigma_{\mbox{\tiny{MC}}}$ metrics indicate that results obtained using $\mathscr{E}_{\varepsilon}$ exhibit superior smoothness when compared to $\mathscr{W}_{\varepsilon}$. 
    This is attributed to the ability of $\mathscr{E}_{\varepsilon}$ to automatically handle topologically complex geometries, which overcomes the limitations of $\mathscr{W}_{\varepsilon}$. 
    Note that a lower value in these curvature computations signifies better smoothness, indicating less variability in the curvature. 
    Besides, as the number of input slices decreases for the same object, both $\sigma_{\mbox{\tiny{GC}}}$ and $\sigma_{\mbox{\tiny{MC}}}$ values also decrease using the same formulation. 
    This trend implies that a reduced number of input slices leads to the loss of finer details in the reconstruction. 
    Even though our framework visually demonstrates its capacity for reconstructed results using limited input slices, we still require accuracy verification to confirm the fidelity compared to the reconstructed results using full input slices by the same formulation. 
    To achieve this accuracy verification, we employ the multi-scale structural similarity index measure (MS-SSIM) to ascertain the precision of reconstructions from sparse slices (details in~\hyperref[supp]{Sec. 4B of Supplement 1}). 
    Then, the result obtained with full input slices serves as the benchmark for these evaluations, which is indicated as one in~\Cref{tab:std_GC_MC}. 
    When comparing results using the same formulation with different input slices, the MS-SSIM values consistently fall within the range of 0.97 to 0.99. 
    These high MS-SSIM scores underscore the 3D structural fidelity of our reconstructions, even when using only $1/6$ of the original slices in our BLIss.}
    Additionally, we present the average elapsed time for each iteration during the reconstruction of the three objects in~\Cref{tab:std_GC_MC}. 
    Here, our framework achieves an average elapsed time per iteration within the scale of seconds to show the efficiency of our proposed BLIss. 

    \pagebreak
    
    %% GC & MC & MS-SSIM
    \begin{table}[htbp]
    \caption{Comparison metrics for three reconstructed objects Deer, Polarbear, and Skull with various input slices and gap(s) by Willmore-based $\mathscr{W}_{\varepsilon}$ and Euler-Elastica-based $\mathscr{E}_{\varepsilon}$ formulations: the standard deviation of Gaussian curvatures $\sigma_{\mbox{\tiny{GC}}}$ and of mean curvatures $\sigma_{\mbox{\tiny{MC}}}$; the average elapsed time of each iteration (seconds/iteration or s/iter); and the multi-scale structural similarity index measure (MS-SSIM) where $\uparrow$ ($\downarrow$): higher (lower) is better. }
    \vspace{-0.5cm}
    \label{tab:std_GC_MC}
    \begin{center}
    \begin{adjustbox}{width=.98\textwidth}
    \begin{tabular}{ccccc|cc|cc|cc}
    \toprule
    & \multirow{2}{*}{Slices} & \multirow{2}{*}{Gap(s)} & \multicolumn{2}{c|}{$\sigma_{\mbox{\tiny{GC}}}$ ($\downarrow$)} & \multicolumn{2}{c|}{$\sigma_{\mbox{\tiny{MC}}}$ ($\downarrow$)} & \multicolumn{2}{c|}{s/iter} & \multicolumn{2}{c}{MS-SSIM ($\uparrow$)} \\ \cline{4-11} 
     &  &  & $\mathscr{W}_{\varepsilon}$ & $\mathscr{E}_{\varepsilon}$ & $\mathscr{W}_{\varepsilon}$ & $\mathscr{E}_{\varepsilon}$ & $\mathscr{W}_{\varepsilon}$ & $\mathscr{E}_{\varepsilon}$ & \multicolumn{1}{c|}{$\mathscr{W}_{\varepsilon}$} & $\mathscr{E}_{\varepsilon}$ \\ \hline
    \multirow{4}{*}{Deer} & 218 & 0 & 6639.6581 & 561.2450 & 29.8726 & 15.9714 & 0.7083 & 0.7163 & \multicolumn{1}{c|}{1} & 1 \\
     & 109 & 1 & 1208.2426 & 520.0736 & 20.8238 & 15.3831 & 0.7394 & 0.7372 & \multicolumn{1}{c|}{0.9996} & 0.9991 \\
     & 55 & 3 & 890.3065 & 511.6518 & 20.2839 & 15.3646 & 0.7149 & 0.6943 & \multicolumn{1}{c|}{0.9976} & 0.9964 \\
     & 37 & 5 & 860.5110 & 488.6312 & 19.3400 & 14.9879 & 0.7109 & 0.6571 & \multicolumn{1}{c|}{0.9946} & 0.9913 \\ \hline
    \multirow{4}{*}{Polarbear} & 258 & 0 & 3158.3526 & 480.5970 & 22.8052 & 14.0061 & 0.9686 & 0.5814 & \multicolumn{1}{c|}{1} & 1 \\
     & 129 & 1 & 1537.6481 & 449.3327 & 17.8706 & 13.2327 & 0.6762 & 0.5755 & \multicolumn{1}{c|}{0.9992} & 0.9982 \\
     & 65 & 3 & 1065.3076 & 359.3619 & 16.7654 & 12.1427 & 0.6113 & \multicolumn{1}{c|}{0.5669} & \multicolumn{1}{c|}{0.9946} & 0.9916 \\
     & 43 & 5 & 915.8394 & 326.9832 & 16.3786 & 12.0142 & 0.7066 & 0.5635 & \multicolumn{1}{c|}{0.9864} & 0.9810 \\ \hline
    \multirow{4}{*}{Skull} & 138 & 0 & 3659.2833 & 1492.1212 & 18.3170 & 17.7190 & 0.1843 & 0.1524 & \multicolumn{1}{c|}{1} & 1 \\
     & 69 & 1 & 1359.3779 & 1326.7824 & 18.0419 & 15.1627 & 0.2094 & 0.1668 & \multicolumn{1}{c|}{0.9996} & 0.9945 \\
     & 35 & 3 & 1174.7408 & 1132.0628 & 16.7263 & 13.1944 & 0.2164 & 0.1484 & \multicolumn{1}{c|}{0.9952} & 0.9862 \\
     & 23 & 5 & 924.0850 & 708.6946 & 15.9234 & 12.8667 & 0.2062 & 0.1527 & \multicolumn{1}{c|}{0.9855} & 0.9709 \\
    \bottomrule
    \end{tabular}
    \end{adjustbox}
    \end{center}
    \end{table}
    \vspace{-0.3cm}

\begin{figure}[htbp]
    \centering
        \begin{subfigure}[b]{0.3\textwidth}
        \includegraphics[width=\textwidth]{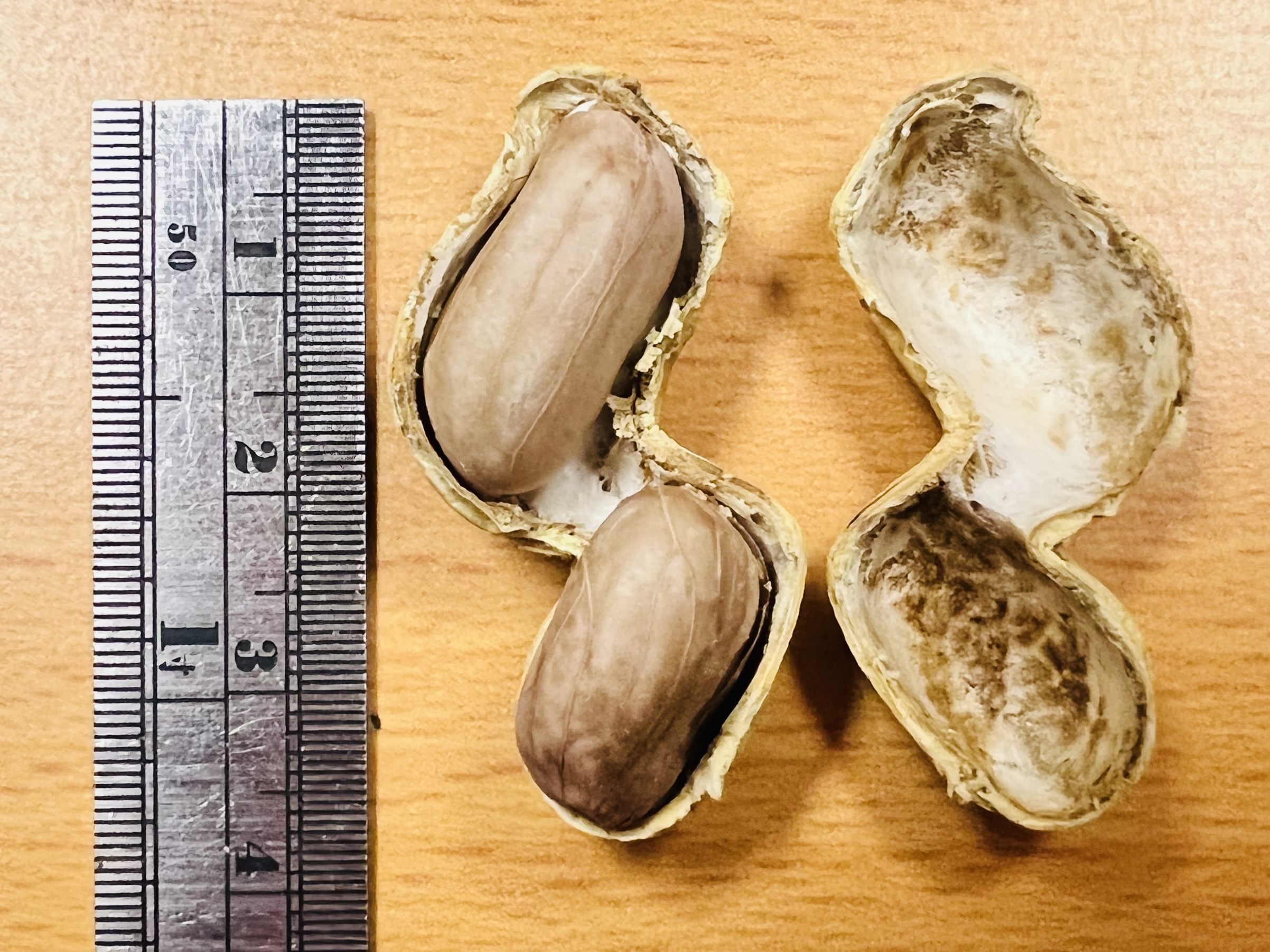}
        \caption{}
        \label{4_Peanut2}
        \end{subfigure}
        \quad
        \begin{subfigure}[b]{0.23\textwidth}
        \includegraphics[width=\textwidth]{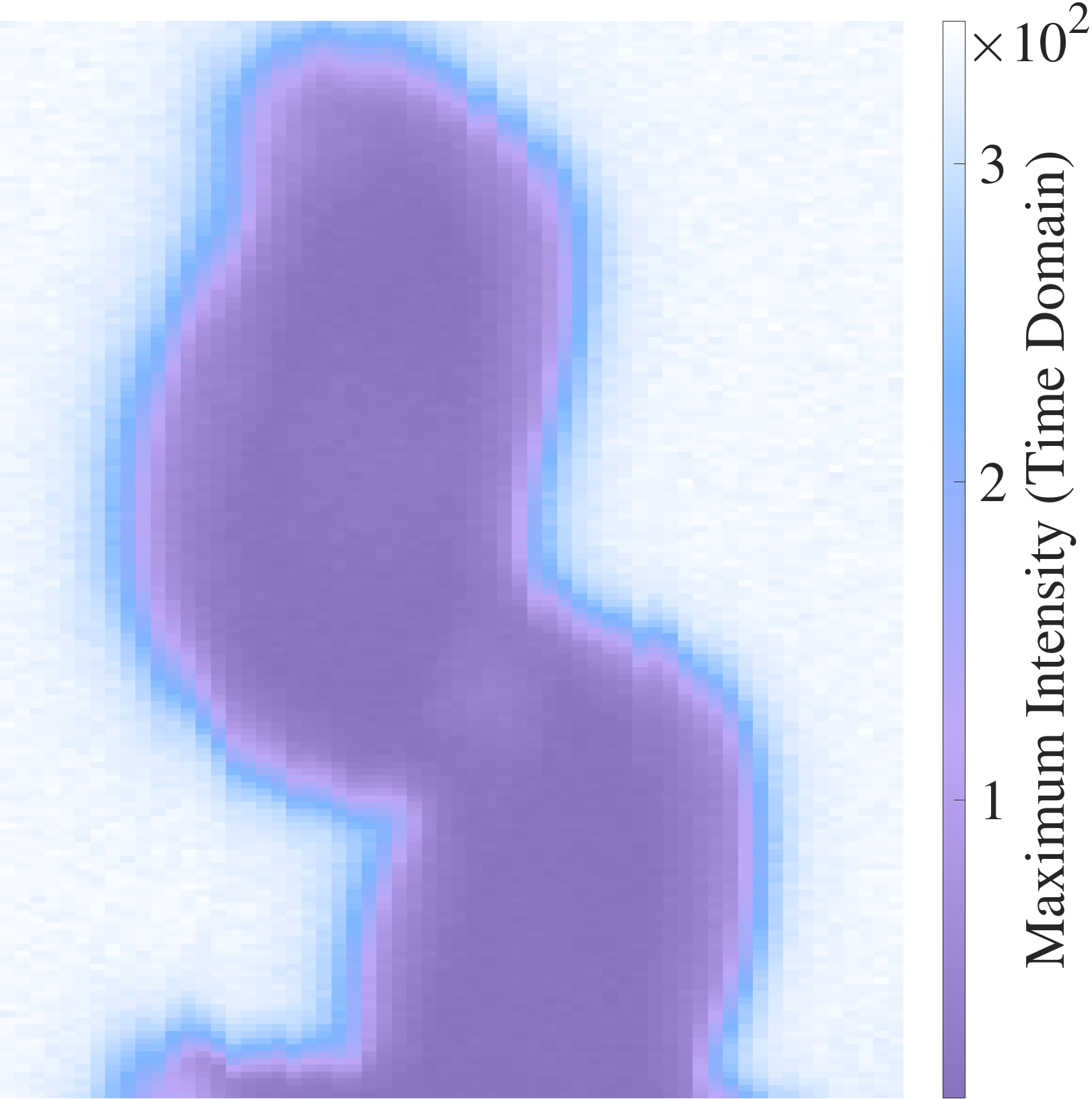}
        \caption{}
        \label{p20230810_THz_Peanut2_2D_1}
        \end{subfigure}
        \quad
        \begin{subfigure}[b]{0.18 \textwidth}\includegraphics[width=\textwidth]{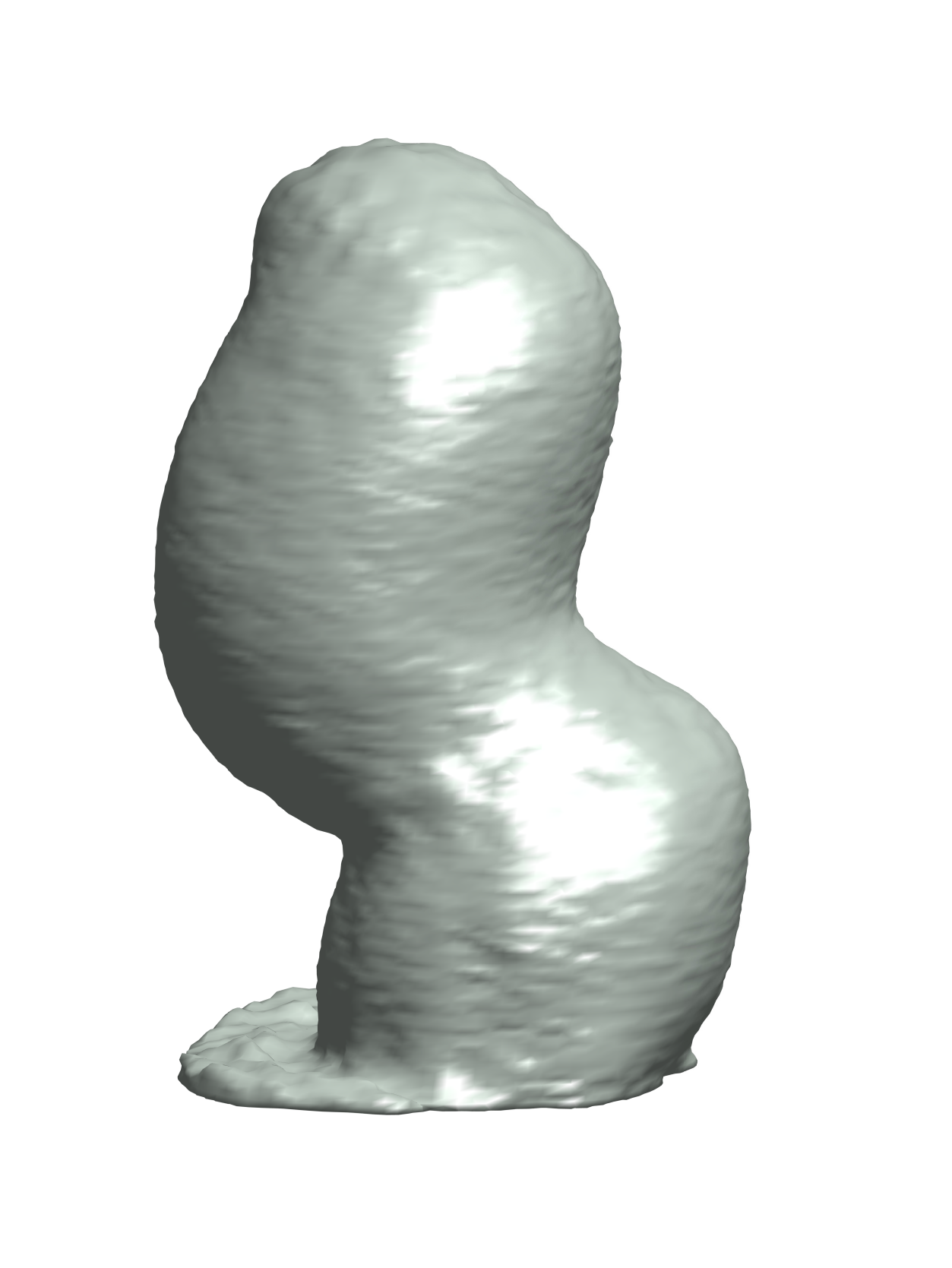}
        \caption{}
        \label{p20230810_THz_Peanut2_3D_IRT_3}
        \end{subfigure}
        \\
        \begin{subfigure}[b]{0.23\textwidth}
        \includegraphics[width=\textwidth]{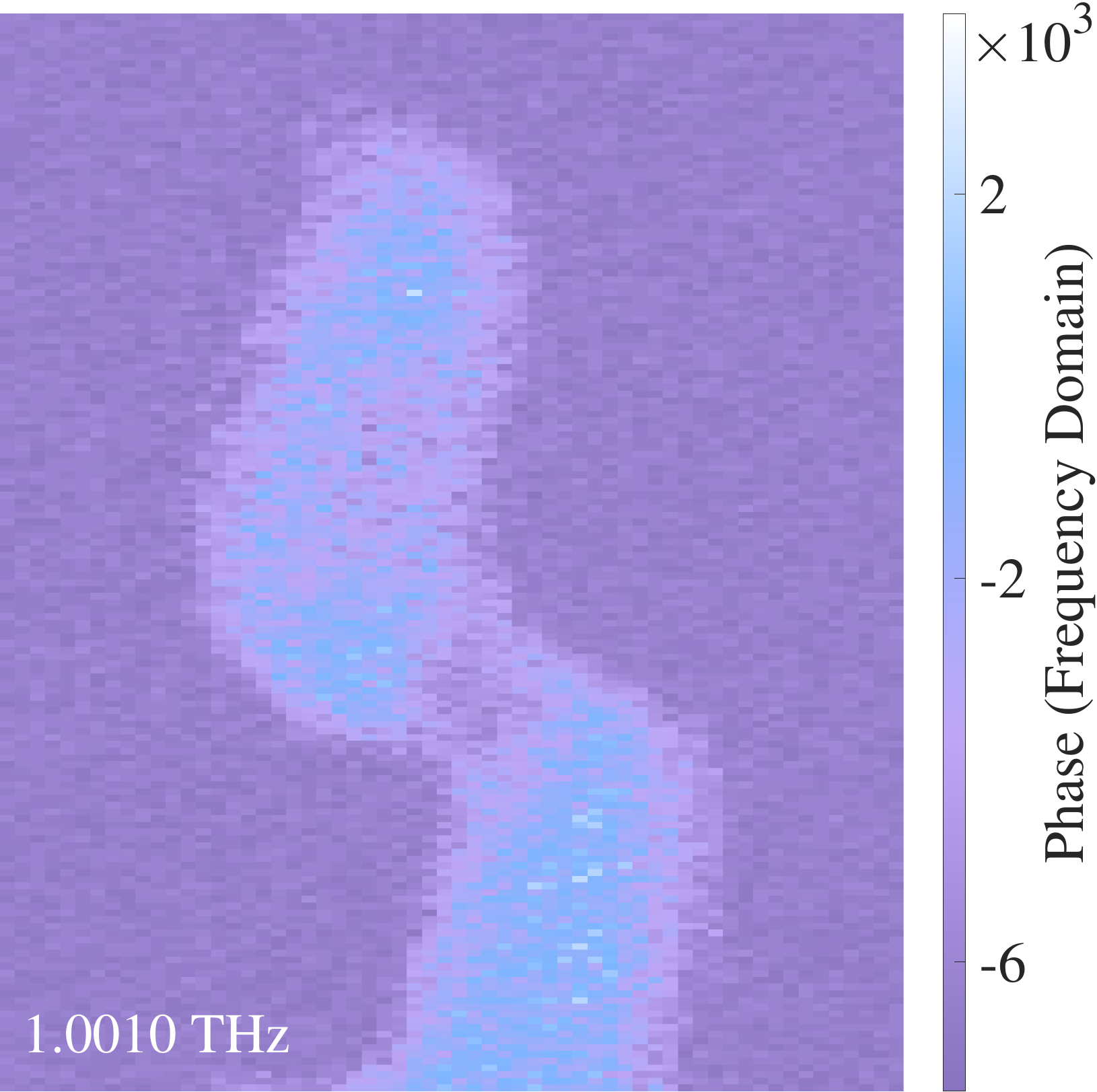}
        \caption{}
        \label{p20230810_THz_Peanut2_2D_2}
        \end{subfigure}
        \,
        \begin{subfigure}[b]{0.23\textwidth}
        \includegraphics[width=\textwidth]{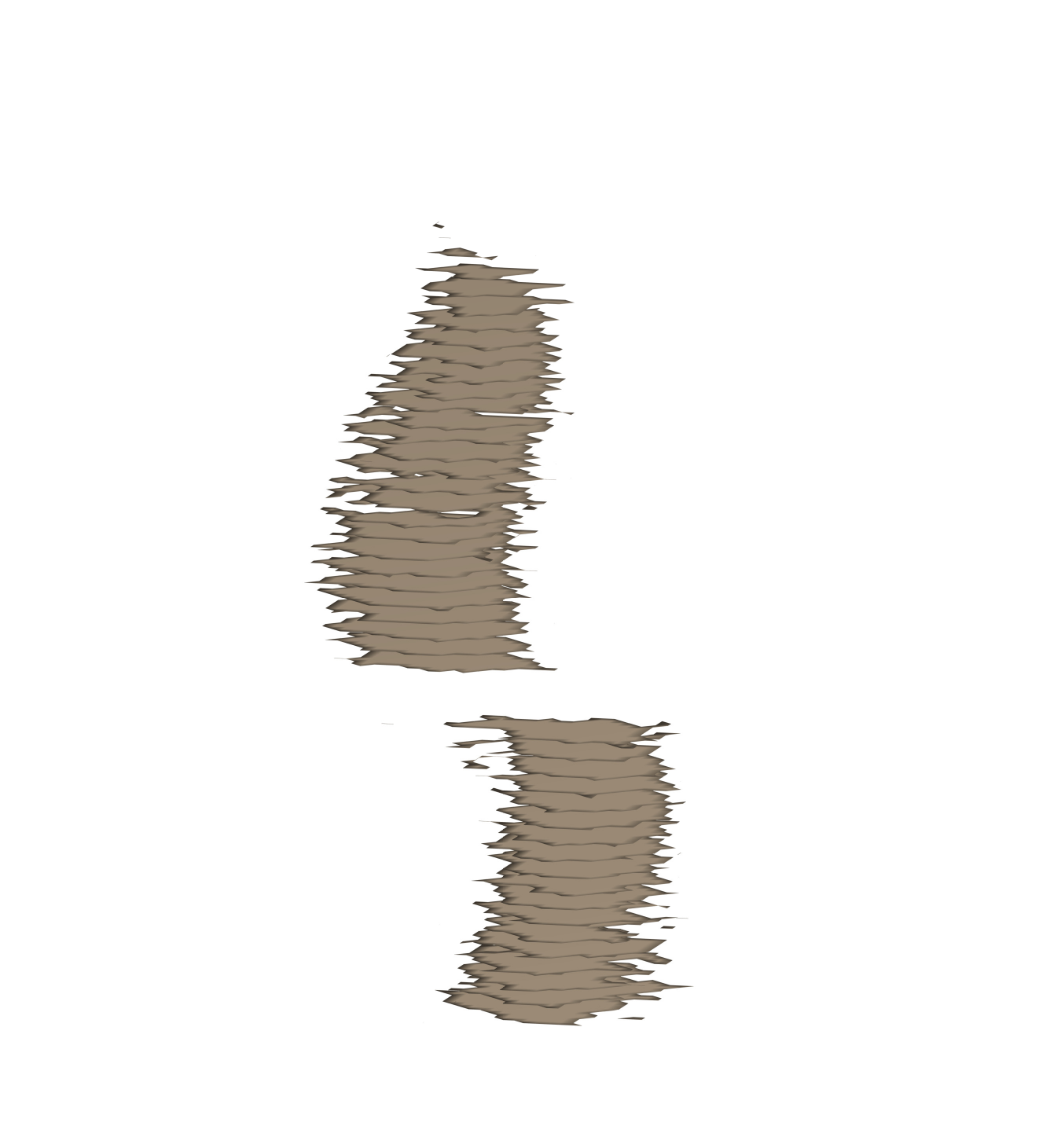}
        \caption{}
        \label{p20230810_THz_Peanut2_hyper_S160G0_Evo_text_2}
        \end{subfigure}
        \,
        \begin{subfigure}[b]{0.23\textwidth}\includegraphics[width=\textwidth]{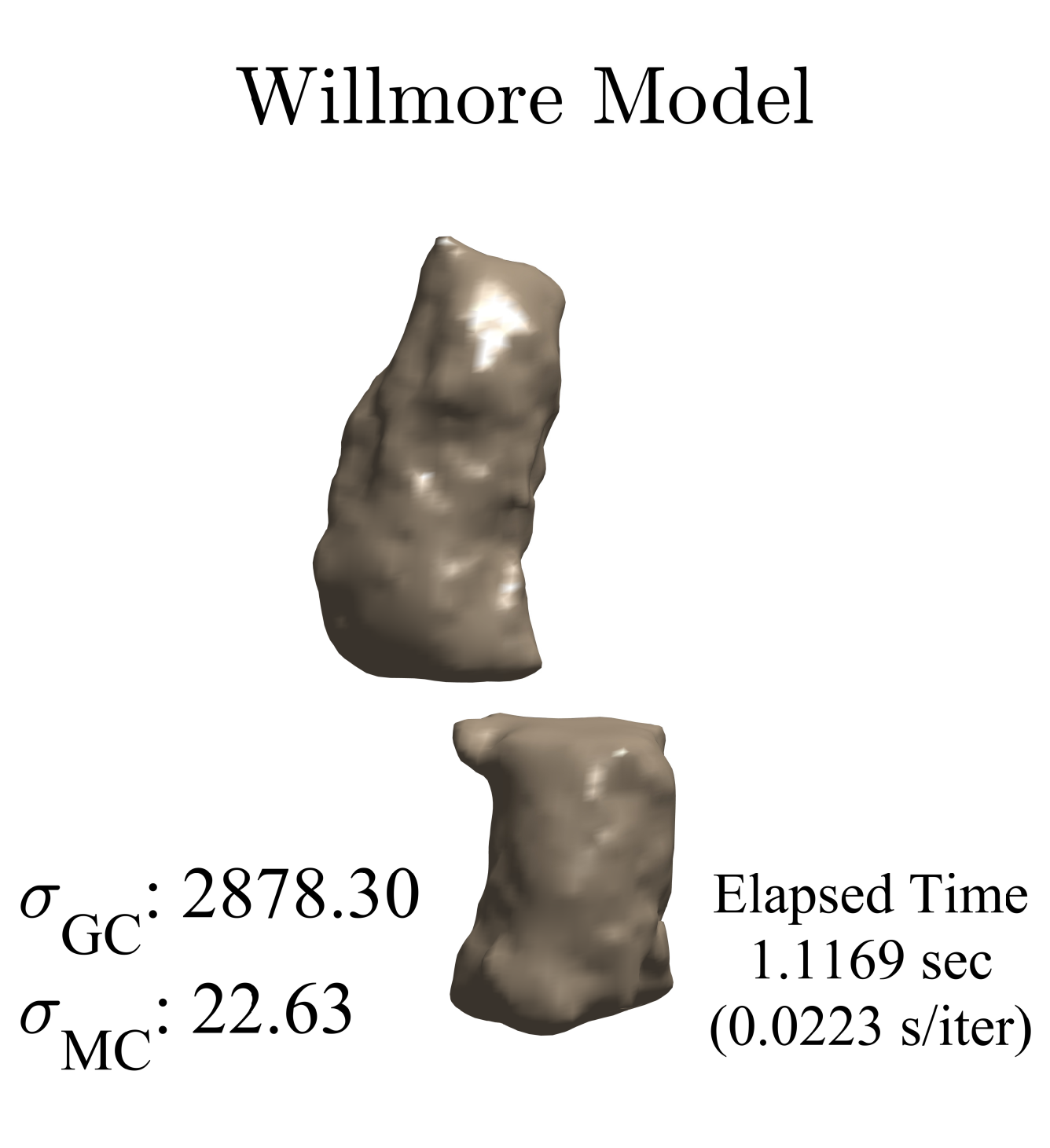}
        \caption{}
        \label{p20230810_THz_Peanut2_hyper_S160G0_Evo_text_6}
        \end{subfigure}
        \,
        \begin{subfigure}[b]{0.23\textwidth}\includegraphics[width=\textwidth]{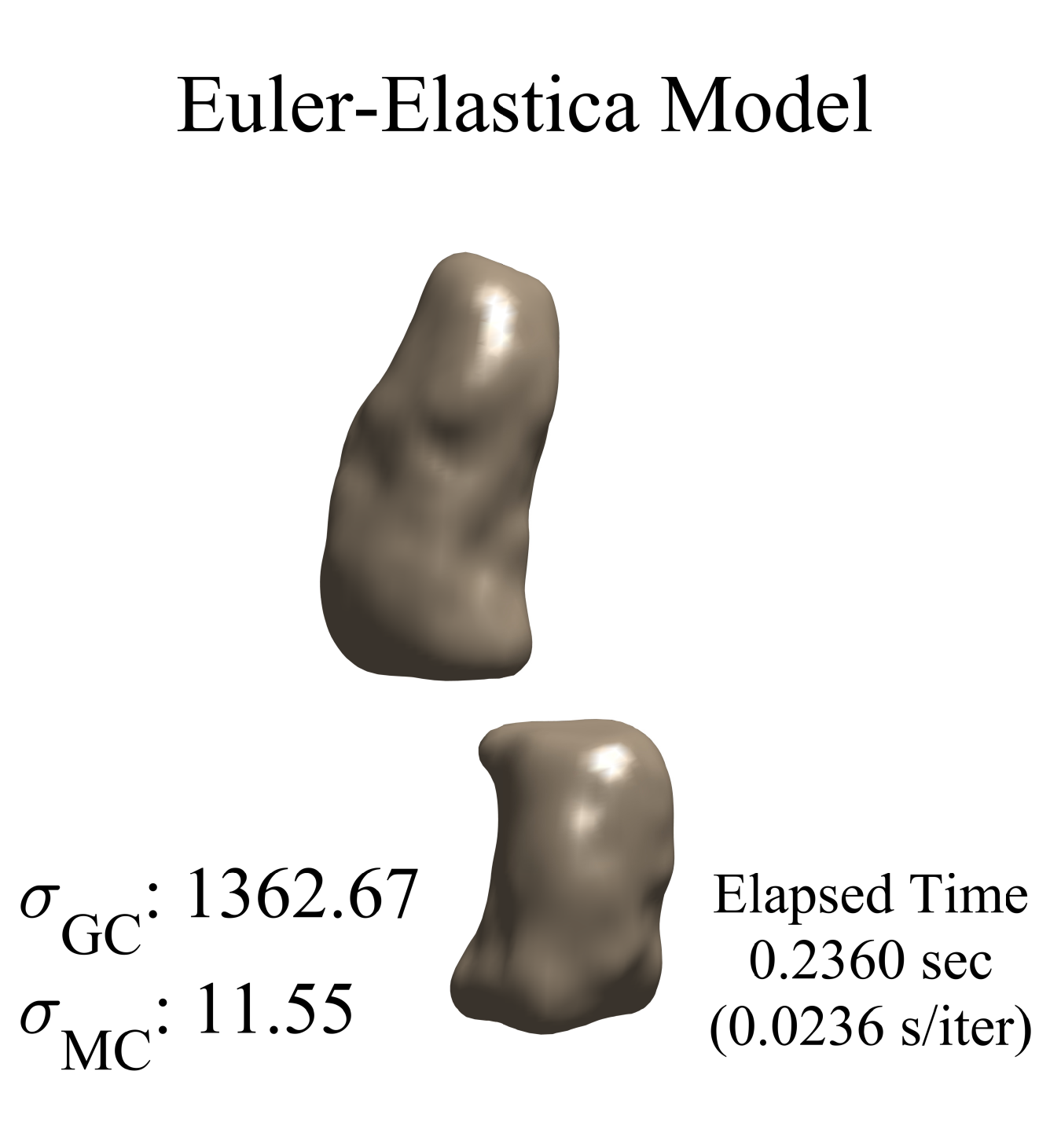}
        \caption{}
        \label{p20230810_THz_Peanut2_hyper_S160G0_Evo_text_7}
        \end{subfigure}
        \caption{Illustrations of~\subref{4_Peanut2} the optical image of the unveiled peanut kernels. ~\subref{p20230810_THz_Peanut2_2D_1} projected Time-MAX THz 2D image. ~\subref{p20230810_THz_Peanut2_3D_IRT_3} reconstructed THz 3D image of peanuts from~\subref{p20230810_THz_Peanut2_2D_1}.~\subref{p20230810_THz_Peanut2_2D_2} representative projection of unwrapped phase at 1.0010 THz.~\subref{p20230810_THz_Peanut2_hyper_S160G0_Evo_text_2} 46 slices extracted from~\subref{p20230810_THz_Peanut2_2D_2} by IRT.~\subref{p20230810_THz_Peanut2_hyper_S160G0_Evo_text_6} reconstruction of peanut kernels by Willmore model $\mathscr{W}_{\varepsilon}$ from~\subref{p20230810_THz_Peanut2_hyper_S160G0_Evo_text_2} with higher computational time (more iterations) and lower smoothness. ~\subref{p20230810_THz_Peanut2_hyper_S160G0_Evo_text_7} optimized reconstruction of kernels by Euler-Elastica model $\mathscr{E}_{\varepsilon}$ from~\subref{p20230810_THz_Peanut2_hyper_S160G0_Evo_text_2} with lower computational time (less iterations) and higher smoothness. }
        \label{4_Peanut}
    \end{figure}

    \textcolor{revision_black}{
    In expanding the discussion on our findings, we emphasize the real-world applications and contributions of our research to the field of THz imaging. 
    Our results pave the way for advanced diagnostic and analytical techniques in non-invasive medical imaging, material characterization, and security screening~\cite{Cooper_2008_security, Cooper_2011_security, Cheng_2020_Security, Takida_2021_Security, Mittleman_2018_THzImaging_Review, Leitenstorfer_2023_Review_Roadmap}. 
    By enhancing the resolution and fidelity of THz imaging, our approach could significantly improve the detection and analysis of bioinformatics, the multi-scale information inside materials, and the inspection of art and historical artifacts without causing damage. 
    These applications underscore the potential of our findings to revolutionize various industries by providing safer, more accurate, and more detailed imaging capabilities.}
    To illustrate the practical application of this research, we have chosen a peanut as our test subject. 
    The peanut exhibits a textured surface and contains two inner seeds \textcolor{revision_black}{along with an air gap.}
    Refer to the optical image of the halved peanut parts after data acquisition in~\cref{4_Peanut2}. 
    We first present the results in~\cref{p20230810_THz_Peanut2_2D_1}-\ref{p20230810_THz_Peanut2_3D_IRT_3}, using the Time-MAX information to reconstruct the outer shell. 
    Furthermore, our approach can be adapted to other data types obtained from the time-resolved THz pulse, which contains hyperspectral information in a broad frequency range.
    One notable aspect is the phase in the frequency domain, which offers distinct advantages, such as depth information, material properties, and better image contrast. 
    Therefore, we derive the phase information in the frequency domain by 
    $
    \Phi(\omega) = \tan^{-1}(b/a), 
    $
    from the Fourier amplitude 
    $
    \mathcal{F}(f(t)) = F(\omega) = a(\omega) + ib(\omega) = \int f(t) e^{-2 \pi i \omega t} \, \mathrm{d}t
    $
    of the time-resolved signals $f(t)$ by Fourier Transform (FT) $\mathcal{F}$. 
    A representative projection of the unwrapped phase at a chosen frequency of 1.0010 THz is illustrated in~\cref{p20230810_THz_Peanut2_2D_2}. 
    The selected frequency is determined by the aim of enhancing material contrast, focusing on fat content for the peanut kernels and fiber for the outer shell, within the THz frequency range. 
    When applying our framework with only $1/3$ of the input slices, reducing 3-fold data acquisition time, as shown in~\cref{p20230810_THz_Peanut2_hyper_S160G0_Evo_text_2}, we obtain the SR results in~\cref{p20230810_THz_Peanut2_hyper_S160G0_Evo_text_6} using $\mathscr{W}_{\varepsilon}$ and~\cref{p20230810_THz_Peanut2_hyper_S160G0_Evo_text_7} using $\mathscr{E}_{\varepsilon}$. 
    Notably, results produced by $\mathscr{E}_{\varepsilon}$ exhibit superior smoothness, as indicated by lower values of $\sigma_{\mbox{\tiny{GC}}}$ and $\sigma_{\mbox{\tiny{MC}}}$. 
    Additionally, the reconstruction process can be completed in seconds using a  commonplace laptop, reaffirming the precision and efficiency of our framework with very limited dataset inputs. 
    
    %% Peanut2

\section{Conclusion}
    \textcolor{revision_black}{
    In our pursuit of advancing pulsed THz 3D SR reconstruction, we propose a promising variational framework based on the $\mathscr{W}_{\varepsilon}$ and an adapted Euler-Elastica-based formulation $\mathscr{E}_{\varepsilon}$. 
    With our 3D representation utilizing the implicit phase-field function in conjunction with the designed variational framework, we achieve super-resolution results, surpassing the resolution limitations imposed by coarse input slices due to scanning resolution (see~\hyperref[supp]{Sec. 5 of Supplement 1}). 
    Furthermore, this framework is versatile, capable of processing a reduced number of LR input slices, speeding up both data acquisition and computational processes. 
    The adaptability to a range of data types boosts its value, particularly in the context of pulsed THz imaging.}
    To substantiate our results and provide a quantitative assessment of their quality, we employ two metrics: global smoothness, measured by the standard deviation of Gaussian curvatures $\sigma_{\mbox{\tiny{GC}}}$ and of mean curvatures $\sigma_{\mbox{\tiny{MC}}}$, and accuracy verification using MS-SSIM. 
    \textcolor{revision_black}{
    The capabilities of this framework are illustrated through} its application to non-destructive THz imaging of peanuts, where it significantly improved the precision and clarity of capturing the internal structure. 
    
    Our work makes substantial strides in addressing long-standing challenges within THz 3D imaging, which have been intricately linked to wave propagation physics and hardware limitations. 
    This advancement has the potential to transform the landscape of THz imaging, expanding its utility across both research and industry. 
    Moreover, \textcolor{revision_black}{its implications extend to} other domains, including X-ray and MRI imaging. 
    By unlocking the capability to achieve higher resolution with sparse inputs, we are paving the way to unveil previously hidden details, offering profound insights into materials and processes that were \textcolor{revision_black}{once} enigmatic. 

    \pagebreak
    
    While our findings lay a foundation for improved THz 3D reconstruction, the voyage of discovery is far from complete. 
    One promising avenue involves adapting our framework to a broader range of materials and applications, extending its universal applicability. 
    Beyond this, \textcolor{revision_black}{integrating deep learning approaches holds untapped potential to} further optimize the imaging process, making it more adaptive and intuitive. 
    Additionally, as hardware technology advances, it could be seamlessly integrated with our approach, further elevating resolution and image clarity (see~\hyperref[supp]{Sec. 6 of Supplement 1}). 
    In the ever-evolving realm of THz imaging, consider our work as the cornerstone, setting the stage for future layers of transformative innovation. 

\begin{backmatter}
\bmsection{Funding}National Science and Technology Council, Taiwan (NSTC 112-2221-E-007-089-MY3). 
% S.-H. Yang expresses thanks for the support from the National Science and Technology Council (NSTC 112-2221-E-007-089-MY3). 
\bmsection{Acknowledgments}The first author is grateful for partial support from the UoL-NTHU Dual PhD Programme. 

\bmsection{Disclosures}The authors declare no conflicts of interest. 

\bmsection{Data Availability}Data underlying the results presented in this paper are not publicly available at this time but may be obtained from the authors upon reasonable request. 
\textcolor{revision_black}{Demo codes for BLIss are available in Ref.~\cite{Zhang_2024_BLIss_Github}. }

\bmsection{Supplemental document}\phantomsection\label{supp}See \href{https://doi.org/10.6084/m9.figshare.24455206}{Supplement 1} for supporting content. 

\end{backmatter}

%%%%%%%%%%%%%%%%%%%%%%% References %%%%%%%%%%%%%%%%%%%%%%%%%

%%%%%%%%%% If using BibTeX:
\bibliography{2_refs.bib}

\end{document}